\documentclass[12pt,a4paper]{article}

\usepackage{amssymb,amsfonts,amsmath}
\usepackage{cite,enumerate,float}
\usepackage{color}
\usepackage{mathrsfs}
\usepackage{cancel}

\usepackage{tikz}
\usetikzlibrary{arrows}
\usetikzlibrary{arrows.meta}
\usetikzlibrary{positioning}
\usetikzlibrary{shapes,snakes}
\usetikzlibrary{fit}
\usetikzlibrary{decorations.pathmorphing,decorations.pathreplacing,decorations.markings}
\usetikzlibrary{calc}

\usepackage{xcolor}
\definecolor{burgundy}{rgb}{0.5, 0.0, 0.13}
\definecolor{olive}{rgb}{0.50, 0.50, 0.0}
\usepackage[linktocpage=true,colorlinks=true,linkcolor=burgundy,citecolor=black!20!blue,urlcolor=black!20!violet]{hyperref}

\usepackage[most]{tcolorbox}

\def\CL {{\cal L}}

\def\CN {{\cal N}}

\def\IC{\mathbb{C}}

\def\IR{{\mathbb{R}}}

\def\IZ{{\mathbb{Z}}}

\def\I{{\rm i}}

\def\bpi{{\boldsymbol{\pi}}}

\def\lm{\limits}
\def\nn{\nonumber}

\def\p{\partial}
\def\Tr{{\rm Tr}}

\DeclareSymbolFont{bbsymbol}{U}{bbold}{m}{n}
\DeclareMathSymbol{\bbzero}{\mathbin}{bbsymbol}{"30}
\DeclareMathSymbol{\bbone}{\mathbin}{bbsymbol}{"31}
\DeclareMathSymbol{\bbtwo}{\mathbin}{bbsymbol}{"32}
\DeclareMathSymbol{\bbthree}{\mathbin}{bbsymbol}{"33}
\DeclareMathSymbol{\bbfour}{\mathbin}{bbsymbol}{"34}
\DeclareMathSymbol{\bbfive}{\mathbin}{bbsymbol}{"35}
\DeclareMathSymbol{\bbsix}{\mathbin}{bbsymbol}{"36}
\DeclareMathSymbol{\bbseven}{\mathbin}{bbsymbol}{"37}
\DeclareMathSymbol{\bbeight}{\mathbin}{bbsymbol}{"38}
\DeclareMathSymbol{\bbnine}{\mathbin}{bbsymbol}{"39}

\newcommand\sqbox[1]{{
	\setbox0=\hbox{\mbox{$\Box$}}
	\setbox1=\hbox{\mbox{\raisebox{0.35ex}{\tiny #1}}}
	\mbox{\raisebox{-0.2ex}{\rlap{\hbox to \wd0{\hss{\box1}\hss}}\box0}}
}}
\newcommand\Kappa{\mathrm{K}}
\def\myblue{white!40!blue}
\def\mygreen{black!60!green}
\def\myh{{\mathsf{h}}}

\tcbset{boxrule=0.6pt,colback=white!97!blue}

\numberwithin{equation}{section}

\begin{document}

\pagenumbering{Alph} 
\begin{titlepage}
	
	\vskip 1.5in
	\begin{center}
		
		{\bf\Large{Charging solid partitions}}
		\vskip 1cm 
		\renewcommand{\thefootnote}{\fnsymbol{footnote}}
		{Dmitry Galakhov$^{1,3,}$\footnote[2]{e-mail: d.galakhov.pion@gmail.com; galakhov@itep.ru} and Wei Li$^{2,}$\footnote[3]{e-mail: weili@mail.itp.ac.cn}} 
		\vskip 0.2in 
		\renewcommand{\thefootnote}{\roman{footnote}}
		{\small{ 
				\textit{$^1$Institute for Information Transmission Problems,}\vskip -.4cm
				\textit{ Moscow, 127994, Russia}
				\vskip 0 cm 
				\textit{$^2$Institute for Theoretical Physics, Chinese Academy of Sciences}\vskip -.4cm
				\textit{ Beijing, 100190, China}
				\vskip 0 cm 
				\textit{$^3$NRC ``Kurchatov Institute'', }\vskip -.4cm
				\textit{Moscow, 123182, Russia}
		}}
	\end{center}
	
	\vskip 0.5in
	\baselineskip 16pt
	
	\begin{abstract}
Solid partitions are the 4D generalization of the plane partitions in 3D and Young diagrams in 2D, and they can be visualized as stacking of 4D unit-size boxes in the positive corner of a 4D room.
		Physically, solid partitions arise naturally as 4D molten crystals that count equivariant D-brane BPS states on the simplest toric Calabi-Yau fourfold, $\IC^4$, generalizing the 3D statement that plane partitions count equivariant D-brane BPS states on $\IC^3$. 
		In the construction of BPS algebras for toric Calabi-Yau threefolds, the so-called charge function on the 3D molten crystal is an important ingredient -- it is the generating function for the eigenvalues of an infinite tower of Cartan elements of the algebra.
		In this paper, we derive the charge function for solid partitions. 
		Compared to the 3D case, the new feature is the appearance of contributions from certain 4-box and 5-box clusters, which will make the construction of the corresponding BPS algebra much more complicated than in the 3D. 
		
	\end{abstract}
	
	\date{August, 2021}
\end{titlepage}
\pagenumbering{arabic} 

\newpage
\tableofcontents
\newpage

\section{Introduction and summary}
\label{sec:IntroSummary}
It has been proposed by Harvey and Moore that it is possible to endow the space of BPS states (for systems with enough supersymmetries) with an algebraic structure \cite{Harvey:1995fq, Harvey:1996gc} and define a \textit{BPS algebra}.
For the IIA string on toric Calabi-Yau threefolds, the BPS algebra of the $\frac{1}{2}$-BPS states was first constructed in the form of a cohomological Hall algebra (CoHA) in \cite{Kontsevich:2010px}. 
Using the description of these BPS states in terms of 3D molten crystals \cite{Ooguri:2009ijd,Ooguri:2010yk}, one can also bootstrap their BPS algebras in the form of quiver Yangians \cite{Li:2020rij}.
For simple cases, it has been demonstrated that the latter is the Drinfeld double of the former \cite{Rapcak:2018nsl,Rapcak:2021hdh}.

A natural question is how to generalize to toric Calabi-Yau four-folds (CY${}_4$), namely, to construct BPS algebras describing BPS D-branes on CY${}_4$.
However, this is a much harder question. 
To illustrate the difficulty, it is enough to compare the simplest cases in these two dimensions: $\mathbb{C}^3$ and $\mathbb{C}^4$.
First of all, the D6-D0-branes in $\mathbb{C}^3$ can be described in terms of plane partitions (a 3D generalization of the 2D Young diagrams), whereas the D8-D0-branes in $\mathbb{C}^4$ are described by the 4D version and are called \textrm{solid partitions}.

To start with, solid (4D) partitions are notorious for the difficulty of writing down their generating function, which is presumably related to the BPS partition function of D8-D0 branes in $\mathbb{C}^4$,  if we assume that each configuration is weighted by $q^{|\bpi|}$ with $|\bpi|$ being the number of 4D boxes in the solid partition $\bpi$.
Recall that the conjectured generating function of solid partitions (with a similar structure) of MacMahon famously failed at level 6.
As a comparison, ordinary (2D) partitions and plane (3D) partition both have rather simple generating functions, which can be written as the plethystic exponents of even simpler expressions.

This problem has been clarified in \cite{Nekrasov:2017cih}. 
In the ADHM construction that describes D-branes wrapping toric Calabi-Yau varieties \cite{Douglas:1995bn, Douglas:1996sw, Awata:2005fa, Kanno:2020ybd}, solid partitions correspond to fixed points on the ADHM space for 8D instantons, and the measure should be given by the loop-correction determinants in the instanton background, instead of the simple one in which each configuration is weighted by $q^{|\bpi|}$. 
Adopting this measure, \cite{Nekrasov:2017cih} proposed a closed-form formula for the generating function of solid partitions, as a plethystic exponent.\footnote{
	For subsequent development see \cite{Nekrasov:2018xsb, Bonelli:2020gku, Szabo:2023ixw, Kimura:2022zsm, Piazzalunga:2023qik, Nekrasov:2023nai, Kimura:2023bxy}; for recent progress on counting BPS states of D-branes wrapping toric Calabi-Yau fourfolds (CY${}_4$) and associated geometric structures see \cite{Cao:2017swr, Cao:2018rbp, Cao:2019tnw, Cao:2023lon}.}

The next difficulty appears at an early stage of the bootstrap procedure.
The bootstrapping procedure of \cite{Li:2020rij} starts with a certain ansatz on the actions of operators of the BPS algebra (in the Chevalley basis) on the representation vector $|\bpi\rangle$ labeled by a 3D crystal (a plane partition for the $\mathbb{C}^3$ case). 
We first organize all the Cartan/creation/annihilation operators into three fields of $z$, using a complex spectral parameter $z$. 
The eigen-values of each crystal configuration under the action of the Cartan operators are correspondingly organized into a function of $z$, referred to as the ``charge function" in \cite{Li:2020rij}.
The action of the creation and annihilation operators depends on the charge function.
To fulfil some basic requirement on the BPS algebra, the charge function has the important property that for each crystal configuration, each pole of its charge function either corresponds to a position where the creation operators of the BPS algebra can add a new atom, or a position where an existing atom of the crystal can be removed by the annihilation operators of the BPS algebra.\footnote{
	In this paper, since we will only deal with D-dimensional partitions, henceforward we will refer to these atoms by ``boxes'', which seem to be the standard terminology for D-dimensional partitions.}
In 3D, this property then determines the charge function and furthermore the action of the BPS algebra on the representation space, which then allows us to determine the algebraic relations of the BPS algebra itself. 

Therefore one of the most important steps in determining the BPS algebra is to construct the charge function with the important property explained above. 
In the 3D case, the charge function that has these properties actually takes a very simple form: each box in the partition contributes one rational factor, which is defined in terms of a basic rational function (the ``bonding factor") that eventually enters the definition of the BPS algebra.
However, this simple form of charge function does not generalize to 4D. 
Already for the simplest case of $\mathbb{C}^4$, a charge function with this simple factorized form does not have the correct pole structure that can capture the adding and removing 4D boxes from a given solid partition. 

The aim of this note is to construct the charge function for solid partitions, as a first step towards constructing the BPS algebra describing D8-D0 branes on $\mathbb{C}^4$.
We first arrive at a conjectured charge function by examining all solid partitions with up to 15 boxes, and then prove that it indeed obeys all the requirements  for all solid partitions, by analyzing local shapes of the boundaries of the solid partitions.
Relative to the 3D counterpart, the crucial new feature of the final result \eqref{ch_f} is the appearance of the contributions from certain 4-box and 5-box clusters, apart from the contributions from single boxes. 
We expect that this construction can be straightforwardly generalized to other toric Calabi-Yau fourfolds and the new features will be crucial in constructing the BPS algebra for $\IC^4$.\footnote{
	We are not aware of an algebra with representations furnished by solid partitions. This suggests that the resulting BPS algebra for solid partitions could turn out to be a \emph{new} algebraic structure, e.g.\ a ``\emph{Mama}-algebra'' in the sense of \cite{Nekrasov:2017cih}.} 

This paper is organized as follows. 
In Sec.~\ref{s:CY4} we give a brief review of the construction of the effective theory of D-brane systems on $\IC^4$ and the counting problem for BPS states in this system.
In Sec.~\ref{sec:solid} we review the construction of solid partitions. 
The main result is in Sec.~\ref{sec:ChargeFunction}, where we construct the charge function for the solid partitions.
Then in Sec.~\ref{S:discussion} we propose some interesting problems for further research.
Finally, in App. \ref{app:warmup} we review the Young diagrams (2D) and plane partitions (3D) together with the construction of their charge functions; and App. \ref{s:gauge} contains material for the effective theory on the D8 brane world-volume.

\section{BPS D-branes on toric Calabi-Yau fourfolds}\label{s:CY4}

The effective theory of D-branes wrapping toric Calabi-Yau varieties is known to flow to an effective super-Yang-Mills-Higgs theory whose gauge-matter-interaction content is encoded in a quiver \cite{Douglas:1996sw}.

In the case of toric Calabi-Yau fourfolds (CY${}_4$), the effective theory is given by a 2D $\CN=(0,2)$ theory (see App. \ref{s:gauge} for details).
The corresponding quiver can be lifted to  a periodic extended quiver on a 3-torus \cite{Franco:2015tna,Franco:2015tya,Franco:2019bmx,Franco:2021elb}.
In addition to the usual gauge and chiral multiplets encoded in the nodes and oriented edges of the quiver, the 2d $\CN=(0,2)$ theory has a Fermi multiplet encoded in the unoriented edges of the extended quiver.
The usual F-term of $\CN=1$ 4d theories (see App. \ref{s:gauge}) is replaced by $EJ$-terms defined by a signed sum over faces with unoriented edges.

In this note, we concentrate on the simplest toric CY${}_4$: $\IC^4$.
We follow \cite{Franco:2015tna} and consider a quiver with canonical framing $I$, which represents a canonical D8-brane wrapping the whole $\IC^4$:
\begin{equation}
	\begin{array}{c}
		\begin{tikzpicture}
			\begin{scope}[rotate=135]
				\draw[thick, postaction={decorate},decoration={markings, 
					mark= at position 0.7 with {\arrow{stealth}}}] (0,0) to[out=350,in=270] (1.2,0) to[out=90,in=10] (0,0);
				\node[left] at (1.2,0) {$\scriptstyle B_1$}; 
			\end{scope}
			\begin{scope}[rotate=90]
				\draw[thick, burgundy] (0,0) to[out=350,in=270] (1.2,0) to[out=90,in=10] (0,0);
				\node[above, burgundy] at (1.2,0) {$\scriptstyle \Lambda^{(1)}$}; 
			\end{scope}
			\begin{scope}[rotate=45]
				\draw[thick, postaction={decorate},decoration={markings, 
					mark= at position 0.7 with {\arrow{stealth}}}] (0,0) to[out=350,in=270] (1.2,0) to[out=90,in=10] (0,0);
				\node[right] at (1.2,0) {$\scriptstyle B_2$}; 
			\end{scope}
			\begin{scope}[rotate=0]
				\draw[thick, burgundy] (0,0) to[out=350,in=270] (1.2,0) to[out=90,in=10] (0,0);
				\node[right, burgundy] at (1.2,0) {$\scriptstyle \Lambda^{(2)}$}; 
			\end{scope}
			\begin{scope}[rotate=-45]
				\draw[thick, postaction={decorate},decoration={markings, 
					mark= at position 0.7 with {\arrow{stealth}}}] (0,0) to[out=350,in=270] (1.2,0) to[out=90,in=10] (0,0);
				\node[right] at (1.2,0) {$\scriptstyle B_3$}; 
			\end{scope}
			\begin{scope}[rotate=-90]
				\draw[thick, burgundy] (0,0) to[out=350,in=270] (1.2,0) to[out=90,in=10] (0,0);
				\node[below, burgundy] at (1.2,0) {$\scriptstyle \Lambda^{(3)}$}; 
			\end{scope}
			\begin{scope}[rotate=-135]
				\draw[thick, postaction={decorate},decoration={markings, 
					mark= at position 0.7 with {\arrow{stealth}}}]  (0,0) to[out=350,in=270] (1.2,0) to[out=90,in=10] (0,0);
				\node[left] at (1.2,0) {$\scriptstyle B_4$}; 
			\end{scope}
			\draw[thick, postaction={decorate},decoration={markings, 
				mark= at position 0.7 with {\arrow{stealth}}}] (-2,0) -- (0,0) node[pos=0.25,above] {$\scriptstyle I$};
			\begin{scope}[shift={(-2,0)}]
				\draw[fill=gray] (-0.1,-0.1) -- (-0.1,0.1) -- (0.1,0.1) -- (0.1,-0.1) -- cycle;
			\end{scope}
			\draw[fill=\myblue] (0,0) circle (0.1);
		\end{tikzpicture}
	\end{array},\quad\begin{array}{c}
		E_k=\left[B_4,B_k\right],\;\; k=1,2,3\,;\\
		W_{J}=\sum\lm_{i,j,k=1}^3\epsilon_{ijk}\Tr \left(\Lambda^{(i)}B_j B_k \right)\,.
	\end{array}
\end{equation}

For the case of $\IC^4$, the periodic quiver and the $EJ$-vacuum equations are:
\begin{equation}
	\begin{array}{c}
		\begin{tikzpicture}
			\draw[burgundy, ultra thick] (0,0) -- (2,2) (0,0) -- (1.5,-1) (0,0) -- (-0.5,1);
			\begin{scope}[thick, decoration={markings, mark= at position 0.6 with {\arrow{stealth}}}]
				\draw[postaction = {decorate}] (0,0) -- (2,0) node[pos=0.5,below] {$\scriptstyle B_2$};
				\draw[postaction = {decorate}] (0,0) -- (0,2) node[pos=0.7,right] {$\scriptstyle B_3$};
				\draw[postaction = {decorate}] (0,0) -- (-0.5,-1) node[pos=0.5,below right] {$\scriptstyle B_1$};
				\draw[postaction = {decorate}] (1.5,1) -- (0,0);
			\end{scope}
			\draw[dashed] (0,2) -- (2,2) -- (1.5,1) -- (-0.5,1) -- (0,2) (2,0) -- (1.5,-1) -- (-0.5,-1) (2,0) -- (2,2) (1.5,-1) -- (1.5,1) (-0.5,-1) -- (-0.5,1);
			\draw[fill=\myblue] (0,0) circle (0.08) (2,0) circle (0.08) (0,2) circle (0.08) (2,2) circle (0.08) (-0.5,1) circle (0.08) (-0.5,-1) circle (0.08) (1.5,1) circle (0.08) (1.5,-1) circle (0.08);
			\node[above left] at (-0.5,1) {$\scriptstyle \Lambda^{(2)}$}; 
			\node[right] at (2,2) {$\scriptstyle \Lambda^{(1)}$}; 
			\node[right] at (1.5,-1) {$\scriptstyle \Lambda^{(1)}$};
		\end{tikzpicture}
	\end{array}\quad 
	\renewcommand{\arraystretch}{1.2}\begin{array}{ccc}
		& J & E\\
		\Lambda^{(1)}: & B_2\cdot B_3-B_3\cdot B_2=0\,, & B_4\cdot B_1-B_1\cdot B_4=0\\
		\Lambda^{(2)}: & B_3\cdot B_1-B_1\cdot B_3=0\,, & B_4\cdot B_2-B_2\cdot B_4=0\\
		\Lambda^{(3)}: & B_1\cdot B_2-B_2\cdot B_1=0\,, & B_4\cdot B_3-B_3\cdot B_4=0
	\end{array}
\end{equation}
The vacuum equations for the fields $B_a$ and $I$ consist of the canonical $D$-term and the $EJ$-term discussed above:
\begin{equation}
	\begin{split}
		\mbox{$D$-term: }\qquad  & \sum\lm_{i=1}^{4}\left[B_i,B_i^{\dagger}\right]+II^{\dagger}=\zeta\bbone\,,\\
		\mbox{$EJ$-term: } \qquad & \left[B_i,B_j\right]=0,\;\quad i,j=1,2,3,4\,.
	\end{split}
\end{equation}
Note that these vacuum equations are equivalent to the ADHM description of instantons on $\IR^8$ \cite{Nekrasov:2017cih, Bonelli:2020gku, Szabo:2022zyn}.

Classical vacua are fixed points on this variety that correspond to equivariant 8D instantons in the $\Omega$-background.
We can count them in a way similar to the 4D or 6D case \cite{Szendroi:2007nu}, using a theorem of King \cite{King}.
Consider the equivariant monomials in the quiver path algebra:
\begin{equation}
	\mathscr{A}=\IC[B_1,B_2,B_3,B_4]/\langle [B_a,B_b]=0\rangle\,,
\end{equation}
which act on a cyclic vector $I$.
All such monomials can be enumerated by points of the positive sedecant $\IZ_{\geq 0}^4$ in a 4D integral lattice.
We denote the canonical basis in the 4D space by
\begin{equation}
	\begin{split}
		\vec e_1&=\left(\begin{array}{cccc}
			1\,,&0\,, &0\,, & 0\\
		\end{array}\right)\,,\\
		\vec e_2&=\left(\begin{array}{cccc}
			0\,,&1\,, &0\,, & 0\\
		\end{array}\right)\,,\\
		\vec e_3&=\left(\begin{array}{cccc}
			0\,,&0\,, &1\,, & 0\\
		\end{array}\right)\,,\\
		\vec e_4&=\left(\begin{array}{cccc}
			0\,,&0\,, &0\,, & 1\\
		\end{array}\right)\,.\\
	\end{split}
\end{equation}
A monomial is defined by the coordinates of a point in this lattice:
\begin{equation}\label{mono}
	(i,j,k,l)\in\IZ_{\geq 0}\;\longleftrightarrow\; B_1^iB_2^jB_3^kB_4^l\cdot I\,.
\end{equation}

Some of these operators acquire vacuum expectation values (vevs) when acting on the vacuum state.
A set of operators with vevs forms a subset of points (which will correspond to boxes in a solid partition) in $\IZ_{\geq 0}^4$, which we call a 4D crystal.
The stability conditions on the fixed points of the quiver moduli space, together with the $EJ$-term constraints, impose a convexity constraint on possible crystal configurations,  known in the literature as the melting rule \cite{Okounkov:2003sp,Iqbal:2003ds,Szendroi:2007nu,MR2836398,MR2592501,Ooguri:2009ijd,Yamazaki:2010fz}.
As a result, the BPS vacua of D-brane systems wrapping toric CY${}_4$'s are described by a melting crystal model.

In the current case, the melting rule is:
\begin{tcolorbox}
	\begin{equation}\label{melting}
		\begin{aligned}
			&\mbox{For any box $\Box\in\IZ_{\geq 0}^4$,}\\
			&\mbox{if there is another box $\Box'\in \Kappa$ at $\vec x(\Box')=\vec x(\Box)+\vec e_k$ for any $k=1,2,3,4$\,,\quad}\\
			&		\mbox{then $\Box\in \Kappa$.}    
		\end{aligned}
	\end{equation}
\end{tcolorbox}
\noindent $\Kappa$ is called a molten crystal if it satisfies the melting rule.

\section{Solid partitions}\label{sec:solid}

A solid partition of an integer $n$ is a three-dimensional array $\bpi$ of non-negative integers $\bpi_{i,j,k}\geq 0$ satisfying:
\begin{equation}
	\sum\lm_{i,j,k}\bpi_{i,j,k}=n,\quad \bpi_{i+1,j,k}\leq\bpi_{i,j,k},\;\bpi_{i,j+1,k}\leq\bpi_{i,j,k},\;\bpi_{i,j,k+1}\leq\bpi_{i,j,k},\quad \forall\; i,j,k\in\IZ_{\geq 0}.
\end{equation}
The (naive) generating series for solid partitions, which is defined as
\begin{equation}
	P(q)\equiv \sum_{\bpi} q^{|\bpi|}    
\end{equation}
is given by \cite{OEIS}:
\begin{equation}
	\begin{split}
		P(q)=&1+q+4q^2+ 10q^3+ 26q^4+ 59q^5+ 140q^6+ 307q^7+ 684q^8+ 1464q^9\\
		&+3122q^{10} +6500q^{11} +13426q^{12}+27248q^{13} + 54804q^{14}+\ldots.
	\end{split}
\end{equation}
Unlike the 2D and 3D counterparts, the coefficients in its plethystic logarithm are neither monotonic nor all positive:
\begin{equation}
	\begin{aligned}
		f(q)=\textrm{PL}[P](q)=&1 + q + 3 q^2 + 6 q^3 + 10 q^4 + 15 q^5 + 20 q^6 + 26 q^7+ 34 q^8\\
		&+46 q^9 + 68 q^{10} + 97 q^{11} + 120 q^{12} + 112 q^{13} + 23 q^{14}\\
		& -186 q^{15} - 496 q^{16} - 735 q^{17} - 531 q^{18}+779 q^{19}+\dots \,,
	\end{aligned}
\end{equation}
where the non-monotonicity first appears at $q^{13}$ and the first negative term appears at  $q^{15}$.\footnote{
	This is what makes it impossible to interpret the naive generating series as the multi-particle partition function of a physics system, in which the single-particle partition function is given by its plethystic logarithm. To relate the solid partitions to a physical system, one needs to adopt a non-trivial measure instead, such as done in  \cite{Nekrasov:2017cih}.}

A solid partition can be visualized as stacking 4D unit-size boxes in the corner of a 4D room.
Namely, a solid partition $\bpi$ can be represented by a subset of nodes (boxes in the solid partition) of $\IZ_{\geq 0}^4$, with coordinates:
\begin{equation}
	(i,j,k,l),\quad 0\leq l<\bpi_{i,j,k}.
\end{equation}
A set of solid partitions is equivalent to a set of 4D crystals in $\IZ_{\geq 0}^4$ satisfying the melting rule \eqref{melting}, thus we will use these two terms interchangeably.

\begin{figure}
	\begin{center}
		\begin{tikzpicture}
			\begin{scope}
				\begin{scope}[scale=0.3]
					\foreach \x/\y/\z/\w in {0./7./-0.866025/6.5, 0./7./0.866025/6.5, -0.866025/4.5/-0.866025/5.5, -0.866025/4.5/-1.73205/4., -0.866025/4.5/0./4., -0.866025/5.5/-0.866025/6.5, -0.866025/5.5/0./5., -0.866025/6.5/0./6., -1.73205/2./-1.73205/3., -1.73205/2./-2.59808/1.5, -1.73205/2./-0.866025/1.5, -1.73205/3./-1.73205/4., -1.73205/3./-0.866025/2.5, -1.73205/4./-0.866025/3.5, -2.59808/-0.5/-2.59808/0.5, -2.59808/-0.5/-3.4641/-1., -2.59808/-0.5/-1.73205/-1., -2.59808/0.5/-2.59808/1.5, -2.59808/0.5/-1.73205/0., -2.59808/1.5/-1.73205/1., -3.4641/-2./-3.4641/-1., -3.4641/-2./-2.59808/-2.5, -3.4641/-1./-2.59808/-1.5, 0.866025/4.5/0.866025/5.5, 0.866025/4.5/0./4., 0.866025/4.5/1.73205/4., 0.866025/5.5/0.866025/6.5, 0.866025/5.5/0./5., 0.866025/6.5/0./6., 0./4./0./5., 0./4./-0.866025/3.5, 0./4./0.866025/3.5, 0./5./0./6., -0.866025/1.5/-0.866025/2.5, -0.866025/1.5/-1.73205/1., -0.866025/1.5/0./1., -0.866025/2.5/-0.866025/3.5, -0.866025/2.5/0./2., -0.866025/3.5/0./3., -1.73205/-1./-1.73205/0., -1.73205/-1./-2.59808/-1.5, -1.73205/-1./-0.866025/-1.5, -1.73205/0./-1.73205/1., -1.73205/0./-0.866025/-0.5, -1.73205/1./-0.866025/0.5, -2.59808/-2.5/-2.59808/-1.5, -2.59808/-2.5/-1.73205/-3., -2.59808/-1.5/-1.73205/-2., 1.73205/3./1.73205/4., 1.73205/3./0.866025/2.5, 1.73205/3./2.59808/2.5, 1.73205/4./0.866025/3.5, 0.866025/2.5/0.866025/3.5, 0.866025/2.5/0./2., 0.866025/2.5/1.73205/2., 0.866025/3.5/0./3., 0./1./0./2., 0./1./-0.866025/0.5, 0./1./0.866025/0.5, 0./2./0./3., 0./2./0.866025/1.5, -0.866025/-2.5/-0.866025/-1.5, -0.866025/-2.5/-1.73205/-3., -0.866025/-2.5/0./-3., -0.866025/-1.5/-0.866025/-0.5, -0.866025/-1.5/-1.73205/-2., -0.866025/-1.5/0./-2., -0.866025/-0.5/-0.866025/0.5, -0.866025/-0.5/0./-1., -0.866025/0.5/0./0., -1.73205/-3./-1.73205/-2., 2.59808/1.5/2.59808/2.5, 2.59808/1.5/1.73205/1., 2.59808/1.5/3.4641/1., 2.59808/2.5/1.73205/2., 1.73205/1./1.73205/2., 1.73205/1./0.866025/0.5, 1.73205/1./2.59808/0.5, 1.73205/2./0.866025/1.5, 0.866025/-1.5/0.866025/-0.5, 0.866025/-1.5/0./-2., 0.866025/-1.5/1.73205/-2., 0.866025/-0.5/0.866025/0.5, 0.866025/-0.5/0./-1., 0.866025/-0.5/1.73205/-1., 0.866025/0.5/0.866025/1.5, 0.866025/0.5/0./0., 0.866025/0.5/1.73205/0., 0./-3./0./-2., 0./-3./0.866025/-3.5, 0./-2./0./-1., 0./-2./0.866025/-2.5, 0./-1./0./0., 3.4641/0./3.4641/1., 3.4641/0./2.59808/-0.5, 3.4641/0./4.33013/-0.5, 3.4641/1./2.59808/0.5, 2.59808/-0.5/2.59808/0.5, 2.59808/-0.5/1.73205/-1., 2.59808/-0.5/3.4641/-1., 2.59808/0.5/1.73205/0., 1.73205/-3./1.73205/-2., 1.73205/-3./0.866025/-3.5, 1.73205/-3./2.59808/-3.5, 1.73205/-2./1.73205/-1., 1.73205/-2./0.866025/-2.5, 1.73205/-2./2.59808/-2.5, 1.73205/-1./1.73205/0., 1.73205/-1./2.59808/-1.5, 0.866025/-3.5/0.866025/-2.5, 4.33013/-0.5/3.4641/-1., 4.33013/-0.5/5.19615/-1., 3.4641/-2./3.4641/-1., 3.4641/-2./2.59808/-2.5, 3.4641/-2./4.33013/-2.5, 3.4641/-1./2.59808/-1.5, 3.4641/-1./4.33013/-1.5, 2.59808/-3.5/2.59808/-2.5, 2.59808/-3.5/3.4641/-4., 2.59808/-2.5/2.59808/-1.5, 2.59808/-2.5/3.4641/-3., 5.19615/-2./5.19615/-1., 5.19615/-2./4.33013/-2.5, 5.19615/-2./6.06218/-2.5, 5.19615/-1./4.33013/-1.5, 4.33013/-3.5/4.33013/-2.5, 4.33013/-3.5/3.4641/-4., 4.33013/-3.5/5.19615/-4., 4.33013/-2.5/4.33013/-1.5, 4.33013/-2.5/3.4641/-3., 4.33013/-2.5/5.19615/-3., 3.4641/-4./3.4641/-3., 6.06218/-3.5/6.06218/-2.5, 6.06218/-3.5/5.19615/-4., 6.06218/-2.5/5.19615/-3., 5.19615/-4./5.19615/-3.}
					{
						\draw (\x,\y) -- (\z,\w);
					}
				\end{scope}
				\node at (0,-2) {a) $\bpi$}; 
			\end{scope}
			\begin{scope}[shift={(4.5,0)}]
				\begin{scope}[scale=0.3]
					\foreach \x/\y/\z/\w in {0./7./-0.866025/6.5, 0./7./0.866025/6.5, -0.866025/4.5/-0.866025/5.5, -0.866025/4.5/-1.73205/4., -0.866025/4.5/0./4., -0.866025/5.5/-0.866025/6.5, -0.866025/5.5/0./5., -0.866025/6.5/0./6., -1.73205/2./-1.73205/3., -1.73205/2./-2.59808/1.5, -1.73205/2./-0.866025/1.5, -1.73205/3./-1.73205/4., -1.73205/3./-0.866025/2.5, -1.73205/4./-0.866025/3.5, -2.59808/-0.5/-2.59808/0.5, -2.59808/-0.5/-3.4641/-1., -2.59808/-0.5/-1.73205/-1., -2.59808/0.5/-2.59808/1.5, -2.59808/0.5/-1.73205/0., -2.59808/1.5/-1.73205/1., -3.4641/-2./-3.4641/-1., -3.4641/-2./-2.59808/-2.5, -3.4641/-1./-2.59808/-1.5, 0.866025/4.5/0.866025/5.5, 0.866025/4.5/0./4., 0.866025/4.5/1.73205/4., 0.866025/5.5/0.866025/6.5, 0.866025/5.5/0./5., 0.866025/6.5/0./6., 0./4./0./5., 0./4./-0.866025/3.5, 0./4./0.866025/3.5, 0./5./0./6., -0.866025/1.5/-0.866025/2.5, -0.866025/1.5/-1.73205/1., -0.866025/1.5/0./1., -0.866025/2.5/-0.866025/3.5, -0.866025/2.5/0./2., -0.866025/3.5/0./3., -1.73205/-1./-1.73205/0., -1.73205/-1./-2.59808/-1.5, -1.73205/-1./-0.866025/-1.5, -1.73205/0./-1.73205/1., -1.73205/0./-0.866025/-0.5, -1.73205/1./-0.866025/0.5, -2.59808/-2.5/-2.59808/-1.5, -2.59808/-2.5/-1.73205/-3., -2.59808/-1.5/-1.73205/-2., 1.73205/3./1.73205/4., 1.73205/3./0.866025/2.5, 1.73205/3./2.59808/2.5, 1.73205/4./0.866025/3.5, 0.866025/2.5/0.866025/3.5, 0.866025/2.5/0./2., 0.866025/2.5/1.73205/2., 0.866025/3.5/0./3., 0./1./0./2., 0./1./-0.866025/0.5, 0./1./0.866025/0.5, 0./2./0./3., 0./2./0.866025/1.5, -0.866025/-2.5/-0.866025/-1.5, -0.866025/-2.5/-1.73205/-3., -0.866025/-2.5/0./-3., -0.866025/-1.5/-0.866025/-0.5, -0.866025/-1.5/-1.73205/-2., -0.866025/-1.5/0./-2., -0.866025/-0.5/-0.866025/0.5, -0.866025/-0.5/0./-1., -0.866025/0.5/0./0., -1.73205/-3./-1.73205/-2., 2.59808/1.5/2.59808/2.5, 2.59808/1.5/1.73205/1., 2.59808/1.5/3.4641/1., 2.59808/2.5/1.73205/2., 1.73205/1./1.73205/2., 1.73205/1./0.866025/0.5, 1.73205/1./2.59808/0.5, 1.73205/2./0.866025/1.5, 0.866025/-1.5/0.866025/-0.5, 0.866025/-1.5/0./-2., 0.866025/-1.5/1.73205/-2., 0.866025/-0.5/0.866025/0.5, 0.866025/-0.5/0./-1., 0.866025/-0.5/1.73205/-1., 0.866025/0.5/0.866025/1.5, 0.866025/0.5/0./0., 0.866025/0.5/1.73205/0., 0./-3./0./-2., 0./-3./0.866025/-3.5, 0./-2./0./-1., 0./-2./0.866025/-2.5, 0./-1./0./0., 3.4641/0./3.4641/1., 3.4641/0./2.59808/-0.5, 3.4641/0./4.33013/-0.5, 3.4641/1./2.59808/0.5, 2.59808/-0.5/2.59808/0.5, 2.59808/-0.5/1.73205/-1., 2.59808/-0.5/3.4641/-1., 2.59808/0.5/1.73205/0., 1.73205/-3./1.73205/-2., 1.73205/-3./0.866025/-3.5, 1.73205/-3./2.59808/-3.5, 1.73205/-2./1.73205/-1., 1.73205/-2./0.866025/-2.5, 1.73205/-2./2.59808/-2.5, 1.73205/-1./1.73205/0., 1.73205/-1./2.59808/-1.5, 0.866025/-3.5/0.866025/-2.5, 4.33013/-0.5/3.4641/-1., 4.33013/-0.5/5.19615/-1., 3.4641/-2./3.4641/-1., 3.4641/-2./2.59808/-2.5, 3.4641/-2./4.33013/-2.5, 3.4641/-1./2.59808/-1.5, 3.4641/-1./4.33013/-1.5, 2.59808/-3.5/2.59808/-2.5, 2.59808/-3.5/3.4641/-4., 2.59808/-2.5/2.59808/-1.5, 2.59808/-2.5/3.4641/-3., 5.19615/-2./5.19615/-1., 5.19615/-2./4.33013/-2.5, 5.19615/-2./6.06218/-2.5, 5.19615/-1./4.33013/-1.5, 4.33013/-3.5/4.33013/-2.5, 4.33013/-3.5/3.4641/-4., 4.33013/-3.5/5.19615/-4., 4.33013/-2.5/4.33013/-1.5, 4.33013/-2.5/3.4641/-3., 4.33013/-2.5/5.19615/-3., 3.4641/-4./3.4641/-3., 6.06218/-3.5/6.06218/-2.5, 6.06218/-3.5/5.19615/-4., 6.06218/-2.5/5.19615/-3., 5.19615/-4./5.19615/-3.}
					{
						\draw (\x,\y) -- (\z,\w);
					}
					\tikzset{sty1/.style={fill=white!30!red}}
					\draw[sty1] (0.866025,6.5) -- (0.866025,7.5) -- (0.,7.) -- (0.,6.) -- cycle;
					\draw[sty1] (0.,4.) -- (0.,5.) -- (-0.866025,4.5) -- (-0.866025,3.5) -- cycle;
					\draw[sty1] (-0.866025,1.5) -- (-0.866025,2.5) -- (-1.73205,2.) -- (-1.73205,1.) -- cycle;
					\draw[sty1] (-1.73205,-1.) -- (-1.73205,0.) -- (-2.59808,-0.5) -- (-2.59808,-1.5) -- cycle;
					\draw[sty1] (-2.59808,-2.5) -- (-2.59808,-1.5) -- (-3.4641,-2.) -- (-3.4641,-3.) -- cycle;
					\draw[sty1] (1.73205,4.) -- (1.73205,5.) -- (0.866025,4.5) -- (0.866025,3.5) -- cycle;
					\draw[sty1] (2.59808,2.5) -- (2.59808,3.5) -- (1.73205,3.) -- (1.73205,2.) -- cycle;
					\draw[sty1] (0.,-3.) -- (0.,-2.) -- (-0.866025,-2.5) -- (-0.866025,-3.5) -- cycle;
					\draw[sty1] (3.4641,1.) -- (3.4641,2.) -- (2.59808,1.5) -- (2.59808,0.5) -- cycle;
					\draw[sty1] (1.73205,-2.) -- (1.73205,-1.) -- (0.866025,-1.5) -- (0.866025,-2.5) -- cycle;
					\draw[sty1] (4.33013,-0.5) -- (4.33013,0.5) -- (3.4641,0.) -- (3.4641,-1.) -- cycle;
					\draw[sty1] (2.59808,-3.5) -- (2.59808,-2.5) -- (1.73205,-3.) -- (1.73205,-4.) -- cycle;
					\draw[sty1] (4.33013,-2.5) -- (4.33013,-1.5) -- (3.4641,-2.) -- (3.4641,-3.) -- cycle;
					\draw[sty1] (6.06218,-2.5) -- (6.06218,-1.5) -- (5.19615,-2.) -- (5.19615,-3.) -- cycle;
					\draw[sty1] (5.19615,-4.) -- (5.19615,-3.) -- (4.33013,-3.5) -- (4.33013,-4.5) -- cycle;
					\draw[sty1] (6.9282,-4.) -- (6.9282,-3.) -- (6.06218,-3.5) -- (6.06218,-4.5) -- cycle;
					\draw[sty1] (-0.866025,6.5) -- (-0.866025,7.5) -- (0.,7.) -- (0.,6.) -- cycle;
					\draw[sty1] (-1.73205,4.) -- (-1.73205,5.) -- (-0.866025,4.5) -- (-0.866025,3.5) -- cycle;
					\draw[sty1] (-2.59808,1.5) -- (-2.59808,2.5) -- (-1.73205,2.) -- (-1.73205,1.) -- cycle;
					\draw[sty1] (-3.4641,-1.) -- (-3.4641,0.) -- (-2.59808,-0.5) -- (-2.59808,-1.5) -- cycle;
					\draw[sty1] (-4.33013,-2.5) -- (-4.33013,-1.5) -- (-3.4641,-2.) -- (-3.4641,-3.) -- cycle;
					\draw[sty1] (0.,4.) -- (0.,5.) -- (0.866025,4.5) -- (0.866025,3.5) -- cycle;
					\draw[sty1] (0.866025,2.5) -- (0.866025,3.5) -- (1.73205,3.) -- (1.73205,2.) -- cycle;
					\draw[sty1] (-1.73205,-3.) -- (-1.73205,-2.) -- (-0.866025,-2.5) -- (-0.866025,-3.5) -- cycle;
					\draw[sty1] (1.73205,1.) -- (1.73205,2.) -- (2.59808,1.5) -- (2.59808,0.5) -- cycle;
					\draw[sty1] (0.,-2.) -- (0.,-1.) -- (0.866025,-1.5) -- (0.866025,-2.5) -- cycle;
					\draw[sty1] (2.59808,-0.5) -- (2.59808,0.5) -- (3.4641,0.) -- (3.4641,-1.) -- cycle;
					\draw[sty1] (0.866025,-3.5) -- (0.866025,-2.5) -- (1.73205,-3.) -- (1.73205,-4.) -- cycle;
					\draw[sty1] (2.59808,-2.5) -- (2.59808,-1.5) -- (3.4641,-2.) -- (3.4641,-3.) -- cycle;
					\draw[sty1] (4.33013,-2.5) -- (4.33013,-1.5) -- (5.19615,-2.) -- (5.19615,-3.) -- cycle;
					\draw[sty1] (3.4641,-4.) -- (3.4641,-3.) -- (4.33013,-3.5) -- (4.33013,-4.5) -- cycle;
					\draw[sty1] (5.19615,-4.) -- (5.19615,-3.) -- (6.06218,-3.5) -- (6.06218,-4.5) -- cycle;
					\draw[sty1] (0.,8.) -- (-0.866025,7.5) -- (0.,7.) -- (0.866025,7.5) -- cycle;
					\draw[sty1] (-0.866025,5.5) -- (-1.73205,5.) -- (-0.866025,4.5) -- (0.,5.) -- cycle;
					\draw[sty1] (-1.73205,3.) -- (-2.59808,2.5) -- (-1.73205,2.) -- (-0.866025,2.5) -- cycle;
					\draw[sty1] (-2.59808,0.5) -- (-3.4641,0.) -- (-2.59808,-0.5) -- (-1.73205,0.) -- cycle;
					\draw[sty1] (-3.4641,-1.) -- (-4.33013,-1.5) -- (-3.4641,-2.) -- (-2.59808,-1.5) -- cycle;
					\draw[sty1] (0.866025,5.5) -- (0.,5.) -- (0.866025,4.5) -- (1.73205,5.) -- cycle;
					\draw[sty1] (1.73205,4.) -- (0.866025,3.5) -- (1.73205,3.) -- (2.59808,3.5) -- cycle;
					\draw[sty1] (-0.866025,-1.5) -- (-1.73205,-2.) -- (-0.866025,-2.5) -- (0.,-2.) -- cycle;
					\draw[sty1] (2.59808,2.5) -- (1.73205,2.) -- (2.59808,1.5) -- (3.4641,2.) -- cycle;
					\draw[sty1] (0.866025,-0.5) -- (0.,-1.) -- (0.866025,-1.5) -- (1.73205,-1.) -- cycle;
					\draw[sty1] (3.4641,1.) -- (2.59808,0.5) -- (3.4641,0.) -- (4.33013,0.5) -- cycle;
					\draw[sty1] (1.73205,-2.) -- (0.866025,-2.5) -- (1.73205,-3.) -- (2.59808,-2.5) -- cycle;
					\draw[sty1] (3.4641,-1.) -- (2.59808,-1.5) -- (3.4641,-2.) -- (4.33013,-1.5) -- cycle;
					\draw[sty1] (5.19615,-1.) -- (4.33013,-1.5) -- (5.19615,-2.) -- (6.06218,-1.5) -- cycle;
					\draw[sty1] (4.33013,-2.5) -- (3.4641,-3.) -- (4.33013,-3.5) -- (5.19615,-3.) -- cycle;
					\draw[sty1] (6.06218,-2.5) -- (5.19615,-3.) -- (6.06218,-3.5) -- (6.9282,-3.) -- cycle;
				\end{scope}
				\node at (0,-2) {b) ${\rm Add}(\bpi)$}; 
			\end{scope}
			\begin{scope}[shift={(9,0)}]
				\begin{scope}[scale=0.3]
					\foreach \x/\y/\z/\w in {0./7./-0.866025/6.5, 0./7./0.866025/6.5, -0.866025/4.5/-0.866025/5.5, -0.866025/4.5/-1.73205/4., -0.866025/4.5/0./4., -0.866025/5.5/-0.866025/6.5, -0.866025/5.5/0./5., -0.866025/6.5/0./6., -1.73205/2./-1.73205/3., -1.73205/2./-2.59808/1.5, -1.73205/2./-0.866025/1.5, -1.73205/3./-1.73205/4., -1.73205/3./-0.866025/2.5, -1.73205/4./-0.866025/3.5, -2.59808/-0.5/-2.59808/0.5, -2.59808/-0.5/-3.4641/-1., -2.59808/-0.5/-1.73205/-1., -2.59808/0.5/-2.59808/1.5, -2.59808/0.5/-1.73205/0., -2.59808/1.5/-1.73205/1., -3.4641/-2./-3.4641/-1., -3.4641/-2./-2.59808/-2.5, -3.4641/-1./-2.59808/-1.5, 0.866025/4.5/0.866025/5.5, 0.866025/4.5/0./4., 0.866025/4.5/1.73205/4., 0.866025/5.5/0.866025/6.5, 0.866025/5.5/0./5., 0.866025/6.5/0./6., 0./4./0./5., 0./4./-0.866025/3.5, 0./4./0.866025/3.5, 0./5./0./6., -0.866025/1.5/-0.866025/2.5, -0.866025/1.5/-1.73205/1., -0.866025/1.5/0./1., -0.866025/2.5/-0.866025/3.5, -0.866025/2.5/0./2., -0.866025/3.5/0./3., -1.73205/-1./-1.73205/0., -1.73205/-1./-2.59808/-1.5, -1.73205/-1./-0.866025/-1.5, -1.73205/0./-1.73205/1., -1.73205/0./-0.866025/-0.5, -1.73205/1./-0.866025/0.5, -2.59808/-2.5/-2.59808/-1.5, -2.59808/-2.5/-1.73205/-3., -2.59808/-1.5/-1.73205/-2., 1.73205/3./1.73205/4., 1.73205/3./0.866025/2.5, 1.73205/3./2.59808/2.5, 1.73205/4./0.866025/3.5, 0.866025/2.5/0.866025/3.5, 0.866025/2.5/0./2., 0.866025/2.5/1.73205/2., 0.866025/3.5/0./3., 0./1./0./2., 0./1./-0.866025/0.5, 0./1./0.866025/0.5, 0./2./0./3., 0./2./0.866025/1.5, -0.866025/-2.5/-0.866025/-1.5, -0.866025/-2.5/-1.73205/-3., -0.866025/-2.5/0./-3., -0.866025/-1.5/-0.866025/-0.5, -0.866025/-1.5/-1.73205/-2., -0.866025/-1.5/0./-2., -0.866025/-0.5/-0.866025/0.5, -0.866025/-0.5/0./-1., -0.866025/0.5/0./0., -1.73205/-3./-1.73205/-2., 2.59808/1.5/2.59808/2.5, 2.59808/1.5/1.73205/1., 2.59808/1.5/3.4641/1., 2.59808/2.5/1.73205/2., 1.73205/1./1.73205/2., 1.73205/1./0.866025/0.5, 1.73205/1./2.59808/0.5, 1.73205/2./0.866025/1.5, 0.866025/-1.5/0.866025/-0.5, 0.866025/-1.5/0./-2., 0.866025/-1.5/1.73205/-2., 0.866025/-0.5/0.866025/0.5, 0.866025/-0.5/0./-1., 0.866025/-0.5/1.73205/-1., 0.866025/0.5/0.866025/1.5, 0.866025/0.5/0./0., 0.866025/0.5/1.73205/0., 0./-3./0./-2., 0./-3./0.866025/-3.5, 0./-2./0./-1., 0./-2./0.866025/-2.5, 0./-1./0./0., 3.4641/0./3.4641/1., 3.4641/0./2.59808/-0.5, 3.4641/0./4.33013/-0.5, 3.4641/1./2.59808/0.5, 2.59808/-0.5/2.59808/0.5, 2.59808/-0.5/1.73205/-1., 2.59808/-0.5/3.4641/-1., 2.59808/0.5/1.73205/0., 1.73205/-3./1.73205/-2., 1.73205/-3./0.866025/-3.5, 1.73205/-3./2.59808/-3.5, 1.73205/-2./1.73205/-1., 1.73205/-2./0.866025/-2.5, 1.73205/-2./2.59808/-2.5, 1.73205/-1./1.73205/0., 1.73205/-1./2.59808/-1.5, 0.866025/-3.5/0.866025/-2.5, 4.33013/-0.5/3.4641/-1., 4.33013/-0.5/5.19615/-1., 3.4641/-2./3.4641/-1., 3.4641/-2./2.59808/-2.5, 3.4641/-2./4.33013/-2.5, 3.4641/-1./2.59808/-1.5, 3.4641/-1./4.33013/-1.5, 2.59808/-3.5/2.59808/-2.5, 2.59808/-3.5/3.4641/-4., 2.59808/-2.5/2.59808/-1.5, 2.59808/-2.5/3.4641/-3., 5.19615/-2./5.19615/-1., 5.19615/-2./4.33013/-2.5, 5.19615/-2./6.06218/-2.5, 5.19615/-1./4.33013/-1.5, 4.33013/-3.5/4.33013/-2.5, 4.33013/-3.5/3.4641/-4., 4.33013/-3.5/5.19615/-4., 4.33013/-2.5/4.33013/-1.5, 4.33013/-2.5/3.4641/-3., 4.33013/-2.5/5.19615/-3., 3.4641/-4./3.4641/-3., 6.06218/-3.5/6.06218/-2.5, 6.06218/-3.5/5.19615/-4., 6.06218/-2.5/5.19615/-3., 5.19615/-4./5.19615/-3.}
					{
						\draw (\x,\y) -- (\z,\w);
					}
					\tikzset{sty1/.style={fill=\myblue}}
					\draw[sty1] (0.866025,5.5) -- (0.866025,6.5) -- (0.,6.) -- (0.,5.) -- cycle;
					\draw[sty1] (0.866025,2.5) -- (0.866025,3.5) -- (0.,3.) -- (0.,2.) -- cycle;
					\draw[sty1] (-0.866025,-2.5) -- (-0.866025,-1.5) -- (-1.73205,-2.) -- (-1.73205,-3.) -- cycle;
					\draw[sty1] (1.73205,1.) -- (1.73205,2.) -- (0.866025,1.5) -- (0.866025,0.5) -- cycle;
					\draw[sty1] (0.866025,-0.5) -- (0.866025,0.5) -- (0.,0.) -- (0.,-1.) -- cycle;
					\draw[sty1] (2.59808,-0.5) -- (2.59808,0.5) -- (1.73205,0.) -- (1.73205,-1.) -- cycle;
					\draw[sty1] (1.73205,-3.) -- (1.73205,-2.) -- (0.866025,-2.5) -- (0.866025,-3.5) -- cycle;
					\draw[sty1] (3.4641,-2.) -- (3.4641,-1.) -- (2.59808,-1.5) -- (2.59808,-2.5) -- cycle;
					\draw[sty1] (5.19615,-2.) -- (5.19615,-1.) -- (4.33013,-1.5) -- (4.33013,-2.5) -- cycle;
					\draw[sty1] (4.33013,-3.5) -- (4.33013,-2.5) -- (3.4641,-3.) -- (3.4641,-4.) -- cycle;
					\draw[sty1] (6.06218,-3.5) -- (6.06218,-2.5) -- (5.19615,-3.) -- (5.19615,-4.) -- cycle;
					\draw[sty1] (-0.866025,5.5) -- (-0.866025,6.5) -- (0.,6.) -- (0.,5.) -- cycle;
					\draw[sty1] (-0.866025,2.5) -- (-0.866025,3.5) -- (0.,3.) -- (0.,2.) -- cycle;
					\draw[sty1] (-2.59808,-2.5) -- (-2.59808,-1.5) -- (-1.73205,-2.) -- (-1.73205,-3.) -- cycle;
					\draw[sty1] (0.,1.) -- (0.,2.) -- (0.866025,1.5) -- (0.866025,0.5) -- cycle;
					\draw[sty1] (-0.866025,-0.5) -- (-0.866025,0.5) -- (0.,0.) -- (0.,-1.) -- cycle;
					\draw[sty1] (0.866025,-0.5) -- (0.866025,0.5) -- (1.73205,0.) -- (1.73205,-1.) -- cycle;
					\draw[sty1] (0.,-3.) -- (0.,-2.) -- (0.866025,-2.5) -- (0.866025,-3.5) -- cycle;
					\draw[sty1] (1.73205,-2.) -- (1.73205,-1.) -- (2.59808,-1.5) -- (2.59808,-2.5) -- cycle;
					\draw[sty1] (3.4641,-2.) -- (3.4641,-1.) -- (4.33013,-1.5) -- (4.33013,-2.5) -- cycle;
					\draw[sty1] (2.59808,-3.5) -- (2.59808,-2.5) -- (3.4641,-3.) -- (3.4641,-4.) -- cycle;
					\draw[sty1] (4.33013,-3.5) -- (4.33013,-2.5) -- (5.19615,-3.) -- (5.19615,-4.) -- cycle;
					\draw[sty1] (0.,7.) -- (-0.866025,6.5) -- (0.,6.) -- (0.866025,6.5) -- cycle;
					\draw[sty1] (0.,4.) -- (-0.866025,3.5) -- (0.,3.) -- (0.866025,3.5) -- cycle;
					\draw[sty1] (-1.73205,-1.) -- (-2.59808,-1.5) -- (-1.73205,-2.) -- (-0.866025,-1.5) -- cycle;
					\draw[sty1] (0.866025,2.5) -- (0.,2.) -- (0.866025,1.5) -- (1.73205,2.) -- cycle;
					\draw[sty1] (0.,1.) -- (-0.866025,0.5) -- (0.,0.) -- (0.866025,0.5) -- cycle;
					\draw[sty1] (1.73205,1.) -- (0.866025,0.5) -- (1.73205,0.) -- (2.59808,0.5) -- cycle;
					\draw[sty1] (0.866025,-1.5) -- (0.,-2.) -- (0.866025,-2.5) -- (1.73205,-2.) -- cycle;
					\draw[sty1] (2.59808,-0.5) -- (1.73205,-1.) -- (2.59808,-1.5) -- (3.4641,-1.) -- cycle;
					\draw[sty1] (4.33013,-0.5) -- (3.4641,-1.) -- (4.33013,-1.5) -- (5.19615,-1.) -- cycle;
					\draw[sty1] (3.4641,-2.) -- (2.59808,-2.5) -- (3.4641,-3.) -- (4.33013,-2.5) -- cycle;
					\draw[sty1] (5.19615,-2.) -- (4.33013,-2.5) -- (5.19615,-3.) -- (6.06218,-2.5) -- cycle;
				\end{scope}
				\node at (0,-2) {c) ${\rm Rem}(\bpi)$}; 
			\end{scope}
		\end{tikzpicture}
		\caption{An example of plane partition $\bpi$ (white), its set of addable boxes ${\rm Add}(\bpi)$ (red), and its set of removable boxes ${\rm Rem}(\bpi)$ (blue).}\label{fig:AddRem}
	\end{center}
\end{figure}
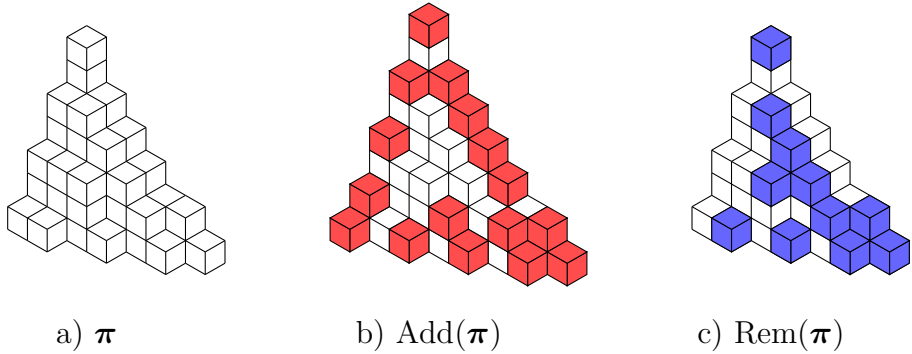

For a solid partition $\bpi$, we define two sets of boxes
\begin{equation}
	{\rm Add}(\bpi)\subset \IZ_{\geq 0}^4
	\qquad \textrm{and} \qquad 
	{\rm Rem}(\bpi)\subset \IZ_{\geq 0}^4    
\end{equation}
as positions in the 4D lattice where  the corresponding boxes can be \emph{added} to $\bpi$ or \emph{removed} from $\bpi$ in such a way that the resulting configurations are again solid partitions. (See Fig.~\ref{fig:AddRem} for the illustration of the 3D counterparts.)

In what follows, we will work with the weight space.
To each field $B_i$, we assign a complex weight (or flavor parameter) $\myh_i$.
These weights measure the equivariant toric action on the CY${}_4$:
\begin{equation}
	(x,y,z,w)\mapsto(e^{\myh_1}x,e^{\myh_2}y,e^{\myh_3}z,e^{\myh_4}w)    
\end{equation} 
and satisfy the Calabi-Yau condition:
\begin{equation}\label{CY}
	\myh_1+\myh_2+\myh_3+\myh_4=0\,.
\end{equation}

Consider a projection:
\begin{equation}
	{\bf prj}:\quad (i,j,k,l)\longmapsto i\,\myh_1+j\,\myh_2+k\,\myh_3+l\,\myh_4\,.
\end{equation}
The physical meaning of this projection is the following: 
an operator $B_1^iB_2^jB_3^kB_4^lI$ which acquires an expectation value in the SUSY vacuum has a flavor charge given by ${\bf prj}(i,j,k,l)$.

We assume that the weights $\myh_k$ are generic complex numbers that satisfy \eqref{CY}.
Therefore three of these numbers, say $\myh_{1,2,3}$, form a non-reduced basis in the module $\IZ\myh_1+\IZ\myh_2+\IZ\myh_3$ over $\IZ$, so that the complex numbers $(i\,\myh_1+j\,\myh_2+k\,\myh_3)$ and $(i'\,\myh_1+j'\myh_2+k'\myh_3)$ for two different tuples $(i,j,k)$ and $(i',j',k')$ are linearly independent over $\IZ$.
In other words, if one has a decomposition of a number $c$:
\begin{equation}\label{eq:decomp}
	c=i\,\myh_1+j\,\myh_2+k\,\myh_3,\quad i,j,k\in\IZ\,,
\end{equation}
then this decomposition is unique.

\begin{figure}[ht!]
	\begin{center}
		\begin{tikzpicture}[scale=2.5]
			\draw (0.,0.) -- (-0.363954,-0.309017) (0.,0.) -- (0.92388,0.) (0.,0.) -- (-0.118256,0.951057) (-0.363954,-0.309017) -- (0.559926,-0.309017) (-0.363954,-0.309017) -- (-0.482209,0.64204) (0.92388,0.) -- (0.559926,-0.309017) (0.92388,0.) -- (0.805624,0.951057) (0.559926,-0.309017) -- (0.44167,0.64204) (-0.118256,0.951057) -- (-0.482209,0.64204) (-0.118256,0.951057) -- (0.805624,0.951057) (-0.482209,0.64204) -- (0.44167,0.64204) (0.805624,0.951057) -- (0.44167,0.64204);
			\begin{scope}[thick,burgundy,decoration={markings, 
					mark= at position 0.6 with {\arrowreversed{stealth}}}]
				\draw[postaction={decorate}] (0.,0.) -- (0.220835,0.32102) node[pos=0.7,left] {$\myh_1$}; \draw[postaction={decorate}] (0.559926,-0.309017) -- (0.220835,0.32102) node[pos=0.75,right] {$\myh_2$}; \draw[postaction={decorate}] (-0.482209,0.64204) -- (0.220835,0.32102) node[pos=0.5,below left] {$\myh_3$}; \draw[postaction={decorate}] (0.805624,0.951057) -- (0.220835,0.32102) node[pos=0.5,below right] {$\myh_4$};
			\end{scope}
			\foreach \x / \y in {0./0., -0.363954/-0.309017, 0.92388/0., 0.559926/-0.309017, -0.118256/0.951057, -0.482209/0.64204, 0.805624/0.951057, 0.44167/0.64204, 0.220835/0.32102}
			{
				\draw[fill=\myblue] (\x,\y) circle (0.05);
			}
		\end{tikzpicture}
		\caption{A parameterization of 3D BCC Bravais lattice $\mathscr{B}$.}\label{fig:voronoi}
	\end{center}
\end{figure}
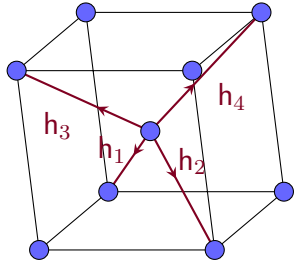

As we have just seen, the projection operator $\bf prj$ maps the positive sedecant (i.e.\ 4D orthant) $\IZ_{\geq 0}^4$, where the 4D crystal associated with the solid partition $\bpi$ is located, into a 3D body-centered cubic (BCC) Bravais lattice, denoted as $\mathscr{B}$ (see Fig.~\ref{fig:voronoi}).
Note that following \eqref{eq:decomp} (which in turn due to the Calabi-Yau constraint \eqref{CY}), the weight space is equivalent to the 3D BCC lattice $\mathscr{B}$, therefore we will use $\myh_{1,2,3,4}$ (subject to $\sum^{4}_{i=1}\myh_i$) to denote both the basis (as complex numbers) of the weight space and the basis (as 3D vectors) of the lattice  $\mathscr{B}$.\footnote{
	For example, we can draw the 3D lattice $\mathscr{B}$ by choosing the following representation for its vectors:
	\begin{equation}\nn
		\myh_1\mapsto\left(
		1, \, -1,\, -1
		\right)
		\,, \quad		
		\myh_2\mapsto\left(
		-1,\, 1,\, -1
		\right)
		\,, \quad
		\myh_3\mapsto\left(
		1,\, 1,\, 1
		\right)
		\,, \quad 
		\myh_4\mapsto\left(
		-1,\, -1,\, 1
		\right)
		\;.
\end{equation}}
Correspondingly, we can use the same letter, such as $z,u,c$ etc, to denote both a point in the 3D lattice $\mathscr{B}$ and a point in the weight space, which the charge function is a meromorphic function of.

Voronoi cells (``bricks") of the BCC lattice are truncated octahedra and furnish a tessellation of $\IR^3$. 
A boundary of a solid partition consist of  plaques of 3D cubes, which can be mapped to parallelepipeds under $\bf prj$. 
Therefore tessellations of $\IR^3$ by parallelepipeds are in 1-to-1 correspondence with solid partitions $\bpi$. See Fig.~\ref{fig:tess} for two examples.

\begin{figure}[ht!]
	\begin{center}
		\begin{tikzpicture}
			\node at (0,0) {$\begin{array}{c}
					\begin{tikzpicture}[scale=1.5]
						\draw[dashed] (0.,0.) -- (0.353553,-0.5) (-0.612372,0.5) -- (-0.258819,0.) (0.353553,-0.5) -- (-0.258819,0.) (0.,-1.) -- (0.353553,-0.5) (-0.612372,-0.5) -- (-0.258819,0.) (0.353553,-0.5) -- (0.965926,0.) (0.,1.) -- (0.353553,0.5) (0.353553,0.5) -- (0.965926,0.) (0.353553,0.5) -- (-0.258819,0.);
						\draw[ultra thick,\myblue] (0.,0.) -- (-0.612372,0.5) (0.,0.) -- (0.353553,-0.5) (0.,0.) -- (-0.353553,-0.5) (0.,0.) -- (0.612372,0.5);
						\draw[thick] (0.,-1.) -- (-0.353553,-0.5) (0.,-1.) -- (-0.612372,-0.5) (0.,-1.) -- (0.612372,-0.5) (0.,1.) -- (-0.353553,0.5) (0.,1.) -- (-0.612372,0.5) (0.,1.) -- (0.612372,0.5) (-0.353553,-0.5) -- (-0.965926,0.) (-0.353553,-0.5) -- (0.258819,0.) (-0.353553,0.5) -- (-0.965926,0.) (-0.353553,0.5) -- (0.258819,0.) (-0.612372,-0.5) -- (-0.965926,0.) (-0.612372,0.5) -- (-0.965926,0.) (0.612372,-0.5) -- (0.258819,0.) (0.612372,-0.5) -- (0.965926,0.) (0.612372,0.5) -- (0.258819,0.) (0.612372,0.5) -- (0.965926,0.);
					\end{tikzpicture}
				\end{array}$};
			\node at (0,-2) {a) Empty 4D corner}; 
			\node at (7,0) {$\begin{array}{c}
					\begin{tikzpicture}[scale=1.5]
						\draw[dashed] (0.,1.) -- (0.353553,0.5) (0.353553,0.5) -- (0.965926,0.) (0.,0.) -- (0.353553,0.5) (-0.612372,0.5) -- (-0.258819,0.) (0.353553,0.5) -- (-0.258819,0.) (-0.612372,-0.5) -- (-0.258819,0.) (0.353553,-0.5) -- (0.965926,0.) (0.353553,-0.5) -- (-0.258819,0.) (0.,-1.) -- (0.353553,-0.5);
						\draw[ultra thick,\myblue] (0.,0.) -- (0.353553,0.5) (0.,0.) -- (-0.353553,0.5) (0.,0.) -- (0.612372,-0.5) (0.,0.) -- (-0.612372,-0.5);
						\draw[thick] (0.,-1.) -- (-0.353553,-0.5) (0.,-1.) -- (-0.612372,-0.5) (0.,-1.) -- (0.612372,-0.5) (0.,1.) -- (-0.353553,0.5) (0.,1.) -- (-0.612372,0.5) (0.,1.) -- (0.612372,0.5) (-0.353553,-0.5) -- (-0.965926,0.) (-0.353553,-0.5) -- (0.258819,0.) (-0.353553,0.5) -- (-0.965926,0.) (-0.353553,0.5) -- (0.258819,0.) (-0.612372,-0.5) -- (-0.965926,0.) (-0.612372,0.5) -- (-0.965926,0.) (0.612372,-0.5) -- (0.258819,0.) (0.612372,-0.5) -- (0.965926,0.) (0.612372,0.5) -- (0.258819,0.) (0.612372,0.5) -- (0.965926,0.);
					\end{tikzpicture}
				\end{array}$};
			\node at (7,-2) {b) One hypercubic box in the 4D corner };    
		\end{tikzpicture}
\caption{Two tessellations using 3D Voronoi cells, whose plane partition counterparts are tessellations (8) and (1) in Fig.\ \ref{loc_pic_plane}, respectively.
		}\label{fig:tess}
	\end{center}
\end{figure}

Let us consider the lift map:
\begin{equation}\label{eq:lift}
	\ell:\quad \mathscr{B}\longrightarrow \IZ^4_{\geq 0}\,,
\end{equation}
where the quadruplet $\ell_{k=1,\ldots,4}$ of non-negative integers is defined uniquely in terms of a vector $c\in\mathscr{B}$ by the following rule:
\begin{equation}
	\exists!\; \{\ell_k\}_{k=1,\ldots,4}:\quad c=\sum\lm_{i=1}^4\ell_i\myh_i,\quad\ell_i\geq 0,\quad 0\in\{\ell_k\}_{k=1,\ldots,4}.
\end{equation}

Given a solid partition $\bpi$, we can translate it to the field of heights $\mu(c)$, defined as  the number of boxes projected to the same point $c\in\mathscr{B}$ by $\bf prj$.

Then using the lift map \eqref{eq:lift}, we can map a field of height $\mu(c)$ back to a subset of nodes of $\IZ_{\geq 0}^4$:
\begin{equation}
	L: \;\mu(c)\mapsto 
	\left\{\left(\begin{array}{cccc}
		\ell_1(c)+u \,, & \ell_2(c)+u \,, & \ell_3(c)+u \,, & \ell_4(c)+u
	\end{array}\right)\right\}_{u=0}^{\mu-1}\,.
\end{equation}

The height field $\mu$ is a section of the $\IZ_{\geq 0}^4\to \mathscr{B}$ fibration with a non-trivial connection.
Let us parallel transport a point $(c\in\mathscr{B},\mu\in\IZ_{\geq 0})$ along the vector $\myh_k$ in the following way.
First, we lift it with $L$ to the top-most box in $\IZ_{\geq 0}^4$, then we move this point with the 4D lift $e_k$ of the vector $\myh_k$, and finally we project back to $\mathscr{B}$ with $\bf prj$.
The result of this parallel-transport operation is:
\begin{equation}
	\mathsf{T}_k^{\rm 3D}={\bf prj}\circ \mathsf{T}_k^{\rm 4D}\circ L:\qquad (c,\mu)\mapsto(c+\myh_k\,,\mu+\Delta_k(c))\,,
\end{equation}
where 
\begin{equation}\label{edge_weight}
	\Delta_k(c):=\delta_{0,\ell_{k}(c)}\,\delta_{0,\ell_{k}(c+\myh_k)}\,.
\end{equation}
As we can see, the connection $\Delta_k(c)$ acquires only the values 0 or 1 and it depends only on the edge $(c\to c+\myh_k)\in\mathscr{B}$, rather than on the fiber value $\mu$, therefore we will call the $\Delta_k(c)$ the edge weight.

Now we translate the melting rule \eqref{melting} for the 4D crystal, which ensures that $\bpi$ is indeed a solid partition, into a constraint on the height fields $\mu(c)$.
The melting rule \eqref{melting} simply means that when one parallel transports the top box hovering over the projection $c\in\mathscr{B}$ along $\myh_k$, it should not get below the top box hovering over $c+\myh_k\in\mathscr{B}$.
Thus we arrive at the following rewriting of the melting rule \eqref{melting} in terms of the height field:
\begin{tcolorbox}
	\begin{equation}\label{ineq}
		\mu\left(c+\myh_k\right)\leq \mu\left(c\right)+\Delta_{k}(c),\qquad \forall c\in\mathscr{B},\;k=1,\ldots,4\,.
	\end{equation}
\end{tcolorbox}

\section{Charge function for solid partitions}
\label{sec:ChargeFunction}

\subsection{Strategy}

As summarized in Sec.\ \ref{sec:IntroSummary}, our ultimate goal is to construct the BPS algebra that acts on the set of solid partitions in a manner similar to the 3D case. 
As reviewed in App.\ \ref{s:3D_par}, the crucial ingredients in the construction of the BPS algebra for the $\mathbb{C}^3$ case is the charge function \eqref{ch_f_3D} that satisfies the properties listed in Sec.\ \ref{list:properties3D}.
Therefore, for the 4D case, we will aim to construct a charge function $\psi_\bpi(z)$ on solid partitions $\bpi$ satisfying the following properties:
\begin{enumerate}
	\item $\psi_\bpi(z)$ is a meromorphic function of $z$.
	\item All the poles of $\psi_\bpi(z)$ are simple
	\item All the poles of $\psi_\bpi(z)$ are in 1-to-1 correspondence with the set of projected coordinates  ${\bf prj} \left(\vec x(\Box)\right)$, with the boxes
	$\Box \in {\rm Add}(\bpi)\cup{\rm Rem}(\bpi)$ and $\vec{x}(\Box)$ their 4D coordinates.
\end{enumerate}

In the 3D case, the plane partitions represent BPS states of D-branes on the simplest toric CY${}_3$: $\IC^3$, and the corresponding BPS algebra is the affine Yangian of $\mathfrak{gl}_1$, whose Cartan generators on vector $|\bpi\rangle$ have eigenvalues that can be packaged in $\psi_{\bpi}(z)$ \cite{Li:2020rij}, see \eqref{ch_f_3D}.
The 4D charge function $\psi_{\bpi}(z)$ that we are trying to define in this section will have an analogous physical meaning in 4D.

As we will see shortly, a simple definition that is analogous to the 3D version \eqref{ch_f_3D}, namely \eqref{ch_f_4D_start_1}, which includes contributions only from individual boxes,
together with the constraint \eqref{CY}, fails already at the two-box level.

Our strategy is to use eq.\ \eqref{eq:phi14D} below as a starting point, and implement the process of growing solid partitions level by level starting from the vacuum. 
At each level, the requirements from the list above will demand introducing additional contributions to the charge function. 
We will see that not only do we have to modify the contribution from each individual 4D box \eqref{eq:phi14D}, but we also need to include contributions from certain 4-box and 5-box clusters. 
We arrive at a conjectured form of the charge function by implementing this process from the vacuum (level-0) to level-15. 
Then we use computer to check higher levels and finally give a proof in Sec.\ \ref{ssec:Proof}.

\subsection{Charge function conjectured}

In this subsection, we will try to construct an expression for the charge function that obeys all the requirements listed above. 
The full solid partition is an uplift of a 3D periodic quiver, namely it can be grown from the vacuum layer by layer according to the 3D periodic quiver.
For example, consider the process that starts from the vacuum and reaches the first box in the 2$^{\textrm{nd}}$ layer:
\begin{equation}
	\textrm{vacuum} 
	\longrightarrow  (0,0,0,0)
	\longrightarrow  \vec{e}_i 
	\longrightarrow \vec{e}_i+\vec{e}_{j\neq i} 
	\longrightarrow \vec{e}_i+\vec{e}_{j\neq i} +\vec{e}_{k\neq j\neq i}
	\longrightarrow \sum^4_{i=1}\vec{e}_i
\end{equation}
In this process, we try to define the charge function in such a way that for each solid partition configuration, the set of poles of its charge function accounts for all the adding poles and removing poles. 
Apart from the condition that it has to provide all the necessary poles, we also require that there are no spurious poles (i.e.\ poles that do not belong to the projection of coordinates ${\bf prj} \left(\vec x(\Box)\right)$ of the boxes $\Box\in {\rm Add}(\bpi)\cup{\rm Rem}(\bpi)$).  
\begin{enumerate}
	\item $\textrm{vacuum} \longrightarrow  (0,0,0,0)$.\\
	From the vacuum, i.e.\ the empty 4D room, we can only add a single box, at coordinate $(0,0,0,0)$, with equivariant weight $\myh_{\square}=0$. 
	In order to add this box, we need the contribution from the vacuum to the charge function to be:
	\begin{equation}
		\psi_{\bpi}(z) \ni \frac{1}{z} \,.
	\end{equation}
	\item $(0,0,0,0) \longrightarrow  \vec{e}_i$.\\
	From the first box at $(0,0,0,0)$, we can add all its four nearest neighbors, namely the boxes $\square_i$ at $\vec{e}_i$, with weight $\myh_{\square}=\myh_i$, where $i=1,2,3,4$.
	In order to add these four boxes, we need:
	\begin{equation}\label{ch_f_4D_start_1}
		\psi_{\bpi}(z)=\frac{1}{z}\prod\lm_{\Phi_1\in\bpi}\varphi_1(z-c(\Phi_1))\,,
	\end{equation}
	where each individual 4D box $\Phi_1$ contributes  a bonding factor
	\begin{equation}\label{eq:phi14D}
		\varphi_1(z)=\prod\lm_{i=1}^4\frac{z+\myh_i}{z-\myh_i}\,,
	\end{equation}
	together with the constraint \eqref{CY}.
	The poles guarantee that we can add the boxes, and they are in one-to-one correspondence with the projections ${\bf prj}({\vec e}_i)$; whereas the zeros are to cancel the pole of the box at the origin so that it can no longer be removed once some of its nearest neighbors have been added. 
	This is parallel to the 3D case \eqref{ch_f_3D}. 
	
	\item $\vec{e}_i \longrightarrow \vec{e}_i+\vec{e}_{j\neq i}$.\\
	This is the first step where we see the need to modify the expression \eqref{ch_f_4D_start_1} with \eqref{eq:phi14D}. 
	First, we know from the melting rule \eqref{melting} that in order to add the box at $\vec{e}_i+\vec{e}_{j\neq i}$, we need both the boxes at $\vec{e}_i$ and $\vec{e}_{j\neq i}$ to be present first.
	
	Consider the configuration with two boxes, one at the origin and one at one of its nearest neighbor sites, say at $\vec{e}_1$. 
	Its charge function is
	\begin{equation}
		\begin{aligned}
			\psi_{\square_0\square_1}(u)&=\frac{1}{u}\cdot \varphi_1(u)\cdot \varphi_1(u-\myh_i)\\
			&=\frac{1}{\cancel{u}}\cdot \frac{(u+\myh_1)(u+\myh_2)(u+\myh_3)(u+\myh_4)}{(u-\myh_1)(u-\myh_2)(u-\myh_3)(u-\myh_4)}\\
			&\qquad  \cdot \frac{\cancel{u}(u+\myh_2-\myh_1)(u+\myh_3-\myh_1)(u+\myh_4-\myh_1)}{(u-2\myh_1)(u-\myh_2-\myh_1)(u-\myh_3-\myh_1)(u-\myh_4-\myh_1)}\,.
		\end{aligned}
	\end{equation}
	We immediately see the difference from the 3D case: now there are three spurious poles at
	\begin{equation}
		\myh_1+\myh_j \qquad\textrm{with} \qquad  j=2,3,4\,,
	\end{equation}
	corresponding to adding of the three boxes at $\vec{e}_1+\vec{e}_{j\neq 1}$,\footnote{
		The reason that these poles are spurious is that we need both the boxes at $\vec{e}_i$ and $\vec{e}_{j\neq i}$ to be present first in order to add the box at $\vec{e}_i+\vec{e}_{j\neq i}$.} which haven't been canceled by the zeros from $\varphi_1(u)$ defined in \eqref{eq:phi14D}.
	To remedy this, we modify the contribution from individual boxes \eqref{eq:phi14D} to\footnote{
		Another choice is to introduce the contribution from the 2-box cluster to cancel this spurious pole, but this is a less efficient route since we would soon need to introduce the contribution from some additional 2-box cluster (with different type of configuration) at level-3.}
	\begin{equation}\label{eq:phi14Dnew}
		\varphi_1(z)=\prod\lm_{i=1}^4\frac{z+\myh_i}{z-\myh_i} \prod_{1\leq i < j \leq 4}(z+\myh_i+\myh_j)
	\end{equation}
	together with the constraint \eqref{CY}.
	This modification cures the problem with spurious poles for two-box configurations.
	\item $\vec{e}_i+\vec{e}_{j\neq i} 
	\longrightarrow \vec{e}_i+\vec{e}_{j\neq i} +\vec{e}_{k\neq j\neq i}$.\\
	At this step, we will see the need to include contributions from certain clusters of 4-boxes, apart from those from individual boxes.
	First, we know from the melting rule \eqref{melting} that in order to add the box at $\vec{e}_i+\vec{e}_{j\neq i} +\vec{e}_{k\neq j\neq i} $, we need all three boxes at 
	\begin{equation}
		\vec{e}_i+\vec{e}_{j\neq i} 
		\,,\qquad 
		\vec{e}_i+\vec{e}_{k\neq j\neq i}
		\,,\qquad 
		\vec{e}_{j\neq i} +\vec{e}_{k\neq j\neq i} 
	\end{equation}
	to be present first, which in turn need the three boxes at
	\begin{equation}
		\vec{e}_i 
		\,,\qquad 
		\vec{e}_{j\neq i} 
		\,,\qquad 
		\vec{e}_{k\neq j\neq i} 
	\end{equation}
	to be present, which in turn need the box at the origin to be present.
	Now consider the $1+3+3=7$-box configuration, with the 7 boxes at the coordinates
	\begin{equation}
		(0,0,0,0)
		\,, \quad  
		\vec{e}_i
		\,,\quad  
		\vec{e}_j
		\,,\quad  
		\vec{e}_k
		\,,\quad  
		\vec{e}_i+ \vec{e}_j
		\,,\quad  
		\vec{e}_j +\vec{e}_k
		\,,\quad  
		\vec{e}_k +\vec{e}_i \,,
	\end{equation}
	with $i\neq j \neq k$.
	Its charge function \eqref{ch_f_4D_start_1} with the contribution from individual boxes given by \eqref{eq:phi14Dnew}, instead of having a desired pole at $\myh_i+\myh_j+\myh_k$, has a zero there instead!
	To remedy this,\footnote{
		Note that it is impossible to cure this new problem by simply modifying the single box contribution, as what we have done for the spurious poles from the 2-box configurations, without spoiling the correct charge function properties for configurations with fewer than 7 boxes.} we introduce a contribution 
	\begin{equation}
		\varphi_{4,\ell}(u)=\frac{1}{(u+\myh_{\ell})^2}
	\end{equation}
	from the 4-box cluster at
	\begin{equation}
		(0,0,0,0)
		\,, \quad  
		\vec{e}_i
		\,,\quad  
		\vec{e}_j
		\,,\quad  
		\vec{e}_k
	\end{equation}
	with $i\neq j \neq k \neq \ell$.
	More generally, consider a 4-box cluster with 
	\begin{equation}
		\begin{aligned}
			&\square 
			\,,\quad  
			&\square+\vec{e}_i  
			\,,\quad  
			\square+\vec{e}_j  
			\,,\quad  
			\square+\vec{e}_k \,.
		\end{aligned}
	\end{equation}
	It contributes a factor 
	\begin{equation}
		\varphi_{4,\ell}(u-\myh_{\square})=\frac{1}{(u+\myh_{\ell}-h_{\square})^2}\,,
	\end{equation}
	where $(i,j,k,\ell)$ are a permutation of $(1,2,3,4)$.
	Namely, now 
	\begin{equation}\label{eq:chargefunction14}
		\psi_{\bpi}(z)=\frac{1}{z}\prod\lm_{\Phi_1\in\bpi} \varphi_1\left(z-c\left(\Phi_1\right)\right)\prod\lm_{\Phi_{4,\ell}\in\bpi} \varphi_{4,\ell}\left(z-c\left(\Phi_{4,\ell}\right)\right)\,.
	\end{equation}
	
	\item $\vec{e}_i+\vec{e}_{j\neq i} +\vec{e}_{k\neq j\neq i}
	\longrightarrow \sum^4_{i=1}\vec{e}_i$\\
	Now we will demonstrate the need of introducing contributions from certain 5-box clusters.
	First, we know from the melting rule \eqref{melting} that in order to add the box at $\sum^4_{i=1}\vec{e}_i $, we need all four boxes at 
	\begin{equation}
		(\sum^4_{i=1} \vec{e}_i) -\vec{e}_j \qquad \textrm{with} \qquad j=1,2,3,4
	\end{equation}
	to be present first, which in turn need the 6 boxes at
	\begin{equation}
		\vec{e}_i+\vec{e}_{j\neq i} \qquad i,j=1,2,3,4
	\end{equation}
	to be present, which need all 4 boxes at 
	\begin{equation}
		\vec{e}_i \qquad\textrm{with} \qquad i=1,2,3,4
	\end{equation}
	to be present, which in turn need the box at the origin to be present. 
	
	Consider the $1+4+6+2=13$-box state
	\begin{equation}
		\begin{aligned}
			&(0,0,0,0) \,,\quad  \\
			&\vec{e}_i \,,\quad i=1,2,3,4 \,,\\
			& \vec{e}_i+\vec{e}_j  \,, \quad i\neq j = 1,2,3,4 \,, \\
			& (\sum^4_{i=1} \vec{e}_i) -\vec{e}_j \,, \quad j \textrm{ takes two value from } \{1,2,3,4\} \,.
		\end{aligned}
	\end{equation}
	For this configuration, the charge function \eqref{eq:chargefunction14} that we have determined so far has a single pole at $z=0$, which we know to be spurious from the melting rule \eqref{melting}. 
	Similarly, the $1+4+6+3=14$-box state
	\begin{equation}
		\begin{aligned}
			&(0,0,0,0)\,, \quad  \\
			&\vec{e}_i \,,\quad i=1,2,3,4\,, \\
			& \vec{e}_i+\vec{e}_j  \,,\quad i\neq j = 1,2,3,4\,, \\
			& (\sum^4_{i=1} \vec{e}_i) -\vec{e}_j \,,\quad j \textrm{ takes three value from } \{1,2,3,4\} 
		\end{aligned}
	\end{equation}
	has a double pole at $z=0$, which is also spurious.
	To remedy the two problems, we introduce a contribution from the 5-box cluster at
	\begin{equation}
		\begin{aligned}
			&(0,0,0,0) \,, \quad  \vec{e}_i \quad i=1,2,3,4 \\
		\end{aligned}
	\end{equation}
	with
	\begin{equation}\label{eq:4D5cluster}
		\varphi_{5}(u)=u^2
	\end{equation}
	Now we check the  $1+4+6+4=15$-box configuration with 
	\begin{equation}
		\begin{aligned}
			&(0,0,0,0)\,, \quad  \\
			&\vec{e}_i\,, \quad i=1,2,3,4\,, \\
			& \vec{e}_i+\vec{e}_j \,, \quad i\neq j = 1,2,3,4 \,, \\
			& (\sum^4_{i=1} \vec{e}_i) -\vec{e}_j \,,\quad j=1,2,3,4 
		\end{aligned}
	\end{equation}
	This is the configuration from which one can add the first box at the 2nd layer, at $(1,1,1,1)$.
	Including the additional contribution \eqref{eq:4D5cluster}, the charge function now has the correct single pole at $z=0$!
	(Otherwise, it would have had a triple pole at $z=0$.)
	More generally, for a 5-box configuration with 
	\begin{equation}
		\begin{aligned}
			&\square\,, \quad  
			&\square+\vec{e}_i \quad i=1,2,3,4 \,,
		\end{aligned}
	\end{equation}
	it contributes a factor 
	\begin{equation}
		\varphi_{5}(u-\myh_{\square})=(u-\myh_{\square})^2
	\end{equation}
\end{enumerate}

Summarizing, these iterative considerations fix the charge function for a solid partition $\bpi$ to be
\begin{tcolorbox}
	\begin{equation}\label{ch_f}
		\psi_{\bpi}(z)=\frac{1}{z}\prod\lm_{\Phi_1\in\bpi} \varphi_1\left(z-c\left(\Phi_1\right)\right)\prod\lm_{\Phi_{4,k}\in\bpi} \varphi_{4,k}\left(z-c\left(\Phi_{4,k}\right)\right)\prod\lm_{\Phi_5\in\bpi} \varphi_5\left(z-c\left(\Phi_5\right)\right)\,,
	\end{equation}
\end{tcolorbox}
\noindent 
with 4 types of contributions:
\begin{itemize}
	\item The first factor $\frac{1}{z}$ is the contribution from the vacuum, with a pole at $0$ that allows the first box to be added.
	\item[\textcolor{\myblue}{\textbullet}]
	A singlet $\Phi_1$ is a single box $\Box\in\bpi$. 
	The contribution is the function 
	\begin{equation}
		\varphi_1(z)=\prod\lm_{i=1}^4\frac{z+\myh_i}{z-\myh_i}\times\prod\lm_{1\leq i<j\leq 4}(z+\myh_i+\myh_j)\,,\\
	\end{equation}
	shifted by the coordinate $c\left(\Phi_1\right)$, which is the projected coordinate of the box ${\bf prj}\left(\vec x\left(\Box\right)\right)$.
	\item[\textcolor{\myblue}{\textbullet}]
	A \emph{fan quadruplet} $\Phi_{4,k}$ is a quadruplet of boxes $\Box_{i=1,2,3,4}\in\bpi$ satisfying the coordinate constraint:
	\begin{equation}
		\vec x(\Box_2)-\vec x(\Box_1)=\vec e_a,\quad \vec x(\Box_3)-\vec x(\Box_1)=\vec e_b,\quad \vec x(\Box_4)-\vec x(\Box_1)=\vec e_c\,,
	\end{equation}
	where the indices $\{a,b,c,k\}$ represent any permutation of $\{1,2,3,4\}$.
	Its contribution is the function
	\begin{equation}
		\varphi_{4,k}(z)=\frac{1}{\left(z+\myh_k\right)^2},\qquad k=1,2,3,4 \,,\\
	\end{equation}
	shifted by the coordinate  $c\left(\Phi_{4}\right)$, which is the projected coordinate of the first box ${\bf prj}\left(\vec x\left(\Box_1\right)\right)$.
	
	\item[\textcolor{\myblue}{\textbullet}]
	A \emph{quintuplet} $\Phi_{5}$ is a quintuplet of boxes $\Box_{i=1,2,3,4,5}\in\bpi$ satisfying the coordinate constraint:
	\begin{equation}
		\vec x(\Box_{i+1})-\vec x(\Box_1)=\vec e_i,\;\;\forall i=1,2,3,4\,.
	\end{equation}
	Its contribution is the function
	\begin{equation}
		\varphi_{5}(z)=z^2\,.
	\end{equation}
	shifted by the coordinate $c\left(\Phi_{5}\right)$, which is the projected coordinate of the first box ${\bf prj}\left(\vec x\left(\Box_1\right)\right)$.
\end{itemize}

We have checked using computer that the charge function \eqref{ch_f} obeys the required properties for all solid partitions up to 19 boxes, and for all solid partitions confined in a $3\times 3\times 3\times 3$ hypercube.
In the next subsection, we will prove that the charge function \eqref{ch_f} obeys the required properties for \emph{all} solid partitions. 

\subsection{Proof}
\label{ssec:Proof}
In this subsection, we give a proof that the charge function \eqref{ch_f} satisfies the requirements for all solid partition configurations.
The proof proceeds in three steps:
\begin{enumerate}
	\item First we translate the expression of the charge function\eqref{ch_f}, which is a function directly of the solid partition configurations, into the potential function $w(c)[\mu]$ of the lattice point $c\in\mathscr{B}$ and the height field $\mu(x)$.
	The two functions contain the same amount of information.
	\item We then show that $w(c)[\mu]$ depends only on the value of the height field $\mu$ at point $c$ and some of its nearest and next-to-nearest neighbor points in the lattice $\mathscr{B}$.
	We call this small collection of neighbor points of the lattice $\mathscr{B}$ a local patch and prove that the set of inequalities \eqref{ineq} has only a  \emph{finite} number of integral solutions on any local patches.
	\item Finally, we simply analyze the behavior of the charge function \eqref{ch_f} via its corresponding potential function $w(c)[\mu]$ for all solutions computed in the previous step and show that the charge function \eqref{ch_f} has all the desired properties on all these solutions and thus conclude the proof.
\end{enumerate}

\subsubsection{Local potential}

Let us introduce the local potential $w$ for a lattice point $c\in\mathscr{B}$: it is defined as the order of the pole of the charge function $\psi_{\bpi}$ at point $c$:
\begin{equation}
	\psi_{\pi}(z)=\frac{\beta}{(z-c)^{w(c)}}\times(1+\mathcal{O}(z-c)),\quad \mbox{ as }z\to c\,.
\end{equation}
Apparently, when $w(c)$ is negative, $\psi_{\bpi}$ has a zero of order ``$-w$" at this point.

From \eqref{ch_f} we can derive the expression for the local potential $w(c)$ for the lattice point $c$ to be
\begin{equation}\label{weight}
	\begin{split}
		w(c)=&\sum\lm_{k=1}^4\left(\mu(c-\myh_k)-\mu(c+\myh_k)\right)-\sum\lm_{1\leq i < j\leq 4}\mu(c+\myh_i+\myh_j)+\\
		&\quad+2\sum_{k=1}^4\mathop{\rm min}\lm_{\substack{1\leq i\leq 4\\ i\neq k}}\big\{\mu(c+\myh_k+\myh_i)-\Delta_i(c+\myh_k)\big\}\\
		&\quad\quad-2\;\mathop{\rm min}\lm_{1\leq i\leq 4}\big\{\mu(c+\myh_i)-\Delta_i(c)\big\}+\delta_{c,0}\,,
	\end{split}
\end{equation}
where the first two terms come from individual boxes $\Phi_1$, the next two terms are the contributions from the quadruplets $\Phi_{4,k}$ and the quintuplets $\Phi_5$, respectively, and the last term is from the vacuum contribution in \eqref{ch_f}.
While the contribution from $\Phi_1$ is easy to understand, let us explain the other two in more detail. 

To capture the contribution from a cluster of boxes such as $\Phi_{4,k}$ and $\Phi_5$, we first need to determine all the situations when the position of the box $\Box$ and all its neighboring positions along the required edges $\vec e_{k}$ are filled.
Consider a row of $\mu(c)$ boxes hovering over some position $c$:
\begin{equation}
	\left\{
	\vec\ell(c)
	\,,\,
	\vec\ell(c)+\vec s 
	\,,\,
	\ldots
	\,,\,
	\vec\ell(c)+(\mu(c)-1)\vec s
	\right\}
\end{equation}
where
\begin{equation}
	\vec\ell(c):=\sum\lm_{k=1}^4\,\ell_k(c)\vec e_k
	\qquad\textrm{and}\qquad
	\vec s:=\sum\lm_{k=1}^4\vec e_k\,.
\end{equation}
Now we shift all these boxes by a vector $\vec e_k$.
This is equivalent to the parallel transport of the corresponding sections by $\mathsf{T}_k^{\rm 3D}$:
\begin{equation}\label{eq:Lambda1}
	\begin{split}
		\Lambda_1:=\Big\{\vec\ell(c+\myh_k)+\Delta_k(c)\vec s
		\,,\,
		\vec\ell(c+\myh_k)+(\Delta_k(c)+1)\vec s
		\,,\,
		&\ldots\\
		,\,
		&\vec\ell(c+\myh_k)+(\Delta_k(c)+\mu(c)-1)\vec s\Big\}\,.
	\end{split}
\end{equation}
Then we compare it with the row of boxes hovering over $c+\myh_k$:
\begin{equation}\label{eq:Lambda2}
	\Lambda_2:=\left\{\vec\ell(c+\myh_k)
	\,,\,
	\vec\ell(c+\myh_k)+\vec s
	\,,\,
	\ldots
	\,,\,
	\vec\ell(c+\myh_k)+(\mu(c+\myh_k)-1)\vec s\right\}\,.
\end{equation}
Using the set of inequalities \eqref{ineq} and $\Delta_k(c)\geq 0$, we find that the size of the intersection of two sets of boxes \eqref{eq:Lambda1} and \eqref{eq:Lambda2} equals:
\begin{equation}
	|\Lambda_1\cap\Lambda_2|=\mu(c+\myh_k)-\Delta_k(c)\,.
\end{equation}
To compute the size of the intersection of all sets over all required directions $\vec{e}_k$, we only need to apply the minimum function in \eqref{weight}.

\subsubsection{Solving inequalities locally}

Now let us consider a local ``patch'' of nodes in the lattice $\mathscr{B}$ that are close to point $c$:
\begin{equation}
	\Pi_c=\left\{c\right\}\cup\left\{c\pm\myh_k\right\}_{1\leq k\leq 4}\cup\left\{c+(\myh_i+\myh_j)\right\}_{\substack{1\leq i,j\leq 4\\ i\neq j}}\,,
\end{equation}
The first and second subsets of this patch represent the center and the 8 vertices of the cube in the 3D BCC Bravais lattice cell (see Fig.~\ref{fig:voronoi}), respectively, while the last subset represents the 6 boxes sitting at the centers of the 6 neighboring BCC cells, across each face of the cube.
Below we draw this patch with all the oriented edges $c\to c+\myh_k$ in it:
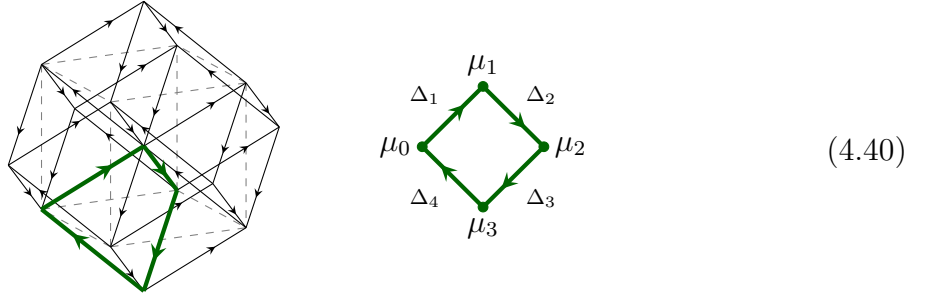
\begin{equation}\label{pic:localpatch}
	\begin{array}{c}
		\begin{tikzpicture}
			\foreach \x/\y/\z/\w in {0.437016/-0.585051/0.437016/1.33554, 0.437016/-0.585051/1.345/-1.08222, 0.437016/-0.585051/-1.345/-0.83837, -0.437016/-1.33554/-0.437016/0.585051, -0.437016/-1.33554/1.345/-1.08222, -0.437016/-1.33554/-1.345/-0.83837, -1.345/1.08222/-0.437016/0.585051, -1.345/1.08222/0.437016/1.33554, -1.345/1.08222/-1.345/-0.83837, 1.345/0.83837/-0.437016/0.585051, 1.345/0.83837/0.437016/1.33554, 1.345/0.83837/1.345/-1.08222}
			{
				\draw[dashed, thin, gray] (\x,\y) -- (\z,\w);
			}
			\foreach \x/\y/\z/\w in {-1.78201/-0.253319/-1.345/-0.83837, -1.78201/-0.253319/-0.437016/0.585051, -1.345/1.08222/-1.78201/-0.253319, -1.345/1.08222/-0.907981/0.497166, -1.345/1.08222/0./1.92059, -0.907981/0.497166/-1.345/-0.83837, -0.907981/0.497166/0.437016/1.33554, -0.437016/-1.33554/-1.78201/-0.253319, -0.437016/-1.33554/0./-1.92059, -0.437016/-1.33554/0.907981/-0.497166, -0.437016/0.585051/0./0., 0./-1.92059/1.345/-1.08222, 0./0./-1.345/1.08222, 0./0./-0.437016/-1.33554, 0./0./1.345/0.83837, 0./1.92059/-0.437016/0.585051, 0./1.92059/0.437016/1.33554, 0.437016/-0.585051/-0.907981/0.497166, 0.437016/-0.585051/1.78201/0.253319, 0.437016/1.33554/0./0., 0.907981/-0.497166/-0.437016/0.585051, 0.907981/-0.497166/1.345/-1.08222, 1.345/-1.08222/0./0., 1.345/0.83837/0./1.92059, 1.345/0.83837/0.907981/-0.497166, 1.345/0.83837/1.78201/0.253319, 1.78201/0.253319/0.437016/1.33554, 1.78201/0.253319/1.345/-1.08222}
			{
				\draw[postaction={decorate},decoration={markings, 
					mark= at position 0.7 with {\arrow{stealth}}}] (\x,\y) -- (\z,\w);
			}
			\foreach \x/\y/\z/\w in {0./0./0.437016/-0.585051, 0.437016/-0.585051/0./-1.92059, 0./-1.92059/-1.345/-0.83837, -1.345/-0.83837/0./0.}
			{
				\draw[ultra thick, \mygreen, postaction={decorate},decoration={markings, 
					mark= at position 0.7 with {\arrow{stealth}}}] (\x,\y) -- (\z,\w);
			}
		\end{tikzpicture}
	\end{array} \quad\quad
	\begin{array}{c}
		\begin{tikzpicture}[scale=0.8]
			\draw[ultra thick, \mygreen, postaction={decorate},decoration={markings, 
				mark= at position 0.7 with {\arrow{stealth}}}] (-1,0) -- (0,1) node[black, pos=0.5,above left] {$\scriptstyle \Delta_1$};
			\draw[ultra thick, \mygreen, postaction={decorate},decoration={markings, 
				mark= at position 0.7 with {\arrow{stealth}}}] (0,1) -- (1,0) node[black, pos=0.5,above right] {$\scriptstyle \Delta_2$};
			\draw[ultra thick, \mygreen, postaction={decorate},decoration={markings, 
				mark= at position 0.7 with {\arrow{stealth}}}] (1,0) -- (0,-1) node[black, pos=0.5,below right] {$\scriptstyle \Delta_3$};
			\draw[ultra thick, \mygreen, postaction={decorate},decoration={markings, 
				mark= at position 0.7 with {\arrow{stealth}}}] (0,-1) -- (-1,0) node[black, pos=0.5,below left] {$\scriptstyle \Delta_4$};
			\draw[\mygreen, fill=\mygreen] (-1,0) circle (0.08) (0,1) circle (0.08) (1,0) circle (0.08) (0,-1) circle (0.08);
			\node[left] at (-1,0) {$\mu_0$};
			\node[above] at (0,1) {$\mu_1$};
			\node[right] at (1,0) {$\mu_2$};
			\node[below] at (0,-1) {$\mu_3$};
		\end{tikzpicture}
	\end{array}
\end{equation}

Let us first consider a fibration of the height field over an oriented cycle in patch $\Pi_c$ that passes through the box at the center $c$, then at one of the four vertices (say at $c+\myh_{1}$), then at the center of the neighboring cell (say at $c+\myh_{1}+\myh_{2}$), then the vertex $c-\myh_{4}$, before coming back to the starting point (see the path marked by the green color in \eqref{pic:localpatch}):
\begin{equation}\label{ranges}
	\left(\begin{array}{c}
		c\\
		\mu_0
	\end{array}\right)\overset{+\myh_1}{\longrightarrow}\left(\begin{array}{c}
		c+\myh_1\\
		\mu_1
	\end{array}\right)\overset{+\myh_2}{\longrightarrow}\left(\begin{array}{c}
		c+\myh_1+\myh_2=c-\myh_3-\myh_4\\
		\mu_2
	\end{array}\right)\overset{+\myh_3}{\longrightarrow}\left(\begin{array}{c}
		c-\myh_4\\
		\mu_3
	\end{array}\right)\overset{+\myh_4}{\longrightarrow}\left(\begin{array}{c}
		c\\
		\mu_0
	\end{array}\right)
\end{equation}
Using the set of inequalities \eqref{ineq}, one can derive the following ranges for the possible heights of the neighboring boxes, $\mu_{1,2,3}$, relative to that of the box at the center, $\mu_0$:
\begin{equation}
	\begin{array}{c}
		\mu_1\leq \mu_0+\Delta_1\\
		\mu_2\leq \mu_1+\Delta_2\\
		\mu_3\leq \mu_2+\Delta_3\\
		\mu_0\leq \mu_3+\Delta_4\\
	\end{array}\;\Rightarrow\;
	\begin{array}{rcccl}
		\mu_0-\Delta_2-\Delta_3-\Delta_4&\leq& \mu_1&\leq&\mu_0+\Delta_1\\
		\mu_0-\Delta_3-\Delta_4&\leq& \mu_2&\leq&\mu_0+\Delta_1+\Delta_2\\
		\mu_0-\Delta_4&\leq& \mu_3&\leq&\mu_0+\Delta_1+\Delta_2+\Delta_3\\
	\end{array}\,,
\end{equation}
where $\Delta_i$ are the corresponding edge weights, which take the values of 0 or 1.
Such cycles can be drawn through all the boxes at the vertices and at the centers of the neighboring cells in patch $\Pi_c$.
It follows that, on the patch $\Pi_c$, the set of inequalities \eqref{ineq} allows only for a \emph{finite} set of integer solutions, for a given $\mu_0$ (the height at the center of the patch $c$, a modulus).
Let us denote such a solution as $M(\Pi_c,\mu_0)$.

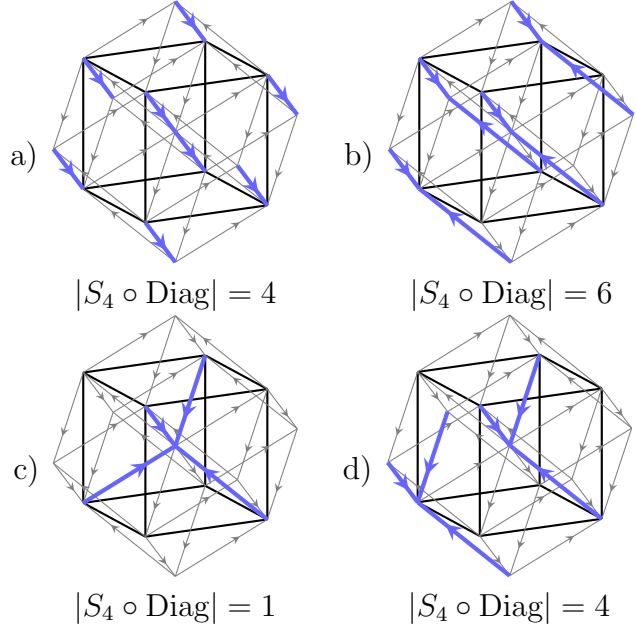
\begin{figure}[ht!]
	\begin{center}
		$\begin{array}{c}
			\mbox{a)}\begin{array}{c}
				\begin{tikzpicture}[scale=0.9]
					\foreach \x/\y/\z/\w in {0.437016/-0.585051/0.437016/1.33554, 0.437016/-0.585051/1.345/-1.08222, 0.437016/-0.585051/-1.345/-0.83837, -0.437016/-1.33554/-0.437016/0.585051, -0.437016/-1.33554/1.345/-1.08222, -0.437016/-1.33554/-1.345/-0.83837, -1.345/1.08222/-0.437016/0.585051, -1.345/1.08222/0.437016/1.33554, -1.345/1.08222/-1.345/-0.83837, 1.345/0.83837/-0.437016/0.585051, 1.345/0.83837/0.437016/1.33554, 1.345/0.83837/1.345/-1.08222}
					{
						\draw[thick] (\x,\y) -- (\z,\w);
					}
					\foreach \x/\y/\z/\w in {0./0./-0.437016/-1.33554, 0./0./-1.345/1.08222, 0./0./1.345/0.83837, 0.437016/-0.585051/0./-1.92059, 0.437016/-0.585051/-0.907981/0.497166, 0.437016/-0.585051/1.78201/0.253319, -0.437016/-1.33554/-1.78201/-0.253319, -0.437016/-1.33554/0.907981/-0.497166, 0.437016/1.33554/0./0., -1.345/1.08222/-1.78201/-0.253319, -1.345/1.08222/0./1.92059, 1.345/-1.08222/0./0., 1.345/0.83837/0.907981/-0.497166, 1.345/0.83837/0./1.92059, -1.345/-0.83837/0./0., 0./-1.92059/1.345/-1.08222, 0./-1.92059/-1.345/-0.83837, -0.907981/0.497166/0.437016/1.33554, -0.907981/0.497166/-1.345/-0.83837, 1.78201/0.253319/0.437016/1.33554, 1.78201/0.253319/1.345/-1.08222, -1.78201/-0.253319/-0.437016/0.585051, 0.907981/-0.497166/-0.437016/0.585051, 0./1.92059/-0.437016/0.585051}
					{
						\draw[gray, postaction={decorate},decoration={markings, 
							mark= at position 0.7 with {\arrow{stealth}}}] (\x,\y) -- (\z,\w);
					}
					\foreach \x/\y/\z/\w in {0./0./0.437016/-0.585051, -0.437016/0.585051/0./0., -0.437016/-1.33554/0./-1.92059, -1.345/1.08222/-0.907981/0.497166, 1.345/0.83837/1.78201/0.253319, -1.78201/-0.253319/-1.345/-0.83837, 0.907981/-0.497166/1.345/-1.08222, 0./1.92059/0.437016/1.33554}
					{
						\draw[ultra thick, \myblue, postaction={decorate},decoration={markings, 
							mark= at position 0.7 with {\arrow{stealth}}}] (\x,\y) -- (\z,\w);
					}
				\end{tikzpicture}\\
				|S_4\circ{\rm Diag}|=4
			\end{array}\quad
			\mbox{b)}\begin{array}{c}
				\begin{tikzpicture}[scale=0.9]
					\foreach \x/\y/\z/\w in {0.437016/-0.585051/0.437016/1.33554, 0.437016/-0.585051/1.345/-1.08222, 0.437016/-0.585051/-1.345/-0.83837, -0.437016/-1.33554/-0.437016/0.585051, -0.437016/-1.33554/1.345/-1.08222, -0.437016/-1.33554/-1.345/-0.83837, -1.345/1.08222/-0.437016/0.585051, -1.345/1.08222/0.437016/1.33554, -1.345/1.08222/-1.345/-0.83837, 1.345/0.83837/-0.437016/0.585051, 1.345/0.83837/0.437016/1.33554, 1.345/0.83837/1.345/-1.08222}
					{
						\draw[thick] (\x,\y) -- (\z,\w);
					}
					\foreach \x/\y/\z/\w in {0./0./0.437016/-0.585051, 0./0./-0.437016/-1.33554, 0./0./-1.345/1.08222, 0./0./1.345/0.83837, 0.437016/-0.585051/0./-1.92059, 0.437016/-0.585051/1.78201/0.253319, -0.437016/-1.33554/0./-1.92059, -0.437016/-1.33554/-1.78201/-0.253319, -0.437016/-1.33554/0.907981/-0.497166, 0.437016/1.33554/0./0., -1.345/1.08222/-1.78201/-0.253319, -1.345/1.08222/0./1.92059, 1.345/0.83837/1.78201/0.253319, 1.345/0.83837/0.907981/-0.497166, 1.345/0.83837/0./1.92059, -1.345/-0.83837/0./0., 0./-1.92059/1.345/-1.08222, -0.907981/0.497166/0.437016/1.33554, -0.907981/0.497166/-1.345/-0.83837, 1.78201/0.253319/1.345/-1.08222, -1.78201/-0.253319/-0.437016/0.585051, 0.907981/-0.497166/-0.437016/0.585051, 0.907981/-0.497166/1.345/-1.08222, 0./1.92059/-0.437016/0.585051}
					{
						\draw[gray, postaction={decorate},decoration={markings, 
							mark= at position 0.7 with {\arrow{stealth}}}] (\x,\y) -- (\z,\w);
					}
					\foreach \x/\y/\z/\w in {0.437016/-0.585051/-0.907981/0.497166, -0.437016/0.585051/0./0., -1.345/1.08222/-0.907981/0.497166, 1.345/-1.08222/0./0., 0./-1.92059/-1.345/-0.83837, 1.78201/0.253319/0.437016/1.33554, -1.78201/-0.253319/-1.345/-0.83837, 0./1.92059/0.437016/1.33554}
					{
						\draw[ultra thick, \myblue, postaction={decorate},decoration={markings, 
							mark= at position 0.7 with {\arrow{stealth}}}] (\x,\y) -- (\z,\w);
					}
				\end{tikzpicture}\\
				|S_4\circ{\rm Diag}|=6
			\end{array}\\
			\mbox{c)}\begin{array}{c}
				\begin{tikzpicture}[scale=0.9]
					\foreach \x/\y/\z/\w in {0.437016/-0.585051/0.437016/1.33554, 0.437016/-0.585051/1.345/-1.08222, 0.437016/-0.585051/-1.345/-0.83837, -0.437016/-1.33554/-0.437016/0.585051, -0.437016/-1.33554/1.345/-1.08222, -0.437016/-1.33554/-1.345/-0.83837, -1.345/1.08222/-0.437016/0.585051, -1.345/1.08222/0.437016/1.33554, -1.345/1.08222/-1.345/-0.83837, 1.345/0.83837/-0.437016/0.585051, 1.345/0.83837/0.437016/1.33554, 1.345/0.83837/1.345/-1.08222}
					{
						\draw[thick] (\x,\y) -- (\z,\w);
					}
					\foreach \x/\y/\z/\w in {0./0./0.437016/-0.585051, 0./0./-0.437016/-1.33554, 0./0./-1.345/1.08222, 0./0./1.345/0.83837, 0.437016/-0.585051/0./-1.92059, 0.437016/-0.585051/-0.907981/0.497166, 0.437016/-0.585051/1.78201/0.253319, -0.437016/-1.33554/0./-1.92059, -0.437016/-1.33554/-1.78201/-0.253319, -0.437016/-1.33554/0.907981/-0.497166, -1.345/1.08222/-0.907981/0.497166, -1.345/1.08222/-1.78201/-0.253319, -1.345/1.08222/0./1.92059, 1.345/0.83837/1.78201/0.253319, 1.345/0.83837/0.907981/-0.497166, 1.345/0.83837/0./1.92059, 0./-1.92059/1.345/-1.08222, 0./-1.92059/-1.345/-0.83837, -0.907981/0.497166/0.437016/1.33554, -0.907981/0.497166/-1.345/-0.83837, 1.78201/0.253319/0.437016/1.33554, 1.78201/0.253319/1.345/-1.08222, -1.78201/-0.253319/-0.437016/0.585051, -1.78201/-0.253319/-1.345/-0.83837, 0.907981/-0.497166/-0.437016/0.585051, 0.907981/-0.497166/1.345/-1.08222, 0./1.92059/-0.437016/0.585051, 0./1.92059/0.437016/1.33554}
					{
						\draw[gray, postaction={decorate},decoration={markings, 
							mark= at position 0.7 with {\arrow{stealth}}}] (\x,\y) -- (\z,\w);
					}
					\foreach \x/\y/\z/\w in {-0.437016/0.585051/0./0., 0.437016/1.33554/0./0., 1.345/-1.08222/0./0., -1.345/-0.83837/0./0.}
					{
						\draw[ultra thick, \myblue, postaction={decorate},decoration={markings, 
							mark= at position 0.7 with {\arrow{stealth}}}] (\x,\y) -- (\z,\w);
					}
				\end{tikzpicture}\\
				|S_4\circ{\rm Diag}|=1
			\end{array}\quad
			\mbox{d)}\begin{array}{c}
				\begin{tikzpicture}[scale=0.9]
					\foreach \x/\y/\z/\w in {0.437016/-0.585051/0.437016/1.33554, 0.437016/-0.585051/1.345/-1.08222, 0.437016/-0.585051/-1.345/-0.83837, -0.437016/-1.33554/-0.437016/0.585051, -0.437016/-1.33554/1.345/-1.08222, -0.437016/-1.33554/-1.345/-0.83837, -1.345/1.08222/-0.437016/0.585051, -1.345/1.08222/0.437016/1.33554, -1.345/1.08222/-1.345/-0.83837, 1.345/0.83837/-0.437016/0.585051, 1.345/0.83837/0.437016/1.33554, 1.345/0.83837/1.345/-1.08222}
					{
						\draw[thick] (\x,\y) -- (\z,\w);
					}
					\foreach \x/\y/\z/\w in {0./0./0.437016/-0.585051, 0./0./-0.437016/-1.33554, 0./0./-1.345/1.08222, 0./0./1.345/0.83837, 0.437016/-0.585051/0./-1.92059, 0.437016/-0.585051/-0.907981/0.497166, 0.437016/-0.585051/1.78201/0.253319, -0.437016/-1.33554/0./-1.92059, -0.437016/-1.33554/-1.78201/-0.253319, -0.437016/-1.33554/0.907981/-0.497166, -1.345/1.08222/-0.907981/0.497166, -1.345/1.08222/-1.78201/-0.253319, -1.345/1.08222/0./1.92059, 1.345/0.83837/1.78201/0.253319, 1.345/0.83837/0.907981/-0.497166, 1.345/0.83837/0./1.92059, -1.345/-0.83837/0./0., 0./-1.92059/1.345/-1.08222, -0.907981/0.497166/0.437016/1.33554, 1.78201/0.253319/0.437016/1.33554, 1.78201/0.253319/1.345/-1.08222, -1.78201/-0.253319/-0.437016/0.585051, 0.907981/-0.497166/-0.437016/0.585051, 0.907981/-0.497166/1.345/-1.08222, 0./1.92059/-0.437016/0.585051, 0./1.92059/0.437016/1.33554}
					{
						\draw[gray, postaction={decorate},decoration={markings, 
							mark= at position 0.7 with {\arrow{stealth}}}] (\x,\y) -- (\z,\w);
					}
					\foreach \x/\y/\z/\w in {-0.437016/0.585051/0./0., 0.437016/1.33554/0./0., 1.345/-1.08222/0./0., 0./-1.92059/-1.345/-0.83837, -0.907981/0.497166/-1.345/-0.83837, -1.78201/-0.253319/-1.345/-0.83837}
					{
						\draw[ultra thick, \myblue, postaction={decorate},decoration={markings, 
							mark= at position 0.7 with {\arrow{stealth}}}] (\x,\y) -- (\z,\w);
					}
				\end{tikzpicture}\\
				|S_4\circ{\rm Diag}|=4
			\end{array}
		\end{array}$
		\caption{Local edge weight configurations on patch $\Pi_c$.}\label{fig:edge_config}
	\end{center}
\end{figure}

Now we can count the number of such solutions. 
First, note that the weight formula \eqref{weight} is symmetric with respect to the permutation group $S_4$ permuting $\myh_k\to\myh_{\sigma(k)}$, $\sigma\in S_4$.
Therefore we only describe representatives of the $S_4$ orbits.
We find {\bf 4} possible edge weight distributions on the patch $\Pi_c$ modulo the action of $S_4$, shown in Fig.~\ref{fig:edge_config}, where we have 
adopted the following color code for the oriented edges:
\begin{equation}\label{edge_cc}
	\Delta({\color{gray}\blacksquare})=0,\quad \Delta({\color{\myblue}\blacksquare})=1\,,
\end{equation}
and also given, below each edge weight configuration, the size of the corresponding $S_4$ orbit.
The number of solutions to the set of inequalities \eqref{ineq} for each local edge weight configuration on patch $\Pi_c$ depicted in Fig.~\ref{fig:edge_config} turns out to be {\bf 166}.

We will call the configuration $M(\Pi_c,\mu_0)$ \emph{open} if a box can be added to or removed from a stack of boxes on top of point $c$, and \emph{closed} otherwise.
Namely, $M(\Pi_c,\mu_0)$ is \emph{open} if either one of $M(\Pi_c,\mu_0\pm 1)$ is also a solution of \eqref{ineq} over $\Pi_c$, and \emph{closed} otherwise.

Therefore, to \emph{prove} that the charge function \eqref{ch_f} has the desired properties, in particular having poles of order 1 at positions where boxes can be added or removed, we only need to prove that the weight function \eqref{weight} \emph{can distinguish} between open and closed solutions $M(\Pi_c,\mu_0)$.
And this can be done directly since the inequivalent $M(\Pi_c,\mu_0)$ form a finite set.
We will return to this computation at the end of this section.

\subsubsection{Analyzing local configurations}
First, to gain some visual intuition about the boundary behavior of the solid partitions, we would like to discuss the structure of the ``local pictures" -- local shapes of the 3D hyper-surface of a solid partition boundary.
To do so, similarly to what is discussed in the 2D and 3D cases in App.\ \ref{app:warmup}, we will only present explicitly the solutions for one configuration, namely case (c) in Fig.\ \ref{fig:edge_config}, which is $S_4$-symmetric. 
The $S_4$-symmetric configuration (c) corresponds to the configuration where the central box is along the diagonal line, similar to the partition and plane partition cases.
The other 3 $S_4$ orbits in Fig.\ \ref{fig:edge_config} then simply correspond to shifting the resulting local pictures obtained from configuration (c) away from the diagonal line. 
And we choose the $S_4$ symmetric configuration (c) in Fig.\ \ref{fig:edge_config} since in this case the solutions to \eqref{ineq} can also be divided into $S_4$-orbits, where it is enough to analyze a single representative of each orbit.

Now, consider the 15 boxes corresponding to the 15 vectors in the local patch $\Pi_c$ in  the configuration (c)  in Fig.\ \ref{fig:edge_config}.
Let us denote their height field values by
\begin{equation}
	\begin{aligned}
		\mu_{0}
		\,, \qquad 
		\mu_{1,2,3,4}
		\,,\qquad
		\mu_{12,13,23,14,24,34}
		\,, \qquad
		\mu_{-1,-2,-3,-4}\,.
	\end{aligned}
\end{equation}
The set of inequalities  \eqref{ineq} translates into 32 inequalities among them: 
\begin{equation}
	\begin{aligned}
		&\mu_i \leq \mu_0 \,, \\
		&\mu_0 \leq \mu_{-i}+1 \,, \\
		&\mu_{ij}\leq \mu_i \,, \quad \mu_{ij}\leq \mu_j \qquad i\neq j\,, \\
		&\mu_{-k}\leq \mu_{ij} \qquad i\neq j \neq k\,,
	\end{aligned}    
\end{equation}
which can be further translated into 
\begin{equation}
	\mu_0-1\leq \mu_{-3,-4}\leq \mu_{12} \leq \mu_{1,2} \leq \mu_0   \,, 
\end{equation}
etc.

Using our strategy of considering only a single representative in each $S_4$ orbit, we obtain 28 $S_4$ symmetry classes of solutions in total, corresponding to {\bf 28} local pictures, shown in Fig.~\ref{fig:ineq_sol}, where we have used the following color coding to label the local height field $\mu(c)$ value:
\begin{equation}\label{vertex_cc}
	\mu\left(\!\!\begin{array}{c}
		\begin{tikzpicture}
			\draw[fill=\myblue] (0,0) circle (0.1);
		\end{tikzpicture}
	\end{array}\!\!\right)=\mu_0,\quad \mu\left(\!\!\begin{array}{c}
		\begin{tikzpicture}
			\draw[fill=white] (0,0) circle (0.1);
		\end{tikzpicture}
	\end{array}\!\!\right)=\mu_0-1\,.
\end{equation}

\begin{figure}
	\begin{center}
		\begingroup
		\fontsize{10}{12}\selectfont 
		\begin{equation}\nn
			\begin{array}{|c|c|c|c|}
				\hline
				\begin{tikzpicture}[scale=0.5]
					\node at (0,2.5) {};
					\node at (0,-2.5) {};
					\foreach \x/\y/\z/\w in {1.345/-1.07125/1.345/0.880585, 1.345/-1.07125/-0.437016/-1.26932, 1.345/-1.07125/0.437016/-0.682515, 1.345/0.880585/-0.437016/0.682515, 1.345/0.880585/0.437016/1.26932, -0.437016/-1.26932/-0.437016/0.682515, -0.437016/-1.26932/-1.345/-0.880585, -0.437016/0.682515/-1.345/1.07125, 0.437016/-0.682515/0.437016/1.26932, 0.437016/-0.682515/-1.345/-0.880585, 0.437016/1.26932/-1.345/1.07125, -1.345/-0.880585/-1.345/1.07125, 0./0./-0.907981/0.388734, 0./0./0.907981/-0.388734, 0./0./-1.78201/-0.19807, 0./0./1.78201/0.19807, 0./0./0./1.95183, 0./0./0./-1.95183}
					{
						\draw[thick] (\x,\y) -- (\z,\w);
					}
					\foreach \x/\y in {0./0.}
					{
						\draw[fill=\myblue] (\x,\y) circle (0.2);
					}
					\foreach \x/\y in {0.437016/-0.682515, -0.437016/0.682515, -0.437016/-1.26932, 0.437016/1.26932, -1.345/1.07125, 1.345/-1.07125, 1.345/0.880585, -1.345/-0.880585, 0./-1.95183, -0.907981/0.388734, 1.78201/0.19807, -1.78201/-0.19807, 0.907981/-0.388734, 0./1.95183}
					{
						\draw[fill=white] (\x,\y) circle (0.2);
					}
					\node at (3.5,1) {(1)};
					\node[\mygreen] at (3.5,0) {Open};
					\node at (3.5,-1) {$w=1$};
				\end{tikzpicture}&\begin{tikzpicture}[scale=0.5]
					\node at (0,2.5) {};
					\node at (0,-2.5) {};
					\foreach \x/\y/\z/\w in {1.345/-1.07125/1.345/0.880585, 1.345/-1.07125/-0.437016/-1.26932, 1.345/-1.07125/0.437016/-0.682515, 1.345/0.880585/-0.437016/0.682515, 1.345/0.880585/0.437016/1.26932, -0.437016/-1.26932/-0.437016/0.682515, -0.437016/-1.26932/-1.345/-0.880585, -0.437016/0.682515/-1.345/1.07125, 0.437016/-0.682515/0.437016/1.26932, 0.437016/-0.682515/-1.345/-0.880585, 0.437016/1.26932/-1.345/1.07125, -1.345/-0.880585/-1.345/1.07125, 0./0./-0.907981/0.388734, 0./0./0.907981/-0.388734, 0./0./-1.78201/-0.19807, 0./0./1.78201/0.19807, 0./0./0./1.95183, 0./0./0./-1.95183}
					{
						\draw[thick] (\x,\y) -- (\z,\w);
					}
					\foreach \x/\y in {0./0., 1.345/0.880585}
					{
						\draw[fill=\myblue] (\x,\y) circle (0.2);
					}
					\foreach \x/\y in {0.437016/-0.682515, -0.437016/0.682515, -0.437016/-1.26932, 0.437016/1.26932, -1.345/1.07125, 1.345/-1.07125, -1.345/-0.880585, 0./-1.95183, -0.907981/0.388734, 1.78201/0.19807, -1.78201/-0.19807, 0.907981/-0.388734, 0./1.95183}
					{
						\draw[fill=white] (\x,\y) circle (0.2);
					}
					\node at (3.5,1) {(2)};
					\node[burgundy] at (3.5,0) {Closed};
					\node at (3.5,-1) {$w=0$};
				\end{tikzpicture}&\begin{tikzpicture}[scale=0.5]
					\node at (0,2.5) {};
					\node at (0,-2.5) {};
					\foreach \x/\y/\z/\w in {1.345/-1.07125/1.345/0.880585, 1.345/-1.07125/-0.437016/-1.26932, 1.345/-1.07125/0.437016/-0.682515, 1.345/0.880585/-0.437016/0.682515, 1.345/0.880585/0.437016/1.26932, -0.437016/-1.26932/-0.437016/0.682515, -0.437016/-1.26932/-1.345/-0.880585, -0.437016/0.682515/-1.345/1.07125, 0.437016/-0.682515/0.437016/1.26932, 0.437016/-0.682515/-1.345/-0.880585, 0.437016/1.26932/-1.345/1.07125, -1.345/-0.880585/-1.345/1.07125, 0./0./-0.907981/0.388734, 0./0./0.907981/-0.388734, 0./0./-1.78201/-0.19807, 0./0./1.78201/0.19807, 0./0./0./1.95183, 0./0./0./-1.95183}
					{
						\draw[thick] (\x,\y) -- (\z,\w);
					}
					\foreach \x/\y in {0./0., -1.345/1.07125, 1.345/0.880585}
					{
						\draw[fill=\myblue] (\x,\y) circle (0.2);
					}
					\foreach \x/\y in {0.437016/-0.682515, -0.437016/0.682515, -0.437016/-1.26932, 0.437016/1.26932, 1.345/-1.07125, -1.345/-0.880585, 0./-1.95183, -0.907981/0.388734, 1.78201/0.19807, -1.78201/-0.19807, 0.907981/-0.388734, 0./1.95183}
					{
						\draw[fill=white] (\x,\y) circle (0.2);
					}
					\node at (3.5,1) {(3)};
					\node[burgundy] at (3.5,0) {Closed};
					\node at (3.5,-1) {$w=-1$};
				\end{tikzpicture}&\begin{tikzpicture}[scale=0.5]
					\node at (0,2.5) {};
					\node at (0,-2.5) {};
					\foreach \x/\y/\z/\w in {1.345/-1.07125/1.345/0.880585, 1.345/-1.07125/-0.437016/-1.26932, 1.345/-1.07125/0.437016/-0.682515, 1.345/0.880585/-0.437016/0.682515, 1.345/0.880585/0.437016/1.26932, -0.437016/-1.26932/-0.437016/0.682515, -0.437016/-1.26932/-1.345/-0.880585, -0.437016/0.682515/-1.345/1.07125, 0.437016/-0.682515/0.437016/1.26932, 0.437016/-0.682515/-1.345/-0.880585, 0.437016/1.26932/-1.345/1.07125, -1.345/-0.880585/-1.345/1.07125, 0./0./-0.907981/0.388734, 0./0./0.907981/-0.388734, 0./0./-1.78201/-0.19807, 0./0./1.78201/0.19807, 0./0./0./1.95183, 0./0./0./-1.95183}
					{
						\draw[thick] (\x,\y) -- (\z,\w);
					}
					\foreach \x/\y in {0./0., -1.345/1.07125, 1.345/0.880585, 0./1.95183}
					{
						\draw[fill=\myblue] (\x,\y) circle (0.2);
					}
					\foreach \x/\y in {0.437016/-0.682515, -0.437016/0.682515, -0.437016/-1.26932, 0.437016/1.26932, 1.345/-1.07125, -1.345/-0.880585, 0./-1.95183, -0.907981/0.388734, 1.78201/0.19807, -1.78201/-0.19807, 0.907981/-0.388734}
					{
						\draw[fill=white] (\x,\y) circle (0.2);
					}
					\node at (3.5,1) {(4)};
					\node[burgundy] at (3.5,0) {Closed};
					\node at (3.5,-1) {$w=-2$};
				\end{tikzpicture}\\
				\hline
				\begin{tikzpicture}[scale=0.5]
					\node at (0,2.5) {};
					\node at (0,-2.5) {};
					\foreach \x/\y/\z/\w in {1.345/-1.07125/1.345/0.880585, 1.345/-1.07125/-0.437016/-1.26932, 1.345/-1.07125/0.437016/-0.682515, 1.345/0.880585/-0.437016/0.682515, 1.345/0.880585/0.437016/1.26932, -0.437016/-1.26932/-0.437016/0.682515, -0.437016/-1.26932/-1.345/-0.880585, -0.437016/0.682515/-1.345/1.07125, 0.437016/-0.682515/0.437016/1.26932, 0.437016/-0.682515/-1.345/-0.880585, 0.437016/1.26932/-1.345/1.07125, -1.345/-0.880585/-1.345/1.07125, 0./0./-0.907981/0.388734, 0./0./0.907981/-0.388734, 0./0./-1.78201/-0.19807, 0./0./1.78201/0.19807, 0./0./0./1.95183, 0./0./0./-1.95183}
					{
						\draw[thick] (\x,\y) -- (\z,\w);
					}
					\foreach \x/\y in {0./0., -0.437016/-1.26932, -1.345/1.07125, 1.345/0.880585}
					{
						\draw[fill=\myblue] (\x,\y) circle (0.2);
					}
					\foreach \x/\y in {0.437016/-0.682515, -0.437016/0.682515, 0.437016/1.26932, 1.345/-1.07125, -1.345/-0.880585, 0./-1.95183, -0.907981/0.388734, 1.78201/0.19807, -1.78201/-0.19807, 0.907981/-0.388734, 0./1.95183}
					{
						\draw[fill=white] (\x,\y) circle (0.2);
					}
					\node at (3.5,1) {(5)};
					\node[burgundy] at (3.5,0) {Closed};
					\node at (3.5,-1) {$w=-2$};
				\end{tikzpicture}&\begin{tikzpicture}[scale=0.5]
					\node at (0,2.5) {};
					\node at (0,-2.5) {};
					\foreach \x/\y/\z/\w in {1.345/-1.07125/1.345/0.880585, 1.345/-1.07125/-0.437016/-1.26932, 1.345/-1.07125/0.437016/-0.682515, 1.345/0.880585/-0.437016/0.682515, 1.345/0.880585/0.437016/1.26932, -0.437016/-1.26932/-0.437016/0.682515, -0.437016/-1.26932/-1.345/-0.880585, -0.437016/0.682515/-1.345/1.07125, 0.437016/-0.682515/0.437016/1.26932, 0.437016/-0.682515/-1.345/-0.880585, 0.437016/1.26932/-1.345/1.07125, -1.345/-0.880585/-1.345/1.07125, 0./0./-0.907981/0.388734, 0./0./0.907981/-0.388734, 0./0./-1.78201/-0.19807, 0./0./1.78201/0.19807, 0./0./0./1.95183, 0./0./0./-1.95183}
					{
						\draw[thick] (\x,\y) -- (\z,\w);
					}
					\foreach \x/\y in {0./0., -0.437016/-1.26932, -1.345/1.07125, 1.345/0.880585, 0./1.95183}
					{
						\draw[fill=\myblue] (\x,\y) circle (0.2);
					}
					\foreach \x/\y in {0.437016/-0.682515, -0.437016/0.682515, 0.437016/1.26932, 1.345/-1.07125, -1.345/-0.880585, 0./-1.95183, -0.907981/0.388734, 1.78201/0.19807, -1.78201/-0.19807, 0.907981/-0.388734}
					{
						\draw[fill=white] (\x,\y) circle (0.2);
					}
					\node at (3.5,1) {(6)};
					\node[burgundy] at (3.5,0) {Closed};
					\node at (3.5,-1) {$w=-3$};
				\end{tikzpicture}&\begin{tikzpicture}[scale=0.5]
					\node at (0,2.5) {};
					\node at (0,-2.5) {};
					\foreach \x/\y/\z/\w in {1.345/-1.07125/1.345/0.880585, 1.345/-1.07125/-0.437016/-1.26932, 1.345/-1.07125/0.437016/-0.682515, 1.345/0.880585/-0.437016/0.682515, 1.345/0.880585/0.437016/1.26932, -0.437016/-1.26932/-0.437016/0.682515, -0.437016/-1.26932/-1.345/-0.880585, -0.437016/0.682515/-1.345/1.07125, 0.437016/-0.682515/0.437016/1.26932, 0.437016/-0.682515/-1.345/-0.880585, 0.437016/1.26932/-1.345/1.07125, -1.345/-0.880585/-1.345/1.07125, 0./0./-0.907981/0.388734, 0./0./0.907981/-0.388734, 0./0./-1.78201/-0.19807, 0./0./1.78201/0.19807, 0./0./0./1.95183, 0./0./0./-1.95183}
					{
						\draw[thick] (\x,\y) -- (\z,\w);
					}
					\foreach \x/\y in {0./0., -0.437016/-1.26932, -1.345/1.07125, 1.345/0.880585, 0.907981/-0.388734, 0./1.95183}
					{
						\draw[fill=\myblue] (\x,\y) circle (0.2);
					}
					\foreach \x/\y in {0.437016/-0.682515, -0.437016/0.682515, 0.437016/1.26932, 1.345/-1.07125, -1.345/-0.880585, 0./-1.95183, -0.907981/0.388734, 1.78201/0.19807, -1.78201/-0.19807}
					{
						\draw[fill=white] (\x,\y) circle (0.2);
					}
					\node at (3.5,1) {(7)};
					\node[burgundy] at (3.5,0) {Closed};
					\node at (3.5,-1) {$w=-4$};
				\end{tikzpicture}&\begin{tikzpicture}[scale=0.5]
					\node at (0,2.5) {};
					\node at (0,-2.5) {};
					\foreach \x/\y/\z/\w in {1.345/-1.07125/1.345/0.880585, 1.345/-1.07125/-0.437016/-1.26932, 1.345/-1.07125/0.437016/-0.682515, 1.345/0.880585/-0.437016/0.682515, 1.345/0.880585/0.437016/1.26932, -0.437016/-1.26932/-0.437016/0.682515, -0.437016/-1.26932/-1.345/-0.880585, -0.437016/0.682515/-1.345/1.07125, 0.437016/-0.682515/0.437016/1.26932, 0.437016/-0.682515/-1.345/-0.880585, 0.437016/1.26932/-1.345/1.07125, -1.345/-0.880585/-1.345/1.07125, 0./0./-0.907981/0.388734, 0./0./0.907981/-0.388734, 0./0./-1.78201/-0.19807, 0./0./1.78201/0.19807, 0./0./0./1.95183, 0./0./0./-1.95183}
					{
						\draw[thick] (\x,\y) -- (\z,\w);
					}
					\foreach \x/\y in {0./0., -0.437016/-1.26932, -1.345/1.07125, 1.345/0.880585, -1.78201/-0.19807, 0.907981/-0.388734, 0./1.95183}
					{
						\draw[fill=\myblue] (\x,\y) circle (0.2);
					}
					\foreach \x/\y in {0.437016/-0.682515, -0.437016/0.682515, 0.437016/1.26932, 1.345/-1.07125, -1.345/-0.880585, 0./-1.95183, -0.907981/0.388734, 1.78201/0.19807}
					{
						\draw[fill=white] (\x,\y) circle (0.2);
					}
					\node at (3.5,1) {(8)};
					\node[burgundy] at (3.5,0) {Closed};
					\node at (3.5,-1) {$w=-5$};
				\end{tikzpicture}\\
				\hline
				\begin{tikzpicture}[scale=0.5]
					\node at (0,2.5) {};
					\node at (0,-2.5) {};
					\foreach \x/\y/\z/\w in {1.345/-1.07125/1.345/0.880585, 1.345/-1.07125/-0.437016/-1.26932, 1.345/-1.07125/0.437016/-0.682515, 1.345/0.880585/-0.437016/0.682515, 1.345/0.880585/0.437016/1.26932, -0.437016/-1.26932/-0.437016/0.682515, -0.437016/-1.26932/-1.345/-0.880585, -0.437016/0.682515/-1.345/1.07125, 0.437016/-0.682515/0.437016/1.26932, 0.437016/-0.682515/-1.345/-0.880585, 0.437016/1.26932/-1.345/1.07125, -1.345/-0.880585/-1.345/1.07125, 0./0./-0.907981/0.388734, 0./0./0.907981/-0.388734, 0./0./-1.78201/-0.19807, 0./0./1.78201/0.19807, 0./0./0./1.95183, 0./0./0./-1.95183}
					{
						\draw[thick] (\x,\y) -- (\z,\w);
					}
					\foreach \x/\y in {0./0., -0.437016/0.682515, -0.437016/-1.26932, -1.345/1.07125, 1.345/0.880585, -1.78201/-0.19807, 0.907981/-0.388734, 0./1.95183}
					{
						\draw[fill=\myblue] (\x,\y) circle (0.2);
					}
					\foreach \x/\y in {0.437016/-0.682515, 0.437016/1.26932, 1.345/-1.07125, -1.345/-0.880585, 0./-1.95183, -0.907981/0.388734, 1.78201/0.19807}
					{
						\draw[fill=white] (\x,\y) circle (0.2);
					}
					\node at (3.5,1) {(9)};
					\node[burgundy] at (3.5,0) {Closed};
					\node at (3.5,-1) {$w=-4$};
				\end{tikzpicture}&
				\begin{tikzpicture}[scale=0.5]
					\node at (0,2.5) {};
					\node at (0,-2.5) {};
					\foreach \x/\y/\z/\w in {1.345/-1.07125/1.345/0.880585, 1.345/-1.07125/-0.437016/-1.26932, 1.345/-1.07125/0.437016/-0.682515, 1.345/0.880585/-0.437016/0.682515, 1.345/0.880585/0.437016/1.26932, -0.437016/-1.26932/-0.437016/0.682515, -0.437016/-1.26932/-1.345/-0.880585, -0.437016/0.682515/-1.345/1.07125, 0.437016/-0.682515/0.437016/1.26932, 0.437016/-0.682515/-1.345/-0.880585, 0.437016/1.26932/-1.345/1.07125, -1.345/-0.880585/-1.345/1.07125, 0./0./-0.907981/0.388734, 0./0./0.907981/-0.388734, 0./0./-1.78201/-0.19807, 0./0./1.78201/0.19807, 0./0./0./1.95183, 0./0./0./-1.95183}
					{
						\draw[thick] (\x,\y) -- (\z,\w);
					}
					\foreach \x/\y in {0./0., 0.437016/-0.682515, -0.437016/-1.26932, -1.345/1.07125, 1.345/0.880585}
					{
						\draw[fill=\myblue] (\x,\y) circle (0.2);
					}
					\foreach \x/\y in {-0.437016/0.682515, 0.437016/1.26932, 1.345/-1.07125, -1.345/-0.880585, 0./-1.95183, -0.907981/0.388734, 1.78201/0.19807, -1.78201/-0.19807, 0.907981/-0.388734, 0./1.95183}
					{
						\draw[fill=white] (\x,\y) circle (0.2);
					}
					\node at (3.5,1) {(10)};
					\node[burgundy] at (3.5,0) {Closed};
					\node at (3.5,-1) {$w=-5$};
				\end{tikzpicture}&\begin{tikzpicture}[scale=0.5]
					\node at (0,2.5) {};
					\node at (0,-2.5) {};
					\foreach \x/\y/\z/\w in {1.345/-1.07125/1.345/0.880585, 1.345/-1.07125/-0.437016/-1.26932, 1.345/-1.07125/0.437016/-0.682515, 1.345/0.880585/-0.437016/0.682515, 1.345/0.880585/0.437016/1.26932, -0.437016/-1.26932/-0.437016/0.682515, -0.437016/-1.26932/-1.345/-0.880585, -0.437016/0.682515/-1.345/1.07125, 0.437016/-0.682515/0.437016/1.26932, 0.437016/-0.682515/-1.345/-0.880585, 0.437016/1.26932/-1.345/1.07125, -1.345/-0.880585/-1.345/1.07125, 0./0./-0.907981/0.388734, 0./0./0.907981/-0.388734, 0./0./-1.78201/-0.19807, 0./0./1.78201/0.19807, 0./0./0./1.95183, 0./0./0./-1.95183}
					{
						\draw[thick] (\x,\y) -- (\z,\w);
					}
					\foreach \x/\y in {0./0., 0.437016/-0.682515, -0.437016/-1.26932, -1.345/1.07125, 1.345/0.880585, 0./1.95183}
					{
						\draw[fill=\myblue] (\x,\y) circle (0.2);
					}
					\foreach \x/\y in {-0.437016/0.682515, 0.437016/1.26932, 1.345/-1.07125, -1.345/-0.880585, 0./-1.95183, -0.907981/0.388734, 1.78201/0.19807, -1.78201/-0.19807, 0.907981/-0.388734}
					{
						\draw[fill=white] (\x,\y) circle (0.2);
					}
					\node at (3.5,1) {(11)};
					\node[burgundy] at (3.5,0) {Closed};
					\node at (3.5,-1) {$w=-6$};	
				\end{tikzpicture}&\begin{tikzpicture}[scale=0.5]
					\node at (0,2.5) {};
					\node at (0,-2.5) {};
					\foreach \x/\y/\z/\w in {1.345/-1.07125/1.345/0.880585, 1.345/-1.07125/-0.437016/-1.26932, 1.345/-1.07125/0.437016/-0.682515, 1.345/0.880585/-0.437016/0.682515, 1.345/0.880585/0.437016/1.26932, -0.437016/-1.26932/-0.437016/0.682515, -0.437016/-1.26932/-1.345/-0.880585, -0.437016/0.682515/-1.345/1.07125, 0.437016/-0.682515/0.437016/1.26932, 0.437016/-0.682515/-1.345/-0.880585, 0.437016/1.26932/-1.345/1.07125, -1.345/-0.880585/-1.345/1.07125, 0./0./-0.907981/0.388734, 0./0./0.907981/-0.388734, 0./0./-1.78201/-0.19807, 0./0./1.78201/0.19807, 0./0./0./1.95183, 0./0./0./-1.95183}
					{
						\draw[thick] (\x,\y) -- (\z,\w);
					}
					\foreach \x/\y in {0./0., 0.437016/-0.682515, -0.437016/-1.26932, -1.345/1.07125, 1.345/0.880585, 0.907981/-0.388734, 0./1.95183}
					{
						\draw[fill=\myblue] (\x,\y) circle (0.2);
					}
					\foreach \x/\y in {-0.437016/0.682515, 0.437016/1.26932, 1.345/-1.07125, -1.345/-0.880585, 0./-1.95183, -0.907981/0.388734, 1.78201/0.19807, -1.78201/-0.19807}
					{
						\draw[fill=white] (\x,\y) circle (0.2);
					}
					\node at (3.5,1) {(12)};
					\node[burgundy] at (3.5,0) {Closed};
					\node at (3.5,-1) {$w=-7$};
				\end{tikzpicture}\\
				\hline
				\begin{tikzpicture}[scale=0.5]
					\node at (0,2.5) {};
					\node at (0,-2.5) {};
					\foreach \x/\y/\z/\w in {1.345/-1.07125/1.345/0.880585, 1.345/-1.07125/-0.437016/-1.26932, 1.345/-1.07125/0.437016/-0.682515, 1.345/0.880585/-0.437016/0.682515, 1.345/0.880585/0.437016/1.26932, -0.437016/-1.26932/-0.437016/0.682515, -0.437016/-1.26932/-1.345/-0.880585, -0.437016/0.682515/-1.345/1.07125, 0.437016/-0.682515/0.437016/1.26932, 0.437016/-0.682515/-1.345/-0.880585, 0.437016/1.26932/-1.345/1.07125, -1.345/-0.880585/-1.345/1.07125, 0./0./-0.907981/0.388734, 0./0./0.907981/-0.388734, 0./0./-1.78201/-0.19807, 0./0./1.78201/0.19807, 0./0./0./1.95183, 0./0./0./-1.95183}
					{
						\draw[thick] (\x,\y) -- (\z,\w);
					}
					\foreach \x/\y in {0./0., 0.437016/-0.682515, -0.437016/-1.26932, -1.345/1.07125, 1.345/0.880585, -1.78201/-0.19807, 0.907981/-0.388734, 0./1.95183}
					{
						\draw[fill=\myblue] (\x,\y) circle (0.2);
					}
					\foreach \x/\y in {-0.437016/0.682515, 0.437016/1.26932, 1.345/-1.07125, -1.345/-0.880585, 0./-1.95183, -0.907981/0.388734, 1.78201/0.19807}
					{
						\draw[fill=white] (\x,\y) circle (0.2);
					}
					\node at (3.5,1) {(13)};
					\node[burgundy] at (3.5,0) {Closed};
					\node at (3.5,-1) {$w=-8$};
				\end{tikzpicture}&\begin{tikzpicture}[scale=0.5]
					\node at (0,2.5) {};
					\node at (0,-2.5) {};
					\foreach \x/\y/\z/\w in {1.345/-1.07125/1.345/0.880585, 1.345/-1.07125/-0.437016/-1.26932, 1.345/-1.07125/0.437016/-0.682515, 1.345/0.880585/-0.437016/0.682515, 1.345/0.880585/0.437016/1.26932, -0.437016/-1.26932/-0.437016/0.682515, -0.437016/-1.26932/-1.345/-0.880585, -0.437016/0.682515/-1.345/1.07125, 0.437016/-0.682515/0.437016/1.26932, 0.437016/-0.682515/-1.345/-0.880585, 0.437016/1.26932/-1.345/1.07125, -1.345/-0.880585/-1.345/1.07125, 0./0./-0.907981/0.388734, 0./0./0.907981/-0.388734, 0./0./-1.78201/-0.19807, 0./0./1.78201/0.19807, 0./0./0./1.95183, 0./0./0./-1.95183}
					{
						\draw[thick] (\x,\y) -- (\z,\w);
					}
					\foreach \x/\y in {0./0., 0.437016/-0.682515, -0.437016/-1.26932, -1.345/1.07125, 1.345/0.880585, 1.78201/0.19807, 0.907981/-0.388734, 0./1.95183}
					{
						\draw[fill=\myblue] (\x,\y) circle (0.2);
					}
					\foreach \x/\y in {-0.437016/0.682515, 0.437016/1.26932, 1.345/-1.07125, -1.345/-0.880585, 0./-1.95183, -0.907981/0.388734, -1.78201/-0.19807}
					{
						\draw[fill=white] (\x,\y) circle (0.2);
					}
					\node at (3.5,1) {(14)};
					\node[burgundy] at (3.5,0) {Closed};
					\node at (3.5,-1) {$w=-6$};
				\end{tikzpicture}&\begin{tikzpicture}[scale=0.5]
					\node at (0,2.5) {};
					\node at (0,-2.5) {};
					\foreach \x/\y/\z/\w in {1.345/-1.07125/1.345/0.880585, 1.345/-1.07125/-0.437016/-1.26932, 1.345/-1.07125/0.437016/-0.682515, 1.345/0.880585/-0.437016/0.682515, 1.345/0.880585/0.437016/1.26932, -0.437016/-1.26932/-0.437016/0.682515, -0.437016/-1.26932/-1.345/-0.880585, -0.437016/0.682515/-1.345/1.07125, 0.437016/-0.682515/0.437016/1.26932, 0.437016/-0.682515/-1.345/-0.880585, 0.437016/1.26932/-1.345/1.07125, -1.345/-0.880585/-1.345/1.07125, 0./0./-0.907981/0.388734, 0./0./0.907981/-0.388734, 0./0./-1.78201/-0.19807, 0./0./1.78201/0.19807, 0./0./0./1.95183, 0./0./0./-1.95183}
					{
						\draw[thick] (\x,\y) -- (\z,\w);
					}
					\foreach \x/\y in {0./0., 0.437016/-0.682515, -0.437016/-1.26932, -1.345/1.07125, 1.345/0.880585, 1.78201/0.19807, -1.78201/-0.19807}
					{
						\draw[fill=\myblue] (\x,\y) circle (0.2);
					}
					\foreach \x/\y in {-0.437016/0.682515, 0.437016/1.26932, 1.345/-1.07125, -1.345/-0.880585, 0./-1.95183, -0.907981/0.388734, 0.907981/-0.388734, 0./1.95183}
					{
						\draw[fill=white] (\x,\y) circle (0.2);
					}
					\node at (3.5,1) {(15)};
					\node[burgundy] at (3.5,0) {Closed};
					\node at (3.5,-1) {$w=-7$};
				\end{tikzpicture}&\begin{tikzpicture}[scale=0.5]
					\node at (0,2.5) {};
					\node at (0,-2.5) {};
					\foreach \x/\y/\z/\w in {1.345/-1.07125/1.345/0.880585, 1.345/-1.07125/-0.437016/-1.26932, 1.345/-1.07125/0.437016/-0.682515, 1.345/0.880585/-0.437016/0.682515, 1.345/0.880585/0.437016/1.26932, -0.437016/-1.26932/-0.437016/0.682515, -0.437016/-1.26932/-1.345/-0.880585, -0.437016/0.682515/-1.345/1.07125, 0.437016/-0.682515/0.437016/1.26932, 0.437016/-0.682515/-1.345/-0.880585, 0.437016/1.26932/-1.345/1.07125, -1.345/-0.880585/-1.345/1.07125, 0./0./-0.907981/0.388734, 0./0./0.907981/-0.388734, 0./0./-1.78201/-0.19807, 0./0./1.78201/0.19807, 0./0./0./1.95183, 0./0./0./-1.95183}
					{
						\draw[thick] (\x,\y) -- (\z,\w);
					}
					\foreach \x/\y in {0./0., 0.437016/-0.682515, -0.437016/-1.26932, -1.345/1.07125, 1.345/0.880585, 1.78201/0.19807, -1.78201/-0.19807, 0./1.95183}
					{
						\draw[fill=\myblue] (\x,\y) circle (0.2);
					}
					\foreach \x/\y in {-0.437016/0.682515, 0.437016/1.26932, 1.345/-1.07125, -1.345/-0.880585, 0./-1.95183, -0.907981/0.388734, 0.907981/-0.388734}
					{
						\draw[fill=white] (\x,\y) circle (0.2);
					}
					\node at (3.5,1) {(16)};
					\node[burgundy] at (3.5,0) {Closed};
					\node at (3.5,-1) {$w=-8$};
				\end{tikzpicture}\\
				\hline
				\begin{tikzpicture}[scale=0.5]
					\node at (0,2.5) {};
					\node at (0,-2.5) {};
					\foreach \x/\y/\z/\w in {1.345/-1.07125/1.345/0.880585, 1.345/-1.07125/-0.437016/-1.26932, 1.345/-1.07125/0.437016/-0.682515, 1.345/0.880585/-0.437016/0.682515, 1.345/0.880585/0.437016/1.26932, -0.437016/-1.26932/-0.437016/0.682515, -0.437016/-1.26932/-1.345/-0.880585, -0.437016/0.682515/-1.345/1.07125, 0.437016/-0.682515/0.437016/1.26932, 0.437016/-0.682515/-1.345/-0.880585, 0.437016/1.26932/-1.345/1.07125, -1.345/-0.880585/-1.345/1.07125, 0./0./-0.907981/0.388734, 0./0./0.907981/-0.388734, 0./0./-1.78201/-0.19807, 0./0./1.78201/0.19807, 0./0./0./1.95183, 0./0./0./-1.95183}
					{
						\draw[thick] (\x,\y) -- (\z,\w);
					}
					\foreach \x/\y in {0./0., 0.437016/-0.682515, -0.437016/-1.26932, -1.345/1.07125, 1.345/0.880585, 1.78201/0.19807, -1.78201/-0.19807, 0.907981/-0.388734, 0./1.95183}
					{
						\draw[fill=\myblue] (\x,\y) circle (0.2);
					}
					\foreach \x/\y in {-0.437016/0.682515, 0.437016/1.26932, 1.345/-1.07125, -1.345/-0.880585, 0./-1.95183, -0.907981/0.388734}
					{
						\draw[fill=white] (\x,\y) circle (0.2);
					}
					\node at (3.5,1) {(17)};
					\node[burgundy] at (3.5,0) {Closed};
					\node at (3.5,-1) {$w=-7$};
				\end{tikzpicture}&\begin{tikzpicture}[scale=0.5]
					\node at (0,2.5) {};
					\node at (0,-2.5) {};
					\foreach \x/\y/\z/\w in {1.345/-1.07125/1.345/0.880585, 1.345/-1.07125/-0.437016/-1.26932, 1.345/-1.07125/0.437016/-0.682515, 1.345/0.880585/-0.437016/0.682515, 1.345/0.880585/0.437016/1.26932, -0.437016/-1.26932/-0.437016/0.682515, -0.437016/-1.26932/-1.345/-0.880585, -0.437016/0.682515/-1.345/1.07125, 0.437016/-0.682515/0.437016/1.26932, 0.437016/-0.682515/-1.345/-0.880585, 0.437016/1.26932/-1.345/1.07125, -1.345/-0.880585/-1.345/1.07125, 0./0./-0.907981/0.388734, 0./0./0.907981/-0.388734, 0./0./-1.78201/-0.19807, 0./0./1.78201/0.19807, 0./0./0./1.95183, 0./0./0./-1.95183}
					{
						\draw[thick] (\x,\y) -- (\z,\w);
					}
					\foreach \x/\y in {0./0., 0.437016/-0.682515, -0.437016/-1.26932, -1.345/1.07125, 1.345/0.880585, -0.907981/0.388734, 1.78201/0.19807, -1.78201/-0.19807, 0.907981/-0.388734}
					{
						\draw[fill=\myblue] (\x,\y) circle (0.2);
					}
					\foreach \x/\y in {-0.437016/0.682515, 0.437016/1.26932, 1.345/-1.07125, -1.345/-0.880585, 0./-1.95183, 0./1.95183}
					{
						\draw[fill=white] (\x,\y) circle (0.2);
					}
					\node at (3.5,1) {(18)};
					\node[burgundy] at (3.5,0) {Closed};
					\node at (3.5,-1) {$w=-9$};
				\end{tikzpicture}&\begin{tikzpicture}[scale=0.5]
					\node at (0,2.5) {};
					\node at (0,-2.5) {};
					\foreach \x/\y/\z/\w in {1.345/-1.07125/1.345/0.880585, 1.345/-1.07125/-0.437016/-1.26932, 1.345/-1.07125/0.437016/-0.682515, 1.345/0.880585/-0.437016/0.682515, 1.345/0.880585/0.437016/1.26932, -0.437016/-1.26932/-0.437016/0.682515, -0.437016/-1.26932/-1.345/-0.880585, -0.437016/0.682515/-1.345/1.07125, 0.437016/-0.682515/0.437016/1.26932, 0.437016/-0.682515/-1.345/-0.880585, 0.437016/1.26932/-1.345/1.07125, -1.345/-0.880585/-1.345/1.07125, 0./0./-0.907981/0.388734, 0./0./0.907981/-0.388734, 0./0./-1.78201/-0.19807, 0./0./1.78201/0.19807, 0./0./0./1.95183, 0./0./0./-1.95183}
					{
						\draw[thick] (\x,\y) -- (\z,\w);
					}
					\foreach \x/\y in {0./0., 0.437016/-0.682515, -0.437016/-1.26932, -1.345/1.07125, 1.345/0.880585, -0.907981/0.388734, 1.78201/0.19807, -1.78201/-0.19807, 0.907981/-0.388734, 0./1.95183}
					{
						\draw[fill=\myblue] (\x,\y) circle (0.2);
					}
					\foreach \x/\y in {-0.437016/0.682515, 0.437016/1.26932, 1.345/-1.07125, -1.345/-0.880585, 0./-1.95183}
					{
						\draw[fill=white] (\x,\y) circle (0.2);
					}
					\node at (3.5,1) {(19)};
					\node[burgundy] at (3.5,0) {Closed};
					\node at (3.5,-1) {$w=-6$};
				\end{tikzpicture}&\begin{tikzpicture}[scale=0.5]
					\node at (0,2.5) {};
					\node at (0,-2.5) {};
					\foreach \x/\y/\z/\w in {1.345/-1.07125/1.345/0.880585, 1.345/-1.07125/-0.437016/-1.26932, 1.345/-1.07125/0.437016/-0.682515, 1.345/0.880585/-0.437016/0.682515, 1.345/0.880585/0.437016/1.26932, -0.437016/-1.26932/-0.437016/0.682515, -0.437016/-1.26932/-1.345/-0.880585, -0.437016/0.682515/-1.345/1.07125, 0.437016/-0.682515/0.437016/1.26932, 0.437016/-0.682515/-1.345/-0.880585, 0.437016/1.26932/-1.345/1.07125, -1.345/-0.880585/-1.345/1.07125, 0./0./-0.907981/0.388734, 0./0./0.907981/-0.388734, 0./0./-1.78201/-0.19807, 0./0./1.78201/0.19807, 0./0./0./1.95183, 0./0./0./-1.95183}
					{
						\draw[thick] (\x,\y) -- (\z,\w);
					}
					\foreach \x/\y in {0./0., 0.437016/-0.682515, -0.437016/-1.26932, -1.345/1.07125, 1.345/0.880585, 0./-1.95183, -0.907981/0.388734, 1.78201/0.19807, -1.78201/-0.19807, 0.907981/-0.388734, 0./1.95183}
					{
						\draw[fill=\myblue] (\x,\y) circle (0.2);
					}
					\foreach \x/\y in {-0.437016/0.682515, 0.437016/1.26932, 1.345/-1.07125, -1.345/-0.880585}
					{
						\draw[fill=white] (\x,\y) circle (0.2);
					}
					\node at (3.5,1) {(20)};
					\node[burgundy] at (3.5,0) {Closed};
					\node at (3.5,-1) {$w=-3$};
				\end{tikzpicture}\\
				\hline
				\begin{tikzpicture}[scale=0.5]
					\node at (0,2.5) {};
					\node at (0,-2.5) {};
					\foreach \x/\y/\z/\w in {1.345/-1.07125/1.345/0.880585, 1.345/-1.07125/-0.437016/-1.26932, 1.345/-1.07125/0.437016/-0.682515, 1.345/0.880585/-0.437016/0.682515, 1.345/0.880585/0.437016/1.26932, -0.437016/-1.26932/-0.437016/0.682515, -0.437016/-1.26932/-1.345/-0.880585, -0.437016/0.682515/-1.345/1.07125, 0.437016/-0.682515/0.437016/1.26932, 0.437016/-0.682515/-1.345/-0.880585, 0.437016/1.26932/-1.345/1.07125, -1.345/-0.880585/-1.345/1.07125, 0./0./-0.907981/0.388734, 0./0./0.907981/-0.388734, 0./0./-1.78201/-0.19807, 0./0./1.78201/0.19807, 0./0./0./1.95183, 0./0./0./-1.95183}
					{
						\draw[thick] (\x,\y) -- (\z,\w);
					}
					\foreach \x/\y in {0./0., 0.437016/-0.682515, -0.437016/-1.26932, -1.345/1.07125, 1.345/0.880585, -1.345/-0.880585, 0./-1.95183, -0.907981/0.388734, -1.78201/-0.19807}
					{
						\draw[fill=\myblue] (\x,\y) circle (0.2);
					}
					\foreach \x/\y in {-0.437016/0.682515, 0.437016/1.26932, 1.345/-1.07125, 1.78201/0.19807, 0.907981/-0.388734, 0./1.95183}
					{
						\draw[fill=white] (\x,\y) circle (0.2);
					}
					\node at (3.5,1) {(21)};
					\node[burgundy] at (3.5,0) {Closed};
					\node at (3.5,-1) {$w=-7$};
				\end{tikzpicture}&\begin{tikzpicture}[scale=0.5]
					\node at (0,2.5) {};
					\node at (0,-2.5) {};
					\foreach \x/\y/\z/\w in {1.345/-1.07125/1.345/0.880585, 1.345/-1.07125/-0.437016/-1.26932, 1.345/-1.07125/0.437016/-0.682515, 1.345/0.880585/-0.437016/0.682515, 1.345/0.880585/0.437016/1.26932, -0.437016/-1.26932/-0.437016/0.682515, -0.437016/-1.26932/-1.345/-0.880585, -0.437016/0.682515/-1.345/1.07125, 0.437016/-0.682515/0.437016/1.26932, 0.437016/-0.682515/-1.345/-0.880585, 0.437016/1.26932/-1.345/1.07125, -1.345/-0.880585/-1.345/1.07125, 0./0./-0.907981/0.388734, 0./0./0.907981/-0.388734, 0./0./-1.78201/-0.19807, 0./0./1.78201/0.19807, 0./0./0./1.95183, 0./0./0./-1.95183}
					{
						\draw[thick] (\x,\y) -- (\z,\w);
					}
					\foreach \x/\y in {0./0., 0.437016/-0.682515, -0.437016/-1.26932, -1.345/1.07125, 1.345/0.880585, -1.345/-0.880585, 0./-1.95183, -0.907981/0.388734, -1.78201/-0.19807, 0./1.95183}
					{
						\draw[fill=\myblue] (\x,\y) circle (0.2);
					}
					\foreach \x/\y in {-0.437016/0.682515, 0.437016/1.26932, 1.345/-1.07125, 1.78201/0.19807, 0.907981/-0.388734}
					{
						\draw[fill=white] (\x,\y) circle (0.2);
					}
					\node at (3.5,1) {(22)};
					\node[burgundy] at (3.5,0) {Closed};
					\node at (3.5,-1) {$w=-6$};
				\end{tikzpicture}&\begin{tikzpicture}[scale=0.5]
					\node at (0,2.5) {};
					\node at (0,-2.5) {};
					\foreach \x/\y/\z/\w in {1.345/-1.07125/1.345/0.880585, 1.345/-1.07125/-0.437016/-1.26932, 1.345/-1.07125/0.437016/-0.682515, 1.345/0.880585/-0.437016/0.682515, 1.345/0.880585/0.437016/1.26932, -0.437016/-1.26932/-0.437016/0.682515, -0.437016/-1.26932/-1.345/-0.880585, -0.437016/0.682515/-1.345/1.07125, 0.437016/-0.682515/0.437016/1.26932, 0.437016/-0.682515/-1.345/-0.880585, 0.437016/1.26932/-1.345/1.07125, -1.345/-0.880585/-1.345/1.07125, 0./0./-0.907981/0.388734, 0./0./0.907981/-0.388734, 0./0./-1.78201/-0.19807, 0./0./1.78201/0.19807, 0./0./0./1.95183, 0./0./0./-1.95183}
					{
						\draw[thick] (\x,\y) -- (\z,\w);
					}
					\foreach \x/\y in {0./0., 0.437016/-0.682515, -0.437016/-1.26932, -1.345/1.07125, 1.345/0.880585, -1.345/-0.880585, 0./-1.95183, -0.907981/0.388734, -1.78201/-0.19807, 0.907981/-0.388734, 0./1.95183}
					{
						\draw[fill=\myblue] (\x,\y) circle (0.2);
					}
					\foreach \x/\y in {-0.437016/0.682515, 0.437016/1.26932, 1.345/-1.07125, 1.78201/0.19807}
					{
						\draw[fill=white] (\x,\y) circle (0.2);
					}
					\node at (3.5,1) {(23)};
					\node[burgundy] at (3.5,0) {Closed};
					\node at (3.5,-1) {$w=-5$};
				\end{tikzpicture}&\begin{tikzpicture}[scale=0.5]
					\node at (0,2.5) {};
					\node at (0,-2.5) {};
					\foreach \x/\y/\z/\w in {1.345/-1.07125/1.345/0.880585, 1.345/-1.07125/-0.437016/-1.26932, 1.345/-1.07125/0.437016/-0.682515, 1.345/0.880585/-0.437016/0.682515, 1.345/0.880585/0.437016/1.26932, -0.437016/-1.26932/-0.437016/0.682515, -0.437016/-1.26932/-1.345/-0.880585, -0.437016/0.682515/-1.345/1.07125, 0.437016/-0.682515/0.437016/1.26932, 0.437016/-0.682515/-1.345/-0.880585, 0.437016/1.26932/-1.345/1.07125, -1.345/-0.880585/-1.345/1.07125, 0./0./-0.907981/0.388734, 0./0./0.907981/-0.388734, 0./0./-1.78201/-0.19807, 0./0./1.78201/0.19807, 0./0./0./1.95183, 0./0./0./-1.95183}
					{
						\draw[thick] (\x,\y) -- (\z,\w);
					}
					\foreach \x/\y in {0./0., 0.437016/-0.682515, -0.437016/-1.26932, -1.345/1.07125, 1.345/0.880585, -1.345/-0.880585, 0./-1.95183, -0.907981/0.388734, 1.78201/0.19807, -1.78201/-0.19807, 0.907981/-0.388734, 0./1.95183}
					{
						\draw[fill=\myblue] (\x,\y) circle (0.2);
					}
					\foreach \x/\y in {-0.437016/0.682515, 0.437016/1.26932, 1.345/-1.07125}
					{
						\draw[fill=white] (\x,\y) circle (0.2);
					}
					\node at (3.5,1) {(24)};
					\node[burgundy] at (3.5,0) {Closed};
					\node at (3.5,-1) {$w=-2$};
				\end{tikzpicture}\\
				\hline
				\begin{tikzpicture}[scale=0.5]
					\node at (0,2.5) {};
					\node at (0,-2.5) {};
					\foreach \x/\y/\z/\w in {1.345/-1.07125/1.345/0.880585, 1.345/-1.07125/-0.437016/-1.26932, 1.345/-1.07125/0.437016/-0.682515, 1.345/0.880585/-0.437016/0.682515, 1.345/0.880585/0.437016/1.26932, -0.437016/-1.26932/-0.437016/0.682515, -0.437016/-1.26932/-1.345/-0.880585, -0.437016/0.682515/-1.345/1.07125, 0.437016/-0.682515/0.437016/1.26932, 0.437016/-0.682515/-1.345/-0.880585, 0.437016/1.26932/-1.345/1.07125, -1.345/-0.880585/-1.345/1.07125, 0./0./-0.907981/0.388734, 0./0./0.907981/-0.388734, 0./0./-1.78201/-0.19807, 0./0./1.78201/0.19807, 0./0./0./1.95183, 0./0./0./-1.95183}
					{
						\draw[thick] (\x,\y) -- (\z,\w);
					}
					\foreach \x/\y in {0./0., 0.437016/-0.682515, -0.437016/-1.26932, -1.345/1.07125, 1.345/-1.07125, 1.345/0.880585, -1.345/-0.880585, 0./-1.95183, -0.907981/0.388734, 1.78201/0.19807, -1.78201/-0.19807, 0.907981/-0.388734}
					{
						\draw[fill=\myblue] (\x,\y) circle (0.2);
					}
					\foreach \x/\y in {-0.437016/0.682515, 0.437016/1.26932, 0./1.95183}
					{
						\draw[fill=white] (\x,\y) circle (0.2);
					}
					\node at (3.5,1) {(25)};
					\node[burgundy] at (3.5,0) {Closed};
					\node at (3.5,-1) {$w=-4$};
				\end{tikzpicture}&\begin{tikzpicture}[scale=0.5]
					\node at (0,2.5) {};
					\node at (0,-2.5) {};
					\foreach \x/\y/\z/\w in {1.345/-1.07125/1.345/0.880585, 1.345/-1.07125/-0.437016/-1.26932, 1.345/-1.07125/0.437016/-0.682515, 1.345/0.880585/-0.437016/0.682515, 1.345/0.880585/0.437016/1.26932, -0.437016/-1.26932/-0.437016/0.682515, -0.437016/-1.26932/-1.345/-0.880585, -0.437016/0.682515/-1.345/1.07125, 0.437016/-0.682515/0.437016/1.26932, 0.437016/-0.682515/-1.345/-0.880585, 0.437016/1.26932/-1.345/1.07125, -1.345/-0.880585/-1.345/1.07125, 0./0./-0.907981/0.388734, 0./0./0.907981/-0.388734, 0./0./-1.78201/-0.19807, 0./0./1.78201/0.19807, 0./0./0./1.95183, 0./0./0./-1.95183}
					{
						\draw[thick] (\x,\y) -- (\z,\w);
					}
					\foreach \x/\y in {0./0., 0.437016/-0.682515, -0.437016/-1.26932, -1.345/1.07125, 1.345/-1.07125, 1.345/0.880585, -1.345/-0.880585, 0./-1.95183, -0.907981/0.388734, 1.78201/0.19807, -1.78201/-0.19807, 0.907981/-0.388734, 0./1.95183}
					{
						\draw[fill=\myblue] (\x,\y) circle (0.2);
					}
					\foreach \x/\y in {-0.437016/0.682515, 0.437016/1.26932}
					{
						\draw[fill=white] (\x,\y) circle (0.2);
					}
					\node at (3.5,1) {(26)};
					\node[burgundy] at (3.5,0) {Closed};
					\node at (3.5,-1) {$w=-1$};
				\end{tikzpicture}&\begin{tikzpicture}[scale=0.5]
					\node at (0,2.5) {};
					\node at (0,-2.5) {};
					\foreach \x/\y/\z/\w in {1.345/-1.07125/1.345/0.880585, 1.345/-1.07125/-0.437016/-1.26932, 1.345/-1.07125/0.437016/-0.682515, 1.345/0.880585/-0.437016/0.682515, 1.345/0.880585/0.437016/1.26932, -0.437016/-1.26932/-0.437016/0.682515, -0.437016/-1.26932/-1.345/-0.880585, -0.437016/0.682515/-1.345/1.07125, 0.437016/-0.682515/0.437016/1.26932, 0.437016/-0.682515/-1.345/-0.880585, 0.437016/1.26932/-1.345/1.07125, -1.345/-0.880585/-1.345/1.07125, 0./0./-0.907981/0.388734, 0./0./0.907981/-0.388734, 0./0./-1.78201/-0.19807, 0./0./1.78201/0.19807, 0./0./0./1.95183, 0./0./0./-1.95183}
					{
						\draw[thick] (\x,\y) -- (\z,\w);
					}
					\foreach \x/\y in {0./0., 0.437016/-0.682515, -0.437016/-1.26932, 0.437016/1.26932, -1.345/1.07125, 1.345/-1.07125, 1.345/0.880585, -1.345/-0.880585, 0./-1.95183, -0.907981/0.388734, 1.78201/0.19807, -1.78201/-0.19807, 0.907981/-0.388734, 0./1.95183}
					{
						\draw[fill=\myblue] (\x,\y) circle (0.2);
					}
					\foreach \x/\y in {-0.437016/0.682515}
					{
						\draw[fill=white] (\x,\y) circle (0.2);
					}
					\node at (3.5,1) {(27)};
					\node[burgundy] at (3.5,0) {Closed};
					\node at (3.5,-1) {$w=0$};
				\end{tikzpicture}& \begin{tikzpicture}[scale=0.5]
					\node at (0,2.5) {};
					\node at (0,-2.5) {};
					\foreach \x/\y/\z/\w in {1.345/-1.07125/1.345/0.880585, 1.345/-1.07125/-0.437016/-1.26932, 1.345/-1.07125/0.437016/-0.682515, 1.345/0.880585/-0.437016/0.682515, 1.345/0.880585/0.437016/1.26932, -0.437016/-1.26932/-0.437016/0.682515, -0.437016/-1.26932/-1.345/-0.880585, -0.437016/0.682515/-1.345/1.07125, 0.437016/-0.682515/0.437016/1.26932, 0.437016/-0.682515/-1.345/-0.880585, 0.437016/1.26932/-1.345/1.07125, -1.345/-0.880585/-1.345/1.07125, 0./0./-0.907981/0.388734, 0./0./0.907981/-0.388734, 0./0./-1.78201/-0.19807, 0./0./1.78201/0.19807, 0./0./0./1.95183, 0./0./0./-1.95183}
					{
						\draw[thick] (\x,\y) -- (\z,\w);
					}
					\foreach \x/\y in {0./0., 0.437016/-0.682515, -0.437016/0.682515, -0.437016/-1.26932, 0.437016/1.26932, -1.345/1.07125, 1.345/-1.07125, 1.345/0.880585, -1.345/-0.880585, 0./-1.95183, -0.907981/0.388734, 1.78201/0.19807, -1.78201/-0.19807, 0.907981/-0.388734, 0./1.95183}
					{
						\draw[fill=\myblue] (\x,\y) circle (0.2);
					}
					\node at (3.5,1) {(28)};
					\node[\mygreen] at (3.5,0) {Open};
					\node at (3.5,-1) {$w=1$};
				\end{tikzpicture}\\
				\hline
			\end{array}
		\end{equation}
		\endgroup
		\caption{Independent local solutions $M(\Pi_c,\mu_0)$ to inequalities \eqref{ineq}.}\label{fig:ineq_sol}
	\end{center}
\end{figure}

One could then lift the local pictures in Fig.\ \ref{fig:ineq_sol} to 3D tessellations similarly to the Young diagrams and the plane partition cases in  App.\ \ref{app:warmup}.  
We will not present them here since these 3D tessellations are rather complicated and not very intelligible in print.\footnote{
	It is easier to view them in e.g. Mathematica since one can then rotate them.} 
In words, each of these tessellations is a unique representative of the $S_4$ group orbit and represents a local picture of a 3D hyper-surface with ``pits'', ``hills'', and ``saddles'' that bound a solid partition configuration.
Therefore, we conclude that the number of inequivalent local pictures for the solid partitions is {\bf 28}, in comparison to the {\bf 3} and {\bf{8}} local pictures for the Young diagram and plane partition cases, respectively.

From Fig.~\ref{fig:ineq_sol}, it is clear that when the configuration of the height fields in the neighborhood of point $c$ is ``open", namely when the solid partition $\bpi$ admits adding or removing of boxes, the potential $w=1$, which corresponds to the order-1 pole in the charge function \eqref{ch_f}. 
On the other hand, in a ``closed" case, $w$ is non-positive, indicating a zero of order $-w$ at the corresponding lattice point.

Since the weight function \eqref{weight} is merely a rewriting of the charge function \eqref{ch_f}, and we have shown, by simply examining all {\bf 28} cases in Fig.\ \ref{fig:ineq_sol}, that the weight function \eqref{weight} correctly distinguishes between open and closed solutions $M(\Pi_c,\mu_0)$, we have effectively proven that the charge function \eqref{ch_f} has the desired properties, in particular having poles of order 1 at positions where boxes can be added or removed, for all the local pictures from configuration (c)  of Fig.\ \ref{fig:edge_config}.

Next, we need to consider the other 3, non-$S_4$-symmetric, configurations in Fig.\ \ref{fig:edge_config} as well, which correspond to  contributions from the boxes near the walls of the 4D room, as one moves away from the diagonal. 
One straightforward way is to simply check all the ${\bf 4} \times {\bf 166}={\bf 664}$  combinations of edge weights in Fig.~\ref{fig:edge_config} and their solutions to the set of inequalities \eqref{ineq}, which is what we did first (using computer) and we confirmed that for open local solutions $w=1$ and for closed ones $w\leq 0$.
Therefore, the charge function \eqref{ch_f} has the desired properties for all these configurations as well.

\begin{figure}[ht!]
	\begin{center}
		\begin{tikzpicture}
			\draw[-stealth] (0,0) -- (0,3);
			\draw[-stealth] (0,0) -- (1,0);
			\draw[-stealth] (0,0) -- (-0.7,-0.7);
			\draw[-stealth] (0,0) -- (-1,0.5);
			\draw[fill=black] (-1,0.5) circle (0.05);
			\node[below left] at (-1,0.5) {$\scriptstyle \vec\ell(c)$};
			\begin{scope}[shift={(-1,0.5)}]
				\begin{scope}[scale=0.6,rotate=45]
					\begin{scope}[xscale=0.4]
						\draw[thick, fill=\mygreen, fill opacity=0.3] (0,0) circle (1);
					\end{scope}
					\begin{scope}[shift={(2,0)}]
						\begin{scope}[xscale=0.4]
							\draw[thick, fill=\myblue, fill opacity=0.3] (0,0) circle (1);
						\end{scope}
					\end{scope}
					\begin{scope}[shift={(4,0)}]
						\begin{scope}[xscale=0.4]
							\draw[thick, fill=burgundy, fill opacity=0.3] (0,0) circle (1);
						\end{scope}
					\end{scope}
					\draw[thick] (0,1) -- (4,1) (0,-1) -- (4,-1);
					\node at (1,0) {$\hat \Pi_c^-$};
					\node at (3,0) {$\hat \Pi_c^+$};
					\node(A) at (0,-1) {};
					\node(B) at (2,-1) {};
					\node(C) at (4,-1) {};
					\draw (0,1.2) to[out=90,in=180] (0.4,1.6) -- (1.6,1.6) to[out=0,in=270] (2,2) to[out=270,in=180] (2.4,1.6) -- (3.6,1.6) to[out=0,in=90] (4,1.2);
					\node[above left] at (2,2) {$\scriptstyle \mu_0$};
				\end{scope}
			\end{scope}
			\node[right] (D) at (2.5,1.5) {\footnotesize {\color{burgundy} top} boundary};
			\node[right] (E) at (2.5,0.5) {\footnotesize canonical {\color{\myblue} cut}};
			\node[right] (F) at (2.5,-0.5) {\footnotesize {\color{\mygreen} bottom} boundary};
			\draw[->, thin] (D.west) to[out=180,in=330] (C);
			\draw[->, thin] (E.west) to[out=180,in=330] (B);
			\draw[->, thin] (F.west) to[out=180,in=330] (A);
			\node[right] at (6,0.5) {$\begin{array}{l}
					\hat\Pi_c=\hat \Pi_c^-\cup g(\hat\Pi_c^+),\;g\in S_4\,,\\
					\\
					\psi_{\hat\Pi_c}(z)=\psi_{\hat\Pi_c^-}(z)\cdot \psi_{\hat\Pi_c^+}(z)\,;
				\end{array}$};
		\end{tikzpicture}
		\caption{The lift $\hat \Pi_c$ of a local patch $\Pi_c$ from the projection to the solid partition.}\label{fig:patch_lift}
	\end{center}
\end{figure}
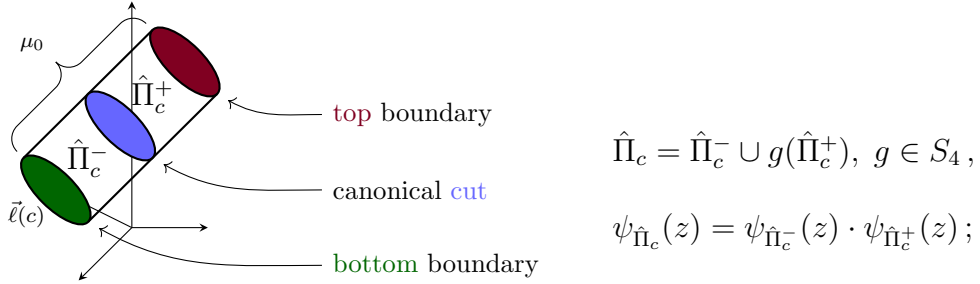

However, let us also give another proof that does not need to involve checking this large number of cases explicitly.
Let us first explain where this large number come from.
Consider a lift $\hat \Pi_c$ of a local patch $\Pi_c$ to $\IZ_{\geq 0}^4$ (see Fig.\ \ref{fig:patch_lift}).
From the stacked boxes in the solid partitions, the lift $\hat \Pi_c$ carves out all layers of the boxes projected with vector $(1,1,1,1)$ to $\mathscr{B}$, therefore $\hat \Pi_c$ can be viewed as a cylinder of height $\mu_0$ and with a 3D cross-section.
As a cylinder, $\hat \Pi_c$ has two boundaries: the top one and the bottom one.
The top boundary belongs to the 3D surface that is part of the outer boundary of the solid partition, whose projection to 3D is called a tessellation, whereas the bottom boundary is at the intersection of $\hat \Pi_c$ with the walls of the 4D room, from the corner of which 4D boxes are stacked to form solid partitions.

When we mod out all the possible local shapes of the top boundaries of these cylinder by $S_4$, we get $\bf 28$ local pictures, shown in Fig.\ \ref{fig:ineq_sol}. 
When we mod out all the local shapes of the bottom boundaries by $S_4$, we get $\bf 4$  edge weight configurations, shown in Fig.\ \ref{fig:edge_config}. 
The fact that the charge function \eqref{ch_f} is symmetric with respect to $S_4$ (which acts by permuting $\myh_{1,2,3,4}$) allows us to consider only individual representatives of the local patch in the $S_4$ conjugacy classes.
However, this does not allow us to mod out by $S_4$ the top and bottom boundaries independently --- the $S_4$ group acts on the cylinder of $\hat\Pi_c$ as a whole!
This is the reason behind the large number of cases mentioned earlier. 

To improve the situation, let us perform a canonical cut somewhere in the middle of the cylinder as it is depicted in Fig.\ \ref{fig:patch_lift}.
This canonical cut is chosen to have the form of the corner of the empty 4D room, transported to the position of the cut, hence is  $S_4$-symmetric.
The cut divides $\hat\Pi_c$ into two haves: $\hat\Pi_c^-$ and $\hat\Pi_c^+$.
Since the shape of the cut is $S_4$-symmetric, we can obtain an arbitrary cylinder $\hat\Pi_c$ by gluing a pair of $S_4$-conjugacy classes $\hat\Pi_c^-$ and $\hat\Pi_c^+$ after first imposing a twist $g\in S_4$ (see Fig.\ \ref{fig:patch_lift}).

Since the charge function \eqref{ch_f} is extensive with respect to size of the solid partition, namely, the products in the expression \eqref{ch_f} decompose into products over $\hat\Pi_c^-$ and $\hat\Pi_c^+$.
Therefore, to prove that the charge function has the right properties, it is sufficient to prove that its $\hat\Pi_c^-$ and $\hat\Pi_c^+$ components have the right properties.
In other words, one only needs to check all the $S_4$-conjugacy classes of the bottom boundaries, which are paired using the canonical cut with the top boundary and then, separately, all the $S_4$-conjugacy classes of the top boundaries, which are paired using the canonical cut with the bottom boundary.
In total, this only gives us ${\bf 4}+{\bf 28}={\bf 32}$  cases to check.

Before we end this section, we mention again that we have also checked explicitly, using computer, all solutions of the set of inequalities \eqref{ineq} for all the possible the edge weight configurations in Fig.\ \ref{fig:edge_config}, without considering the $S_4$ action.
In fact, just to be on the safer side, we even considered a local patch larger than $\Pi_c$, by appending to $\Pi_c$ the lattice nodes located at $\myh_i-\myh_{j\neq i}$, which sit at the centers of the 12 BCC cells located across the 12 edges of the cube, see Fig.\ \ref{fig:voronoi}.
We found that for this larger local patch, there are $\bf 7$ (instead of $\bf 4$) local edge weight configurations and for  each such configuration, there are $\bf 14\, 656$ (instead of $\bf 166$) solutions to the inequalities \eqref{ineq}. 
We have also used computer to check all the ${\bf 7}\times {\bf 14\, 656}={\bf 102\,592}$ cases and verified that  $w=1$ (resp.\ $w\leq 0$) for open (resp.\ closed) local solutions, thus confirmed the validity of the charge function \eqref{ch_f}.

\section{Discussion}\label{S:discussion}
In this note we have constructed the charge function \eqref{ch_f}  for solid partitions that satisfies a set of properties that are natural analogues to the lower dimensional cases and are essential for the charge function to be the ingredient in constructing the corresponding BPS algebra. 
We first obtained a conjectured form by analyzing the solid partitions with the number of boxes up to 15, and then proved that it indeed satisfies the required properties for all possible solid partitions by checking explicitly all local pictures.

Compared to the case for the plane partitions, which describe D-brane BPS states on $\IC^3$, the charge function \eqref{ch_f} of the solid partitions, which describes D-brane BPS states on $\IC^4$, has a somewhat more intricate structure: apart from contributions from individual boxes, the charge function \eqref{ch_f} also receives contributions from  certain 4-box and 5-box clusters.
Therefore, the corresponding BPS algebra would need to have more than just creation and annihilation operators that add and remove single boxes, different from the 3D case in \cite{Li:2020rij}.
Equivalently, if we want to have some effective model description for the solid partition boundary, similar to the anyons organized on a lattice in the case of the plane partitions (see e.g. \cite[Sec.~2.7]{Galakhov:2020vyb}), we would have to supplement the purely pairwise particle interaction with higher order interactions of 4- and 5-particle clusters.

To conclude this note we would like to list some interesting problems for future directions:
\begin{itemize}
	\item[\textcolor{\myblue}{\textbullet}] It would be interesting to generalize the story to dimensions  $D>4$ and construct the charge functions of D-dimensional partitions.
	\item[\textcolor{\myblue}{\textbullet}] One can also generalize to other toric CY${}_4$s, maybe using the effective mechanism of constructing quivers and molten crystal systems for toric CY${}_4$ proposed in \cite{Franco:2015tna,Franco:2015tya,Franco:2019bmx}.\footnote{Note added: the generalization to other toric CY${}_4$s was discussed in \cite{Franco:2023tly} that appeared shortly after our paper.}
	It would be interesting to see how the structure of the charge function varies across the entire toric CY${}_4$ family.

	\item[\textcolor{\myblue}{\textbullet}] The charge function hierarchy problem:
	a $D$-dimensional partition $\bpi^{(D)}$ is a slice in a $(D+1)$-dimensional partition $\bpi^{(D+1)}$.
	Suppose this slice is perpendicular to the $(D+1)$-st axis, then imposing $\myh_{D+1}=0$ produces the CY${}_{D}$ constraint $\sum_{k=1}^D\myh_k=0$ from the CY${}_{D+1}$ constraint.
	Following this hierarchy of partitions across different dimensions, one would expect some corresponding hierarchy of their charge functions, such that they simply reduce to one another:\footnote{
		We would like to thank Alexei Morozov for drawing our attention to this problem.}
	\begin{equation}
		\begin{array}{c}
			\begin{tikzpicture}
				\node(A) at (-2,0) {$\ldots$};
				\node(B) at (0,0) {$\psi_{\bpi}^{(4D)}(z)$};
				\node(C) at (3,0) {$\psi_{\bpi}^{(3D)}(z)$};
				\node(D) at (6,0) {$\psi_{\bpi}^{(2D)}(z)$};
				\path (A) edge[->] (B) (B) edge[->] node[above] {$\scriptstyle \myh_4=0$} (C) (C) edge[->] node[above] {$\scriptstyle \myh_3=0$} (D);
			\end{tikzpicture}
		\end{array}
	\end{equation}
	Note that this doesn't work straightforwardly.
	Indeed if we set either $\myh_3=0$ in the $\varphi_1(z)$ contribution in 3D \eqref{ch_f_3D}, or $\myh_4=0$ in the $\varphi_1(z)$ contribution in 4D \eqref{ch_f}, while keeping the CY constraint $\sum\lm_k\myh_k=0$, then the contribution from the singlet loses poles and the charge function no longer works even for the partition consisting of a single box.
	Therefore it would be very interesting to determine the  mechanism that governs the dimensional reduction at the level of the charge functions.
	
	\item[\textcolor{\myblue}{\textbullet}] 
	As we have seen, the charge function for the $D$-dimensional partitions is a function on the $(D-1)$-dimensional boundaries.
	By enhancing the partitions with statistical weights, one can even reproduce some dynamics of this boundary form and acquire an effective model of $(D-1)$-dimensional gravity \cite{Kenyon:2003uj, Ooguri:2010yk}.
	Thus we wonder whether the simplicity of the charge function in 3D might indicate the emergence of some integrability properties, similar to how 2D gravity turns out to be integrable, and whether some analogous structure can be uncovered in higher dimensions.
	\item[\textcolor{\myblue}{\textbullet}] Finally, the most intriguing problem is whether and how we can bootstrap the BPS algebraic structure using the charge function \eqref{ch_f} that admits contributions from clusters of boxes.
\end{itemize}

\section*{Acknowledgements}

We would like to thank Masahito Yamazaki for the initial collaboration on this project, and also Alexei Morozov, Alexander Popolitov and Nikita Tselousov for helpful discussions. 
We would also like to thank Alexander Popolitov for his invaluable help with handling computer hardware issues.
The work of DG is supported by the Russian Science
Foundation (Grant No.20-12-00195).
The work of WL is supported by NSFC No.\ 11875064, No.\ 12275334, No.\ 11947302, and the Max-Planck Partnergruppen fund; WL is also grateful for the support and hospitality of the Max Planck Institute for Gravitational Physics, the Kavli Institute for Theoretical Physics, Perimeter Institute, and the Issac Newton Institute for Mathematical Sciences (during Program ``Black holes: bridges between number theory and holographic quantum information" with EPSRC Grant ER/R014604/1) when part of this work was carried out.

\appendix

\section{Warm up: Young diagrams and plane partitions}\label{app:warmup}

In this Appendix we review the Young diagrams (2D) and plane partitions (3D), in particular the construction of their charge functions, using the same method that we apply on solid partitions in the main text. 

\subsection{2D: Young diagrams}
The generating function for 2D partitions (Young diagrams) is:
\begin{equation}
	\begin{aligned}
		g(q)&=\prod^{\infty}_{k=1}\frac{1}{(1-q^k)}=\sum^{\infty}_{n=0}\,p(n)q^{n} \\
		&= 1+q + 2 q^2 + 3 q^3 + 5 q^4 + 7 q^5 + 11 q^6 + \dots\,,
	\end{aligned}
\end{equation}
where $p(n)$ counts the number of ways to partition a non-negative integer $n$. 
Its plethystic log is:
\begin{equation}
	f(q)={\rm PL}[g](q)=\frac{1}{1-q}=\sum^{\infty}_{k=0}q^k
\end{equation}

Given a Young diagram, we can translate it into a spin-chain configuration by the following rule, see e.g.\ \cite{Dijkgraaf:2008ua}. 
We position the Young diagram as follows and project it onto a spin-chin, with each vertical strip corresponding to a spin and the spin flipping each time the outer boundary of the Young diagram has a corner (either convex or concave):
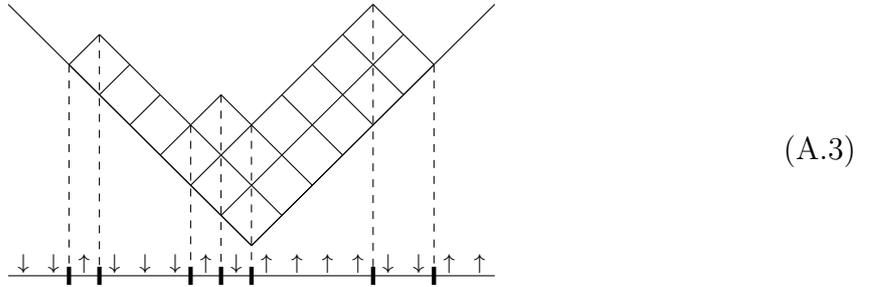
\begin{equation}\label{diagYoung}
	\begin{array}{c}
		\begin{tikzpicture}[scale=0.4]
			\begin{scope}[rotate=45]
				\begin{scope}[scale=1.41421]
					\draw (0,8) -- (0,0) -- (8,0);
					\foreach \x/\y/\z/\w in {0/0/0/1, 0/0/1/0, 0/1/0/2, 0/1/1/1, 0/2/0/3, 0/2/1/2, 0/3/0/4, 0/3/1/3, 0/4/0/5, 0/4/1/4, 0/5/0/6, 0/5/1/5, 0/6/1/6, 1/0/1/1, 1/0/2/0, 1/1/1/2, 1/1/2/1, 1/2/1/3, 1/2/2/2, 1/3/1/4, 1/3/2/3, 1/4/1/5, 1/5/1/6, 2/0/2/1, 2/0/3/0, 2/1/2/2, 2/1/3/1, 2/2/2/3, 2/2/3/2, 3/0/3/1, 3/0/4/0, 3/1/3/2, 3/1/4/1, 3/2/4/2, 4/0/4/1, 4/0/5/0, 4/1/4/2, 4/1/5/1, 4/2/5/2, 5/0/5/1, 5/0/6/0, 5/1/5/2, 5/1/6/1, 5/2/6/2, 6/0/6/1, 6/1/6/2}
					{
						\draw (\x,\y) -- (\z,\w);
					}
				\end{scope}
			\end{scope}
			\draw (-8,-1) -- (8,-1);
			\draw[dashed] (-6,6) -- (-6,-1) (-5,7) -- (-5,-1) (-2,4) -- (-2,-1) (-1,5) -- (-1,-1) (0,4) -- (0,-1) (4,8) -- (4,-1) (6,6) -- (6,-1);
			\foreach \x in {-6,-5,-2,-1,0,4,6}
			{
				\draw[ultra thick] (\x,-1.3) -- (\x,-0.7);
			} 
			\foreach \x in {-8,-7,-5,-4,-3,-1,4,5}
			{
				\node at (\x+0.5,-0.6) {$\scriptstyle \bf\downarrow$};
			}
			\foreach \x in {-6,-2,0,1,2,3,6,7}
			{
				\node at (\x+0.5,-0.6) {$\scriptstyle \bf\uparrow$};
			}
		\end{tikzpicture}
	\end{array}
\end{equation}
Then we view this configuration as a tessellation of the real line by plaques of two types, which we depict as spin-up and spin-down in diagram \eqref{diagYoung}.

There are three possible distributions of the weights of the local edges \eqref{edge_weight}, depending on whether the point of observation is located at the point 0, or to the left or to the right of 0:
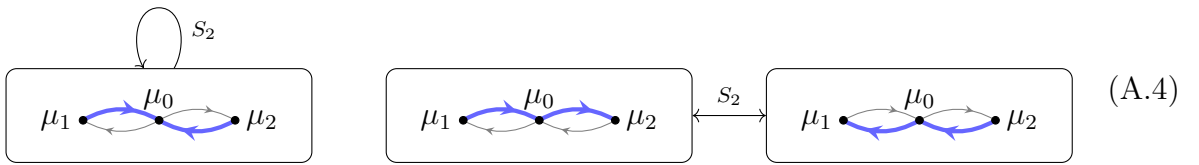
\begin{equation}\label{pic:local2D}
	\begin{array}{c}
		\begin{tikzpicture}
			\node[draw, rounded corners](A) at (0,0) {$\begin{array}{c}
					\begin{tikzpicture}
						\draw[ultra thick, \myblue, postaction={decorate},decoration={markings, 
							mark= at position 0.7 with {\arrow{stealth}}}] (-1,0) to[out=30,in=150] (0,0);
						\draw[gray, postaction={decorate},decoration={markings, 
							mark= at position 0.7 with {\arrow{stealth}}}] (0,0) to[out=30,in=150] (1,0);
						\draw[gray, postaction={decorate},decoration={markings, 
							mark= at position 0.7 with {\arrow{stealth}}}] (0,0) to[out=210,in=330] (-1,0);
						\draw[ultra thick, \myblue, postaction={decorate},decoration={markings, 
							mark= at position 0.7 with {\arrow{stealth}}}] (1,0) to[out=210,in=330] (0,0);
						\draw[fill=black] (-1,0) circle (0.05) (0,0) circle (0.05) (1,0) circle (0.05);
						\node[left] at (-1,0) {$\mu_1$};
						\node[above] at (0,0) {$\mu_0$};
						\node[right] at (1,0) {$\mu_2$};
					\end{tikzpicture}
				\end{array}$};
			\node[draw, rounded corners] (B) at (5,0) {$\begin{array}{c}
					\begin{tikzpicture}
						\draw[ultra thick, \myblue, postaction={decorate},decoration={markings, 
							mark= at position 0.7 with {\arrow{stealth}}}] (-1,0) to[out=30,in=150] (0,0);
						\draw[ultra thick, \myblue, postaction={decorate},decoration={markings, 
							mark= at position 0.7 with {\arrow{stealth}}}] (0,0) to[out=30,in=150] (1,0);
						\draw[gray, postaction={decorate},decoration={markings, 
							mark= at position 0.7 with {\arrow{stealth}}}] (0,0) to[out=210,in=330] (-1,0);
						\draw[gray, postaction={decorate},decoration={markings, 
							mark= at position 0.7 with {\arrow{stealth}}}] (1,0) to[out=210,in=330] (0,0);
						\draw[fill=black] (-1,0) circle (0.05) (0,0) circle (0.05) (1,0) circle (0.05);
						\node[left] at (-1,0) {$\mu_1$};
						\node[above] at (0,0) {$\mu_0$};
						\node[right] at (1,0) {$\mu_2$};
					\end{tikzpicture}
				\end{array}$};
			\node[draw, rounded corners] (C) at (10,0) {$\begin{array}{c}
					\begin{tikzpicture}
						\draw[gray, postaction={decorate},decoration={markings, 
							mark= at position 0.7 with {\arrow{stealth}}}] (-1,0) to[out=30,in=150] (0,0);
						\draw[gray, postaction={decorate},decoration={markings, 
							mark= at position 0.7 with {\arrow{stealth}}}] (0,0) to[out=30,in=150] (1,0);
						\draw[ultra thick, \myblue, postaction={decorate},decoration={markings, 
							mark= at position 0.7 with {\arrow{stealth}}}] (0,0) to[out=210,in=330] (-1,0);
						\draw[ultra thick, \myblue, postaction={decorate},decoration={markings, 
							mark= at position 0.7 with {\arrow{stealth}}}] (1,0) to[out=210,in=330] (0,0);
						\draw[fill=black] (-1,0) circle (0.05) (0,0) circle (0.05) (1,0) circle (0.05);
						\node[left] at (-1,0) {$\mu_1$};
						\node[above] at (0,0) {$\mu_0$};
						\node[right] at (1,0) {$\mu_2$};
					\end{tikzpicture}
				\end{array}$};
			\draw[<->] (B.east) -- (C.west) node[pos=0.5,above] {$\scriptstyle S_2$};
			\draw[->] ([shift={(0.2,0)}]A.north) to[out=60,in=0] node[pos=0.5,right] {$\scriptstyle S_2$} ([shift={(0,0.8)}]A.north) to[out=180,in=120] ([shift={(-0.2,0)}]A.north);
		\end{tikzpicture}
	\end{array}
\end{equation}
where we have adopted the color code \eqref{edge_cc}, and indicted how they transform under the $S_2$ reflection.
Equivalently, the three pictures in \eqref{pic:local2D} correspond to the three cases where the middle box is along, to the right of, and to the left of, the diagonal line, respectively.

Let us express the 2D version of the inequalities \eqref{ineq} explicitly for the first diagram in \eqref{pic:local2D}, which is $S_2$-symmetric.
Using the fact that the middle box is along the diagonal line of the 2D plane, we have
\begin{equation}\label{eq:inequality2D1}
	\begin{aligned}
		&\mu_0\leq \mu_1+1
		\,, \quad 
		\mu_1 \leq \mu_0
		\qquad \textrm{and} \qquad
		\mu_2\leq \mu_0
		\,,\quad 
		\mu_0\leq \mu_2+1\\
		\Longrightarrow\quad 
		&\mu_0-1\leq\mu_1\leq\mu_0
		\qquad \qquad\,\,\,\,\textrm{and} \qquad
		\mu_0-1\leq\mu_2\leq\mu_0\,.    
	\end{aligned}
\end{equation} 
The set of inequalities \eqref{eq:inequality2D1} has  4 solutions:
\begin{equation}\label{eq:solutions2D1}
	\begin{aligned}
		&(1):\mu_1=\mu_2=\mu_0\\
		&(2):\mu_1=\mu_0-1\,, \mu_2=\mu_0
		\quad 
		(3):\mu_1=\mu_0\,, \mu_2=\mu_0-1\\
		&(4):\mu_1=\mu_2=\mu_0-1
	\end{aligned}    
\end{equation}
which correspond to the four cases below:
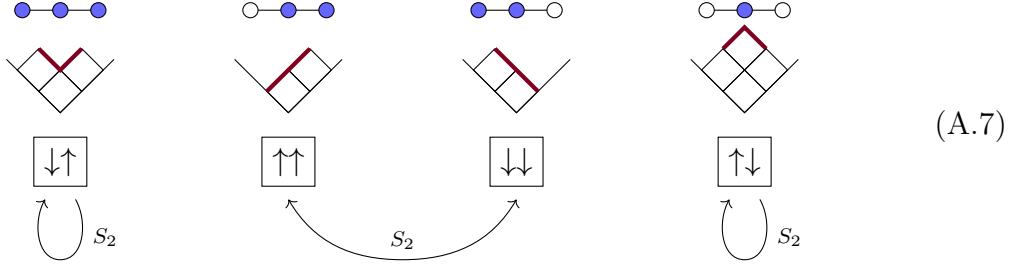
\begin{equation}\label{eq:4cases2D}
	\begin{array}{c}
		\begin{tikzpicture}
			\node at (0,0) {$\begin{array}{c}
					\begin{tikzpicture}
						\draw(-0.5,0) -- (0.5,0);
						\draw[fill=\myblue] (-0.5,0) circle (0.1) (0,0) circle (0.1) (0.5,0) circle (0.1);
					\end{tikzpicture}
				\end{array}$};
			\node at (0,-0.8) {$\begin{array}{c}
					\begin{tikzpicture}[rotate=45, scale=0.4]
						\node at (2,2) {};
						\draw (0,2.5) -- (0,0) -- (2.5,0);
						\foreach \x/\y in {0/0, 0/1, 1/0}
						{
							\draw (\x,\y) -- (\x,\y + 1) -- (\x + 1,\y + 1) -- (\x + 1,\y) -- cycle;
						}
						\draw[ultra thick, burgundy] (1,2) -- (1,1) -- (2,1);
					\end{tikzpicture}
				\end{array}$};
			\node[draw] at (0,-2) {$\downarrow\uparrow$};
			\begin{scope}[shift={(3,0)}]
				\node at (0,0) {$\begin{array}{c}
						\begin{tikzpicture}
							\draw(-0.5,0) -- (0.5,0);
							\draw[fill=\myblue] (0,0) circle (0.1) (0.5,0) circle (0.1);
							\draw[fill=white] (-0.5,0) circle (0.1);
						\end{tikzpicture}
					\end{array}$};
				\node at (0,-0.8) {$\begin{array}{c}
						\begin{tikzpicture}[rotate=45, scale=0.4]
							\node at (2,2) {};
							\draw (0,2.5) -- (0,0) -- (2.5,0);
							\foreach \x/\y in {0/0, 1/0}
							{
								\draw (\x,\y) -- (\x,\y + 1) -- (\x + 1,\y + 1) -- (\x + 1,\y) -- cycle;
							}
							\draw[ultra thick, burgundy] (0,1) -- (2,1);
						\end{tikzpicture}
					\end{array}$};
				\node[draw] at (0,-2) {$\uparrow\uparrow$};
			\end{scope}
			\begin{scope}[shift={(6,0)}]
				\node at (0,0) {$\begin{array}{c}
						\begin{tikzpicture}
							\draw(-0.5,0) -- (0.5,0);
							\draw[fill=\myblue] (0,0) circle (0.1) (-0.5,0) circle (0.1);
							\draw[fill=white] (0.5,0) circle (0.1);
						\end{tikzpicture}
					\end{array}$};
				\node at (0,-0.8) {$\begin{array}{c}
						\begin{tikzpicture}[rotate=45, scale=0.4]
							\node at (2,2) {};
							\draw (0,2.5) -- (0,0) -- (2.5,0);
							\foreach \x/\y in {0/0, 0/1}
							{
								\draw (\x,\y) -- (\x,\y + 1) -- (\x + 1,\y + 1) -- (\x + 1,\y) -- cycle;
							}
							\draw[ultra thick, burgundy] (1,0) -- (1,2);
						\end{tikzpicture}
					\end{array}$};
				\node[draw] at (0,-2) {$\downarrow\downarrow$};
			\end{scope}
			\begin{scope}[shift={(9,0)}]
				\node at (0,0) {$\begin{array}{c}
						\begin{tikzpicture}
							\draw(-0.5,0) -- (0.5,0);
							\draw[fill=\myblue] (0,0) circle (0.1);
							\draw[fill=white] (-0.5,0) circle (0.1) (0.5,0) circle (0.1);
						\end{tikzpicture}
					\end{array}$};
				\node at (0,-0.8) {$\begin{array}{c}
						\begin{tikzpicture}[rotate=45, scale=0.4]
							\node at (2,2) {};
							\draw (0,2.5) -- (0,0) -- (2.5,0);
							\foreach \x/\y in {0/0, 0/1, 1/0, 1/1}
							{
								\draw (\x,\y) -- (\x,\y + 1) -- (\x + 1,\y + 1) -- (\x + 1,\y) -- cycle;
							}
							\draw[ultra thick, burgundy] (1,2) -- (2,2) -- (2,1);
						\end{tikzpicture}
					\end{array}$};
				\node[draw] at (0,-2) {$\uparrow\downarrow$};
			\end{scope}
			\draw[->] (0.2,-2.5) to[out=300,in=0] node[pos=0.5,right] {$\scriptstyle S_2$} (0,-3.3) to[out=180, in=240] (-0.2,-2.5);
			\draw[->] (9.2,-2.5) to[out=300,in=0] node[pos=0.5,right] {$\scriptstyle S_2$} (9,-3.3) to[out=180, in=240] (8.8,-2.5);
			\draw[<->] (3,-2.5) to[out=300,in=180]  (4.5,-3.3) to[out=0, in=240] (6,-2.5);
			\node[above] at (4.5,-3.3) {$\scriptstyle S_2$};
		\end{tikzpicture}
	\end{array}
\end{equation}
with 2 of which in the same orbit of $S_2$.
Thus in this case we have {\bf 3} local pictures (here we have used the color coding for vertices \eqref{vertex_cc}.
Repeating this computation for the other two pictures in \eqref{pic:local2D}, for each of them, we get four solutions as in \eqref{eq:4cases2D}, except that the second (resp.\ third) picture in \eqref{pic:local2D} corresponds to shifting all local pictures in \eqref{eq:4cases2D} to the right (resp.\ left) of the diagonal line.\footnote{
	Note that for the second and third cases in \eqref{pic:local2D}, the inequalities and their solutions are different from \eqref{eq:inequality2D1} and \eqref{eq:solutions2D1}, and only the final configurations are to be compared with \eqref{eq:4cases2D} --- they are the same configurations as  \eqref{eq:4cases2D} shifted to the right and left of the diagonal line, for the second and third cases in \eqref{pic:local2D}.}

For 2D, the Calabi-Yau condition is
\begin{equation}
	\myh_1+\myh_2=0\,.
\end{equation}
Without loss of generality, we can drop the scaling of these parameters and set 
\begin{equation}
	\myh_1=1
	\qquad \textrm{and}\qquad
	\myh_2=-1.    
\end{equation}
In terms of these, the projection operator acquires a very simple form for a box with coordinates $(a,b)$:
\begin{equation}
	{\bf prj}(a,b)=a-b\,.
\end{equation}

The charge function for the 2D partition (Young diagram) $\bpi$ is:
\begin{equation}\label{ch_f_2D}
	\psi_{\bpi}(z)=\frac{1}{z}\prod\lm_{\Phi_1\in\bpi} \varphi_1\left(z-c\left(\Phi_1\right)\right)\prod\lm_{k=1}^2\prod\lm_{\Phi_{2,k}\in\bpi} \varphi_{2,k}\left(z-c\left(\Phi_{2,k}\right)\right)\prod\lm_{\Phi_3\in\bpi} \varphi_3\left(z-c\left(\Phi_3\right)\right)\,,
\end{equation}
where
\begin{itemize}
	\item[\textcolor{\myblue}{\textbullet}]
	A singlet $\Phi_1$ denotes a single box $(a,b)\in\bpi$, contributing
	\begin{equation}
		\varphi_1(z)=\frac{1}{z^2-1}   \,, 
	\end{equation}
	shifted by its projected coordinate $c\left(\Phi_1\right)=a-b$.
	\item[\textcolor{\myblue}{\textbullet}]
	A doublet $\Phi_{2,k=1,2}$ is a pair of boxes at
	\begin{equation}
		(a,b)
		\quad \textrm{and} \quad  (a+\delta_{k,1},b+\delta_{k,2})\,,     
	\end{equation}
	together contributing
	\begin{equation}
		\varphi_{2,k}(z)=z^2
		\,, \qquad k=1,2,    
	\end{equation}
	shifted by its projected coordinate $c\left(\Phi_{4,k}\right)=a-b$.
	\item[\textcolor{\myblue}{\textbullet}]
	A triplet $\Phi_{3}$ is a triplet of boxes at the positions
	\begin{equation}
		(a,b)
		\,,\qquad
		(a+1,b)
		\,,\qquad (a,b+1),  
	\end{equation}
	together contributing
	\begin{equation}
		\varphi_{3}(z)=\frac{1}{z^2}\,,
	\end{equation}
	shifted by its projected coordinate $c\left(\Phi_{3}\right)=a-b$.
\end{itemize}

\subsection{3D: plane partitions} \label{s:3D_par}

The generating function for the plane partition numbers $P(n)$ is the MacMahon function;
\begin{equation}
	\begin{aligned}
		g(q)&=\prod^{\infty}_{k=1}\frac{1}{(1-q^k)^k}=\sum^{\infty}_{n=0}\,P(n)q^{n} \\
		&= 1+q + 3 q^2 + 6 q^3 + 13 q^4 + 24 q^5 + 48 q^6 + \dots\,,
	\end{aligned}
\end{equation}
whose plethystic logarithm is
\begin{equation}
f(q)={\rm PL}[q](q)=\frac{q}{(1-q)^2}=\sum^{\infty}_{k=0}k\,q^k \,.
\end{equation}

The Calabi-Yau constraint in this case reads:
\begin{equation}\label{CY3_constr}
	\myh_1+\myh_2+\myh_3=0\,.
\end{equation}
The projection of the 3D plane partition to the 2D plane \eqref{CY3_constr} works straightforwardly.
The result is a picture of level lines separating areas of different height field values for the dual dimer model \cite{Kenyon:2003uj}:
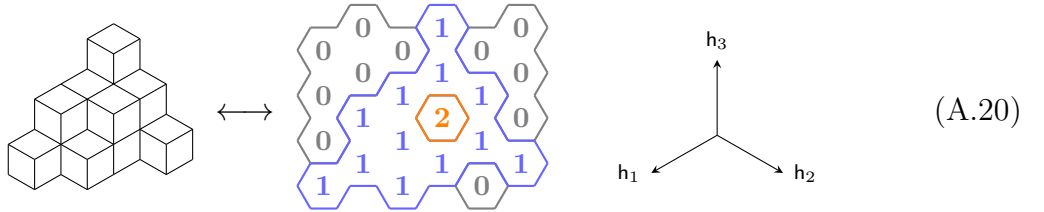
\begin{equation}
	\begin{array}{c}
		\begin{tikzpicture}[scale=0.4]
			\foreach \x/\y/\z/\w in {0./3./0.866025/2.5, 0./3./-0.866025/2.5, 0.866025/1.5/0.866025/2.5, 0.866025/1.5/1.73205/1., 0.866025/1.5/0./1., 0.866025/2.5/0./2., 1.73205/0./1.73205/1., 1.73205/0./2.59808/-0.5, 1.73205/0./0.866025/-0.5, 1.73205/1./0.866025/0.5, 2.59808/-1.5/2.59808/-0.5, 2.59808/-1.5/1.73205/-2., 2.59808/-0.5/1.73205/-1., -0.866025/1.5/-0.866025/2.5, -0.866025/1.5/0./1., -0.866025/1.5/-1.73205/1., -0.866025/2.5/0./2., 0./1./0./2., 0./1./0.866025/0.5, 0./1./-0.866025/0.5, 0.866025/-1.5/0.866025/-0.5, 0.866025/-1.5/1.73205/-2., 0.866025/-1.5/0./-2., 0.866025/-0.5/0.866025/0.5, 0.866025/-0.5/1.73205/-1., 0.866025/-0.5/0./-1., 0.866025/0.5/0./0., 1.73205/-2./1.73205/-1., -1.73205/1./-0.866025/0.5, -1.73205/1./-2.59808/0.5, -0.866025/-0.5/-0.866025/0.5, -0.866025/-0.5/0./-1., -0.866025/-0.5/-1.73205/-1., -0.866025/0.5/0./0., -0.866025/0.5/-1.73205/0., 0./-2./0./-1., 0./-2./-0.866025/-2.5, 0./-1./0./0., 0./-1./-0.866025/-1.5, -2.59808/-0.5/-2.59808/0.5, -2.59808/-0.5/-1.73205/-1., -2.59808/-0.5/-3.4641/-1., -2.59808/0.5/-1.73205/0., -1.73205/-2./-1.73205/-1., -1.73205/-2./-0.866025/-2.5, -1.73205/-2./-2.59808/-2.5, -1.73205/-1./-1.73205/0., -1.73205/-1./-0.866025/-1.5, -1.73205/-1./-2.59808/-1.5, -0.866025/-2.5/-0.866025/-1.5, -3.4641/-2./-3.4641/-1., -3.4641/-2./-2.59808/-2.5, -3.4641/-1./-2.59808/-1.5, -2.59808/-2.5/-2.59808/-1.5}
			{
				\draw (\x,\y) -- (\z,\w);
			}
		\end{tikzpicture}
	\end{array}\longleftrightarrow \begin{array}{c}
		\begin{tikzpicture}[scale=0.6]
			\foreach \x/\y/\z/\w in {-3.17543/-0.5/-2.88675/-1., -3.17543/-0.5/-2.88675/0., -3.17543/0.5/-2.88675/0., -3.17543/0.5/-2.88675/1., -3.17543/1.5/-2.88675/1., -3.17543/1.5/-2.88675/2., -2.88675/-1./-2.3094/-1., -2.88675/2./-2.3094/2., -2.3094/-1./-2.02073/-0.5, -2.3094/0./-2.02073/-0.5, -2.3094/0./-2.02073/0.5, -2.3094/2./-2.02073/2.5, -2.02073/0.5/-1.44338/0.5, -2.02073/2.5/-1.44338/2.5, -1.44338/0.5/-1.1547/1., -1.44338/2.5/-1.1547/2., -1.1547/1./-0.57735/1., -1.1547/2./-0.57735/2., -0.57735/1./-0.288675/1.5, -0.57735/2./-0.288675/1.5, 0.288675/-1.5/0.57735/-2., 0.288675/-1.5/0.57735/-1., 0.288675/1.5/0.57735/1., 0.288675/1.5/0.57735/2., 0.57735/-2./1.1547/-2., 0.57735/-1./1.1547/-1., 0.57735/1./1.1547/1., 0.57735/2./1.1547/2., 1.1547/-2./1.44338/-1.5, 1.1547/-1./1.44338/-1.5, 1.1547/0./1.44338/-0.5, 1.1547/0./1.44338/0.5, 1.1547/1./1.44338/0.5, 1.1547/2./1.44338/2.5, 1.44338/-0.5/2.02073/-0.5, 1.44338/2.5/2.02073/2.5, 2.02073/-0.5/2.3094/0., 2.02073/0.5/2.3094/0., 2.02073/0.5/2.3094/1., 2.02073/1.5/2.3094/1., 2.02073/1.5/2.3094/2., 2.02073/2.5/2.3094/2.}
			{
				\draw[gray,thick] (\x,\y) -- (\z,\w);
			}
			\foreach \x/\y in {0.866025/1.5, 1.73205/0., 1.73205/1., 1.73205/2., -0.866025/1.5, 0.866025/-1.5, -1.73205/1., -1.73205/2., -2.59808/-0.5, -2.59808/0.5, -2.59808/1.5}
			{
				\node[gray] at (\x,\y) {\small\bf 0};
			}
			\foreach \x/\y/\z/\w in {-3.17543/-1.5/-2.88675/-2., -3.17543/-1.5/-2.88675/-1., -2.88675/-2./-2.3094/-2., -2.88675/-1./-2.3094/-1., -2.3094/-2./-2.02073/-1.5, -2.3094/-1./-2.02073/-0.5, -2.3094/0./-2.02073/-0.5, -2.3094/0./-2.02073/0.5, -2.02073/-1.5/-1.44338/-1.5, -2.02073/0.5/-1.44338/0.5, -1.44338/-1.5/-1.1547/-2., -1.44338/0.5/-1.1547/1., -1.1547/-2./-0.57735/-2., -1.1547/1./-0.57735/1., -0.57735/-2./-0.288675/-1.5, -0.57735/0./-0.288675/-0.5, -0.57735/0./-0.288675/0.5, -0.57735/1./-0.288675/1.5, -0.57735/2./-0.288675/1.5, -0.57735/2./-0.288675/2.5, -0.288675/-1.5/0.288675/-1.5, -0.288675/-0.5/0.288675/-0.5, -0.288675/0.5/0.288675/0.5, -0.288675/2.5/0.288675/2.5, 0.288675/-1.5/0.57735/-1., 0.288675/-0.5/0.57735/0., 0.288675/0.5/0.57735/0., 0.288675/1.5/0.57735/1., 0.288675/1.5/0.57735/2., 0.288675/2.5/0.57735/2., 0.57735/-1./1.1547/-1., 0.57735/1./1.1547/1., 1.1547/-1./1.44338/-1.5, 1.1547/0./1.44338/-0.5, 1.1547/0./1.44338/0.5, 1.1547/1./1.44338/0.5, 1.44338/-1.5/2.02073/-1.5, 1.44338/-0.5/2.02073/-0.5, 2.02073/-1.5/2.3094/-1., 2.02073/-0.5/2.3094/-1.}
			{
				\draw[\myblue,thick] (\x,\y) -- (\z,\w);
			}
			\foreach \x/\y in {0./1., 0./2., 0.866025/-0.5, 0.866025/0.5, 1.73205/-1., -0.866025/-0.5, -0.866025/0.5, 0./-1., -1.73205/-1., -1.73205/0., -0.866025/-1.5, -2.59808/-1.5}
			{
				\node[\myblue] at (\x,\y) {\small\bf 1};
			}
			\foreach \x/\y/\z/\w in {-0.57735/0./-0.288675/-0.5, -0.57735/0./-0.288675/0.5, -0.288675/-0.5/0.288675/-0.5, -0.288675/0.5/0.288675/0.5, 0.288675/-0.5/0.57735/0., 0.288675/0.5/0.57735/0.}
			{
				\draw[orange,thick] (\x,\y) -- (\z,\w);
			}
			\foreach \x/\y in {0./0.}
			{
				\node[orange] at (\x,\y) {\small\bf 2};
			}
		\end{tikzpicture}
	\end{array}\quad\begin{array}{c}
		\begin{tikzpicture}
			\draw[-stealth] (0,0) -- (0,1);
			\draw[-stealth] (0,0) -- (-0.866025, -0.5);
			\draw[-stealth] (0,0) -- (0.866025, -0.5);
			\node[above] at (0,1) {$\scriptstyle \myh_3$};
			\node[left] at (-0.866025, -0.5) {$\scriptstyle \myh_1$};
			\node[right] at (0.866025, -0.5) {$\scriptstyle \myh_2$};
		\end{tikzpicture}
	\end{array}
\end{equation}

And the charge function is well-known \cite{Prochazka:2015deb} in this case.
In this case we have only a product over singlets, i.e. over box positions $\vec x(\Box)$ in $\bpi$, with ${\bf prj}\,\vec x(\Box)=\sum^{3}_{i=1}x_i(\Box)\myh_i$:
\begin{equation}\label{ch_f_3D}
	\psi_{\bpi}(z)=\frac{1}{z}\prod\lm_{\Phi_1\in\bpi}\varphi_1(z-c(\Phi_1))\,,
\end{equation}
where the pairwise potential is the bonding factor from \cite{Li:2020rij}:
\begin{equation}
	\varphi_1(z)=\prod\lm_{i=1}^3\frac{z+\myh_i}{z-\myh_i}\,,
\end{equation}
which satisfies the properties
\begin{enumerate}\label{list:properties3D}
	\item $\psi_\bpi(z)$ is a meromorphic function of $z$.
	\item All the poles of $\psi_\bpi(z)$ are simple.
	\item All the poles of $\psi_\bpi(z)$ are in 1-to-1 correspondence with the set of projected coordinates  ${\bf prj} \left(\vec x(\Box)\right)$, with the boxes
	$\Box \in {\rm Add}(\bpi)\cup{\rm Rem}(\bpi)$ and $\vec{x}(\Box)$ their 3D coordinates.
\end{enumerate}
As was shown in \cite{Li:2020rij}, these three properties, in particular the last one, were crucial in bootstrapping the BPS algebra from its action on the 3D crystals (plane partitions in the case of $\mathbb{C}^3$.)

The potential function corresponding to the charge function is
\begin{equation}
	w(c)=\sum\lm_{i=1}^3\left(\mu(c-\myh_i)-\mu(c+\myh_i)\right)+\delta_{c,0}\,.
\end{equation}

Next we enumerate all the $\bf 3$ local edge weight configurations, modulo $S_3$ that permutes $\myh_k$:
\begin{equation}\label{D3_orbits}
	\begin{array}{c}
		\begin{tikzpicture}[scale=0.6]
			\foreach \x/\y/\z/\w in {0./0./-0.866025/-0.5, 0./0./0.866025/-0.5, 0./0./0./1., -0.866025/-0.5/-0.866025/0.5, -0.866025/-0.5/0./-1., 0.866025/-0.5/0.866025/0.5, 0.866025/-0.5/0./-1., 0./1./0.866025/0.5, 0./1./-0.866025/0.5}
			{
				\draw[gray, postaction={decorate},decoration={markings, 
					mark= at position 0.7 with {\arrow{stealth}}}] (\x,\y) -- (\z,\w);
			}
			\foreach \x/\y/\z/\w in {0.866025/0.5/0./0., -0.866025/0.5/0./0., 0./-1./0./0.}
			{
				\draw[ultra thick, \myblue, postaction={decorate},decoration={markings, 
					mark= at position 0.7 with {\arrow{stealth}}}] (\x,\y) -- (\z,\w);
			}
		\end{tikzpicture}\\
		|S_3\cdot{\rm Diag}|=1
	\end{array},\quad\begin{array}{c}
		\begin{tikzpicture}[scale=0.6]
			\foreach \x/\y/\z/\w in {0./0./-0.866025/-0.5, 0./0./0.866025/-0.5, 0./0./0./1., -0.866025/-0.5/-0.866025/0.5, 0.866025/-0.5/0.866025/0.5, 0./1./0.866025/0.5, 0./1./-0.866025/0.5, 0./-1./0./0.}
			{
				\draw[gray, postaction={decorate},decoration={markings, 
					mark= at position 0.7 with {\arrow{stealth}}}] (\x,\y) -- (\z,\w);
			}
			\foreach \x/\y/\z/\w in {-0.866025/-0.5/0./-1., 0.866025/0.5/0./0., 0.866025/-0.5/0./-1., -0.866025/0.5/0./0.}
			{
				\draw[ultra thick, \myblue, postaction={decorate},decoration={markings, 
					mark= at position 0.7 with {\arrow{stealth}}}] (\x,\y) -- (\z,\w);
			}
		\end{tikzpicture}\\
		|S_3\cdot{\rm Diag}|=3
	\end{array},\quad\begin{array}{c}
		\begin{tikzpicture}[scale=0.6]
			\foreach \x/\y/\z/\w in {0./0./0.866025/-0.5, 0./0./0./1., -0.866025/-0.5/-0.866025/0.5, -0.866025/-0.5/0./-1., 0.866025/-0.5/0.866025/0.5, -0.866025/0.5/0./0., 0./1./0.866025/0.5, 0./-1./0./0.}
			{
				\draw[gray, postaction={decorate},decoration={markings, 
					mark= at position 0.7 with {\arrow{stealth}}}] (\x,\y) -- (\z,\w);
			}
			\foreach \x/\y/\z/\w in {0./0./-0.866025/-0.5, 0.866025/0.5/0./0., 0.866025/-0.5/0./-1., 0./1./-0.866025/0.5}
			{
				\draw[ultra thick, \myblue, postaction={decorate},decoration={markings, 
					mark= at position 0.7 with {\arrow{stealth}}}] (\x,\y) -- (\z,\w);
			}
		\end{tikzpicture}\\
		|S_3\cdot{\rm Diag}|=3
	\end{array}\,,
\end{equation}
where below each local edge weight configuration we have also given the size of its orbit under $S_3$.

As in 2D, it is enough to consider the first, $S_3$ symmetric, configuration in \eqref{D3_orbits}.\footnote{
	Similarly to what happened in 2D, the other two configurations in \eqref{D3_orbits} do not give new local pictures but only the shifted versions of the local pictures from the  first configuration.}
For this case, the set of inequalities \eqref{ineq} contains $12$ inequalities (one for each edge) for the  height functions of the $7$ boxes involved.
Denoting them as $\mu_0,\mu_{1,2,3},\mu_{12,23,31}$, we have
\begin{equation}\label{eq:ineq3D}
	\begin{aligned}
		&\mu_0-1\leq \mu_{12}\leq \mu_1 \leq \mu_0 
		\,, \qquad 
		\mu_0-1\leq \mu_{12}\leq \mu_2 \leq \mu_0 \\
		&\mu_0-1\leq \mu_{23}\leq \mu_2 \leq \mu_0 
		\,, \qquad 
		\mu_0-1\leq \mu_{23}\leq \mu_3 \leq \mu_0 \\
		&\mu_0-1\leq \mu_{31}\leq \mu_3 \leq \mu_0 
		\,, \qquad 
		\mu_0-1\leq \mu_{31}\leq \mu_1 \leq \mu_0 \,.
	\end{aligned}
\end{equation}
The set of inequalities \eqref{eq:ineq3D} has {\bf 18} different solutions.
Instead of enumerating all of them here, we only present solutions modulo $S_3$, or equivalently, we only present solutions obeying
\begin{equation}
	\mu_{12}\leq \mu_{31}\leq \mu_{23}\leq   \mu_1  \leq\mu_2\leq \mu_3 \leq \mu_0 \,.
\end{equation}

\begingroup
\renewcommand{\arraystretch}{1.8}
\begin{figure}[ht!]
	\begin{center}
		$\begin{array}{c|c|c|c|c}
			& (1) & (2) & (3) & (4) \\
			\hline
			\mbox{Solutions}
			&
			\begin{array}{c}
				\begin{tikzpicture}[scale=0.6]
					\node at (0,1.2) {};
					\foreach \x/\y/\z/\w in {0./0./-0.866025/-0.5, 0./0./0.866025/-0.5, 0./0./0./1., -0.866025/-0.5/-0.866025/0.5, -0.866025/-0.5/0./-1., 0.866025/-0.5/0.866025/0.5, 0.866025/-0.5/0./-1., 0./1./0.866025/0.5, 0./1./-0.866025/0.5,0.866025/0.5/0./0., -0.866025/0.5/0./0., 0./-1./0./0.}
					{
						\draw (\x,\y) -- (\z,\w);
					}
					\foreach \x/\y in {-0.866025/-0.5, 0.866025/0.5, 0.866025/-0.5, -0.866025/0.5, 0./1., 0./-1.}
					{
						\draw[fill=white] (\x,\y) circle(0.2);
					}
					\foreach \x/\y in {0./0.}
					{
						\draw[fill=\myblue] (\x,\y) circle(0.2);
					}
				\end{tikzpicture}
			\end{array}
			&
			\begin{array}{c}
				\begin{tikzpicture}[scale=0.6]
					\node at (0,1.2) {};
					\foreach \x/\y/\z/\w in {0./0./-0.866025/-0.5, 0./0./0.866025/-0.5, 0./0./0./1., -0.866025/-0.5/-0.866025/0.5, -0.866025/-0.5/0./-1., 0.866025/-0.5/0.866025/0.5, 0.866025/-0.5/0./-1., 0./1./0.866025/0.5, 0./1./-0.866025/0.5,0.866025/0.5/0./0., -0.866025/0.5/0./0., 0./-1./0./0.}
					{
						\draw (\x,\y) -- (\z,\w);
					}
					\foreach \x/\y in {-0.866025/-0.5, 0.866025/0.5, 0.866025/-0.5, -0.866025/0.5, 0./-1.}
					{
						\draw[fill=white] (\x,\y) circle(0.2);
					}
					\foreach \x/\y in {0./0., 0./1.}
					{
						\draw[fill=\myblue] (\x,\y) circle(0.2);
					}
				\end{tikzpicture}
			\end{array}
			&
			\begin{array}{c}
				\begin{tikzpicture}[scale=0.6]
					\node at (0,1.2) {};
					\foreach \x/\y/\z/\w in {0./0./-0.866025/-0.5, 0./0./0.866025/-0.5, 0./0./0./1., -0.866025/-0.5/-0.866025/0.5, -0.866025/-0.5/0./-1., 0.866025/-0.5/0.866025/0.5, 0.866025/-0.5/0./-1., 0./1./0.866025/0.5, 0./1./-0.866025/0.5,0.866025/0.5/0./0., -0.866025/0.5/0./0., 0./-1./0./0.}
					{
						\draw (\x,\y) -- (\z,\w);
					}
					\foreach \x/\y in {-0.866025/-0.5, 0.866025/0.5, -0.866025/0.5, 0./-1.}
					{
						\draw[fill=white] (\x,\y) circle(0.2);
					}
					\foreach \x/\y in {0./0., 0.866025/-0.5, 0./1.}
					{
						\draw[fill=\myblue] (\x,\y) circle(0.2);
					}
				\end{tikzpicture}
			\end{array}
			&
			\begin{array}{c}
				\begin{tikzpicture}[scale=0.6]
					\node at (0,1.2) {};
					\foreach \x/\y/\z/\w in {0./0./-0.866025/-0.5, 0./0./0.866025/-0.5, 0./0./0./1., -0.866025/-0.5/-0.866025/0.5, -0.866025/-0.5/0./-1., 0.866025/-0.5/0.866025/0.5, 0.866025/-0.5/0./-1., 0./1./0.866025/0.5, 0./1./-0.866025/0.5,0.866025/0.5/0./0., -0.866025/0.5/0./0., 0./-1./0./0.}
					{
						\draw (\x,\y) -- (\z,\w);
					}
					\foreach \x/\y in {-0.866025/-0.5, -0.866025/0.5, 0./-1.}
					{
						\draw[fill=white] (\x,\y) circle(0.2);
					}
					\foreach \x/\y in {0./0., 0.866025/0.5, 0.866025/-0.5, 0./1.}
					{
						\draw[fill=\myblue] (\x,\y) circle(0.2);
					}
				\end{tikzpicture}
			\end{array}
			\\
			\hline
			\mbox{Tessellations}
			&
			\begin{array}{c}
				\begin{tikzpicture}[scale=0.3]
					\node at (0,2) {};
					\foreach \x/\y/\z/\w in {0./-2./-0.866025/-1.5, 0./-2./0.866025/-1.5, 0./-1./0./0., 0./-1./-0.866025/-1.5, 0./-1./-0.866025/-0.5, 0./-1./0.866025/-1.5, 0./-1./0.866025/-0.5, 0./0./-0.866025/0.5, 0./0./0.866025/0.5, 0./1./0./2., 0./1./-0.866025/0.5, 0./1./0.866025/0.5, 0./2./-0.866025/1.5, 0./2./0.866025/1.5, -1.73205/-1./-1.73205/0., -1.73205/-1./-0.866025/-1.5, -1.73205/-1./-0.866025/-0.5, -1.73205/0./-1.73205/1., -1.73205/0./-0.866025/0.5, -1.73205/1./-0.866025/1.5, -0.866025/-0.5/-0.866025/0.5, -0.866025/0.5/-0.866025/1.5, 0.866025/-1.5/1.73205/-1., 0.866025/-0.5/0.866025/0.5, 0.866025/-0.5/1.73205/-1., 0.866025/0.5/0.866025/1.5, 0.866025/0.5/1.73205/0., 0.866025/1.5/1.73205/1., 1.73205/-1./1.73205/0., 1.73205/0./1.73205/1.}
					{
						\draw (\x,\y) -- (\z,\w);
					}
					\tikzset{sty1/.style={fill=white!50!red}}
					\draw[sty1] (-0.866025,-0.5) -- (-0.866025,0.5) -- (0.,0.) -- (0.,-1.) -- cycle;
					\draw[sty1] (0.866025,-0.5) -- (0.866025,0.5) -- (0.,0.) -- (0.,-1.) -- cycle;
					\draw[sty1] (0.,1.) -- (0.866025,0.5) -- (0.,0.) -- (-0.866025,0.5) -- cycle;
				\end{tikzpicture}
			\end{array}
			&
			\begin{array}{c}
				\begin{tikzpicture}[scale=0.3]
					\node at (0,2) {};
					\foreach \x/\y/\z/\w in {0./-2./-0.866025/-1.5, 0./-2./0.866025/-1.5, 0./-1./0./0., 0./-1./-0.866025/-1.5, 0./-1./-0.866025/-0.5, 0./-1./0.866025/-1.5, 0./-1./0.866025/-0.5, 0./0./0./1., 0./0./-0.866025/0.5, 0./0./0.866025/0.5, 0./1./-0.866025/1.5, 0./1./0.866025/1.5, 0./2./-0.866025/1.5, 0./2./0.866025/1.5, -1.73205/-1./-1.73205/0., -1.73205/-1./-0.866025/-1.5, -1.73205/-1./-0.866025/-0.5, -1.73205/0./-1.73205/1., -1.73205/0./-0.866025/0.5, -1.73205/1./-0.866025/1.5, -0.866025/-0.5/-0.866025/0.5, -0.866025/0.5/-0.866025/1.5, 0.866025/-1.5/1.73205/-1., 0.866025/-0.5/0.866025/0.5, 0.866025/-0.5/1.73205/-1., 0.866025/0.5/0.866025/1.5, 0.866025/0.5/1.73205/0., 0.866025/1.5/1.73205/1., 1.73205/-1./1.73205/0., 1.73205/0./1.73205/1.}
					{
						\draw (\x,\y) -- (\z,\w);
					}
					\tikzset{sty1/.style={fill=white!50!red}}
					\draw[sty1] (-0.866025,-0.5) -- (-0.866025,0.5) -- (0.,0.) -- (0.,-1.) -- cycle;
					\draw[sty1] (-0.866025,0.5) -- (-0.866025,1.5) -- (0.,1.) -- (0.,0.) -- cycle;
					\draw[sty1] (0.866025,-0.5) -- (0.866025,0.5) -- (0.,0.) -- (0.,-1.) -- cycle;
					\draw[sty1] (0.866025,0.5) -- (0.866025,1.5) -- (0.,1.) -- (0.,0.) -- cycle;
				\end{tikzpicture}
			\end{array}
			&
			\begin{array}{c}
				\begin{tikzpicture}[scale=0.3]
					\node at (0,2) {};
					\foreach \x/\y/\z/\w in {0./-2./-0.866025/-1.5, 0./-2./0.866025/-1.5, 0./-1./0./0., 0./-1./-0.866025/-1.5, 0./-1./-0.866025/-0.5, 0./-1./0.866025/-1.5, 0./0./0./1., 0./0./-0.866025/0.5, 0./0./0.866025/-0.5, 0./0./0.866025/0.5, 0./1./-0.866025/1.5, 0./1./0.866025/1.5, 0./2./-0.866025/1.5, 0./2./0.866025/1.5, -1.73205/-1./-1.73205/0., -1.73205/-1./-0.866025/-1.5, -1.73205/-1./-0.866025/-0.5, -1.73205/0./-1.73205/1., -1.73205/0./-0.866025/0.5, -1.73205/1./-0.866025/1.5, -0.866025/-0.5/-0.866025/0.5, -0.866025/0.5/-0.866025/1.5, 0.866025/-1.5/0.866025/-0.5, 0.866025/-1.5/1.73205/-1., 0.866025/-0.5/1.73205/0., 0.866025/0.5/0.866025/1.5, 0.866025/0.5/1.73205/0., 0.866025/1.5/1.73205/1., 1.73205/-1./1.73205/0., 1.73205/0./1.73205/1.}
					{
						\draw (\x,\y) -- (\z,\w);
					}
					\tikzset{sty1/.style={fill=white!50!red}}
					\draw[sty1] (-0.866025,-0.5) -- (-0.866025,0.5) -- (0.,0.) -- (0.,-1.) -- cycle;
					\draw[sty1] (-0.866025,0.5) -- (-0.866025,1.5) -- (0.,1.) -- (0.,0.) -- cycle;
					\draw[sty1] (0.,-1.) -- (0.,0.) -- (0.866025,-0.5) -- (0.866025,-1.5) -- cycle;
					\draw[sty1] (0.866025,0.5) -- (0.866025,1.5) -- (0.,1.) -- (0.,0.) -- cycle;
					\draw[sty1] (0.866025,0.5) -- (1.73205,0.) -- (0.866025,-0.5) -- (0.,0.) -- cycle;
				\end{tikzpicture}
			\end{array}
			&
			\begin{array}{c}
				\begin{tikzpicture}[scale=0.3]
					\node at (0,2) {};
					\foreach \x/\y/\z/\w in {0./-2./-0.866025/-1.5, 0./-2./0.866025/-1.5, 0./-1./0./0., 0./-1./-0.866025/-1.5, 0./-1./-0.866025/-0.5, 0./-1./0.866025/-1.5, 0./0./0./1., 0./0./-0.866025/0.5, 0./0./0.866025/-0.5, 0./1./-0.866025/1.5, 0./1./0.866025/0.5, 0./1./0.866025/1.5, 0./2./-0.866025/1.5, 0./2./0.866025/1.5, -1.73205/-1./-1.73205/0., -1.73205/-1./-0.866025/-1.5, -1.73205/-1./-0.866025/-0.5, -1.73205/0./-1.73205/1., -1.73205/0./-0.866025/0.5, -1.73205/1./-0.866025/1.5, -0.866025/-0.5/-0.866025/0.5, -0.866025/0.5/-0.866025/1.5, 0.866025/-1.5/0.866025/-0.5, 0.866025/-1.5/1.73205/-1., 0.866025/-0.5/0.866025/0.5, 0.866025/-0.5/1.73205/0., 0.866025/0.5/1.73205/1., 0.866025/1.5/1.73205/1., 1.73205/-1./1.73205/0., 1.73205/0./1.73205/1.}
					{
						\draw (\x,\y) -- (\z,\w);
					}
					\tikzset{sty1/.style={fill=white!50!red}}
					\draw[sty1] (-0.866025,-0.5) -- (-0.866025,0.5) -- (0.,0.) -- (0.,-1.) -- cycle;
					\draw[sty1] (-0.866025,0.5) -- (-0.866025,1.5) -- (0.,1.) -- (0.,0.) -- cycle;
					\draw[sty1] (0.,-1.) -- (0.,0.) -- (0.866025,-0.5) -- (0.866025,-1.5) -- cycle;
					\draw[sty1] (0.,0.) -- (0.,1.) -- (0.866025,0.5) -- (0.866025,-0.5) -- cycle;
				\end{tikzpicture}
			\end{array}
			\\
			\hline
			w & 1 & 0 & -1 & 0 \\
			\hline
			& (5) & (6) & (7) & (8)\\
			\hline
			\mbox{Solutions}&
			\begin{array}{c}
				\begin{tikzpicture}[scale=0.6]
					\node at (0,1.2) {};
					\foreach \x/\y/\z/\w in {0./0./-0.866025/-0.5, 0./0./0.866025/-0.5, 0./0./0./1., -0.866025/-0.5/-0.866025/0.5, -0.866025/-0.5/0./-1., 0.866025/-0.5/0.866025/0.5, 0.866025/-0.5/0./-1., 0./1./0.866025/0.5, 0./1./-0.866025/0.5,0.866025/0.5/0./0., -0.866025/0.5/0./0., 0./-1./0./0.}
					{
						\draw (\x,\y) -- (\z,\w);
					}
					\foreach \x/\y in {0.866025/0.5, -0.866025/0.5, 0./-1.}
					{
						\draw[fill=white] (\x,\y) circle(0.2);
					}
					\foreach \x/\y in {0./0., -0.866025/-0.5, 0.866025/-0.5, 0./1.}
					{
						\draw[fill=\myblue] (\x,\y) circle(0.2);
					}
				\end{tikzpicture}
			\end{array}
			&
			\begin{array}{c}
				\begin{tikzpicture}[scale=0.6]
					\node at (0,1.2) {};
					\foreach \x/\y/\z/\w in {0./0./-0.866025/-0.5, 0./0./0.866025/-0.5, 0./0./0./1., -0.866025/-0.5/-0.866025/0.5, -0.866025/-0.5/0./-1., 0.866025/-0.5/0.866025/0.5, 0.866025/-0.5/0./-1., 0./1./0.866025/0.5, 0./1./-0.866025/0.5,0.866025/0.5/0./0., -0.866025/0.5/0./0., 0./-1./0./0.}
					{
						\draw (\x,\y) -- (\z,\w);
					}
					\foreach \x/\y in {-0.866025/0.5, 0./-1.}
					{
						\draw[fill=white] (\x,\y) circle(0.2);
					}
					\foreach \x/\y in {0./0., -0.866025/-0.5, 0.866025/-0.5, 0./1., 0.866025/0.5}
					{
						\draw[fill=\myblue] (\x,\y) circle(0.2);
					}
				\end{tikzpicture}
			\end{array}
			&
			\begin{array}{c}
				\begin{tikzpicture}[scale=0.6]
					\node at (0,1.2) {};
					\foreach \x/\y/\z/\w in {0./0./-0.866025/-0.5, 0./0./0.866025/-0.5, 0./0./0./1., -0.866025/-0.5/-0.866025/0.5, -0.866025/-0.5/0./-1., 0.866025/-0.5/0.866025/0.5, 0.866025/-0.5/0./-1., 0./1./0.866025/0.5, 0./1./-0.866025/0.5,0.866025/0.5/0./0., -0.866025/0.5/0./0., 0./-1./0./0.}
					{
						\draw (\x,\y) -- (\z,\w);
					}
					\foreach \x/\y in {0./-1.}
					{
						\draw[fill=white] (\x,\y) circle(0.2);
					}
					\foreach \x/\y in {0./0., -0.866025/-0.5, 0.866025/-0.5, -0.866025/0.5, 0./1., 0.866025/0.5}
					{
						\draw[fill=\myblue] (\x,\y) circle(0.2);
					}
				\end{tikzpicture}
			\end{array}
			&
			\begin{array}{c}
				\begin{tikzpicture}[scale=0.6]
					\node at (0,1.2) {};
					\foreach \x/\y/\z/\w in {0./0./-0.866025/-0.5, 0./0./0.866025/-0.5, 0./0./0./1., -0.866025/-0.5/-0.866025/0.5, -0.866025/-0.5/0./-1., 0.866025/-0.5/0.866025/0.5, 0.866025/-0.5/0./-1., 0./1./0.866025/0.5, 0./1./-0.866025/0.5,0.866025/0.5/0./0., -0.866025/0.5/0./0., 0./-1./0./0.}
					{
						\draw (\x,\y) -- (\z,\w);
					}
					\foreach \x/\y in {0./0., -0.866025/-0.5, 0.866025/0.5, 0.866025/-0.5, -0.866025/0.5, 0./1., 0./-1.}
					{
						\draw[fill=\myblue] (\x,\y) circle(0.2);
					}
				\end{tikzpicture}
			\end{array}\\
			\hline
			\mbox{Tessellations}&
			\begin{array}{c}
				\begin{tikzpicture}[scale=0.3]
					\node at (0,2) {};
					\foreach \x/\y/\z/\w in {0./-2./-0.866025/-1.5, 0./-2./0.866025/-1.5, 0./-1./0./0., 0./-1./-0.866025/-1.5, 0./-1./0.866025/-1.5, 0./0./0./1., 0./0./-0.866025/-0.5, 0./0./-0.866025/0.5, 0./0./0.866025/-0.5, 0./0./0.866025/0.5, 0./1./-0.866025/1.5, 0./1./0.866025/1.5, 0./2./-0.866025/1.5, 0./2./0.866025/1.5, -1.73205/-1./-1.73205/0., -1.73205/-1./-0.866025/-1.5, -1.73205/0./-1.73205/1., -1.73205/0./-0.866025/-0.5, -1.73205/0./-0.866025/0.5, -1.73205/1./-0.866025/1.5, -0.866025/-1.5/-0.866025/-0.5, -0.866025/0.5/-0.866025/1.5, 0.866025/-1.5/0.866025/-0.5, 0.866025/-1.5/1.73205/-1., 0.866025/-0.5/1.73205/0., 0.866025/0.5/0.866025/1.5, 0.866025/0.5/1.73205/0., 0.866025/1.5/1.73205/1., 1.73205/-1./1.73205/0., 1.73205/0./1.73205/1.}
					{
						\draw (\x,\y) -- (\z,\w);
					}
					\tikzset{sty1/.style={fill=white!50!red}}
					\draw[sty1] (-0.866025,0.5) -- (-0.866025,1.5) -- (0.,1.) -- (0.,0.) -- cycle;
					\draw[sty1] (0.,-1.) -- (0.,0.) -- (0.866025,-0.5) -- (0.866025,-1.5) -- cycle;
					\draw[sty1] (0.866025,0.5) -- (0.866025,1.5) -- (0.,1.) -- (0.,0.) -- cycle;
					\draw[sty1] (0.,-1.) -- (0.,0.) -- (-0.866025,-0.5) -- (-0.866025,-1.5) -- cycle;
					\draw[sty1] (0.866025,0.5) -- (1.73205,0.) -- (0.866025,-0.5) -- (0.,0.) -- cycle;
					\draw[sty1] (-0.866025,0.5) -- (0.,0.) -- (-0.866025,-0.5) -- (-1.73205,0.) -- cycle;
				\end{tikzpicture}
			\end{array}
			&
			\begin{array}{c}
				\begin{tikzpicture}[scale=0.3]
					\node at (0,1.2) {};
					\begin{scope}[rotate=120]
						\foreach \x/\y/\z/\w in {0./-2./0./-1., 0./-2./-0.866025/-1.5, 0./-2./0.866025/-1.5, 0./-1./-0.866025/-0.5, 0./-1./0.866025/-0.5, 0./0./0./1., 0./0./-0.866025/-0.5, 0./0./-0.866025/0.5, 0./0./0.866025/-0.5, 0./0./0.866025/0.5, 0./1./-0.866025/1.5, 0./1./0.866025/1.5, 0./2./-0.866025/1.5, 0./2./0.866025/1.5, -1.73205/-1./-1.73205/0., -1.73205/-1./-0.866025/-1.5, -1.73205/0./-1.73205/1., -1.73205/0./-0.866025/-0.5, -1.73205/0./-0.866025/0.5, -1.73205/1./-0.866025/1.5, -0.866025/-1.5/-0.866025/-0.5, -0.866025/0.5/-0.866025/1.5, 0.866025/-1.5/0.866025/-0.5, 0.866025/-1.5/1.73205/-1., 0.866025/-0.5/1.73205/0., 0.866025/0.5/0.866025/1.5, 0.866025/0.5/1.73205/0., 0.866025/1.5/1.73205/1., 1.73205/-1./1.73205/0., 1.73205/0./1.73205/1.}
						{
							\draw (\x,\y) -- (\z,\w);
						}
						\tikzset{sty1/.style={fill=white!50!red}}
						\draw[sty1] (-0.866025,0.5) -- (-0.866025,1.5) -- (0.,1.) -- (0.,0.) -- cycle;
						\draw[sty1] (0.866025,0.5) -- (0.866025,1.5) -- (0.,1.) -- (0.,0.) -- cycle;
						\draw[sty1] (0.866025,0.5) -- (1.73205,0.) -- (0.866025,-0.5) -- (0.,0.) -- cycle;
						\draw[sty1] (-0.866025,0.5) -- (0.,0.) -- (-0.866025,-0.5) -- (-1.73205,0.) -- cycle;
						\draw[sty1] (0.,0.) -- (0.866025,-0.5) -- (0.,-1.) -- (-0.866025,-0.5) -- cycle;
					\end{scope}
				\end{tikzpicture}
			\end{array}
			&
			\begin{array}{c}
				\begin{tikzpicture}[scale=0.3]
					\node at (0,1.2) {};
					\begin{scope}[rotate=-120]
						\foreach \x/\y/\z/\w in {0./-2./0./-1., 0./-2./-0.866025/-1.5, 0./-2./0.866025/-1.5, 0./-1./-0.866025/-0.5, 0./-1./0.866025/-0.5, 0./0./0./1., 0./0./-0.866025/-0.5, 0./0./0.866025/-0.5, 0./0./0.866025/0.5, 0./1./-0.866025/0.5, 0./1./-0.866025/1.5, 0./1./0.866025/1.5, 0./2./-0.866025/1.5, 0./2./0.866025/1.5, -1.73205/-1./-1.73205/0., -1.73205/-1./-0.866025/-1.5, -1.73205/0./-1.73205/1., -1.73205/0./-0.866025/-0.5, -1.73205/1./-0.866025/0.5, -1.73205/1./-0.866025/1.5, -0.866025/-1.5/-0.866025/-0.5, -0.866025/-0.5/-0.866025/0.5, 0.866025/-1.5/0.866025/-0.5, 0.866025/-1.5/1.73205/-1., 0.866025/-0.5/1.73205/0., 0.866025/0.5/0.866025/1.5, 0.866025/0.5/1.73205/0., 0.866025/1.5/1.73205/1., 1.73205/-1./1.73205/0., 1.73205/0./1.73205/1.}
						{
							\draw (\x,\y) -- (\z,\w);
						}
						\tikzset{sty1/.style={fill=white!50!red}}
						\draw[sty1] (0.866025,0.5) -- (0.866025,1.5) -- (0.,1.) -- (0.,0.) -- cycle;
						\draw[sty1] (0.,0.) -- (0.,1.) -- (-0.866025,0.5) -- (-0.866025,-0.5) -- cycle;
						\draw[sty1] (0.866025,0.5) -- (1.73205,0.) -- (0.866025,-0.5) -- (0.,0.) -- cycle;
						\draw[sty1] (0.,0.) -- (0.866025,-0.5) -- (0.,-1.) -- (-0.866025,-0.5) -- cycle;
					\end{scope}
				\end{tikzpicture}
			\end{array}
			&
			\begin{array}{c}
				\begin{tikzpicture}[scale=0.3]
					\node at (0,1.2) {};
					\begin{scope}[rotate=180]
						\foreach \x/\y/\z/\w in {0./-2./-0.866025/-1.5, 0./-2./0.866025/-1.5, 0./-1./0./0., 0./-1./-0.866025/-1.5, 0./-1./-0.866025/-0.5, 0./-1./0.866025/-1.5, 0./-1./0.866025/-0.5, 0./0./-0.866025/0.5, 0./0./0.866025/0.5, 0./1./0./2., 0./1./-0.866025/0.5, 0./1./0.866025/0.5, 0./2./-0.866025/1.5, 0./2./0.866025/1.5, -1.73205/-1./-1.73205/0., -1.73205/-1./-0.866025/-1.5, -1.73205/-1./-0.866025/-0.5, -1.73205/0./-1.73205/1., -1.73205/0./-0.866025/0.5, -1.73205/1./-0.866025/1.5, -0.866025/-0.5/-0.866025/0.5, -0.866025/0.5/-0.866025/1.5, 0.866025/-1.5/1.73205/-1., 0.866025/-0.5/0.866025/0.5, 0.866025/-0.5/1.73205/-1., 0.866025/0.5/0.866025/1.5, 0.866025/0.5/1.73205/0., 0.866025/1.5/1.73205/1., 1.73205/-1./1.73205/0., 1.73205/0./1.73205/1.}
						{
							\draw (\x,\y) -- (\z,\w);
						}
						\tikzset{sty1/.style={fill=white!50!red}}
						\draw[sty1] (-0.866025,-0.5) -- (-0.866025,0.5) -- (0.,0.) -- (0.,-1.) -- cycle;
						\draw[sty1] (0.866025,-0.5) -- (0.866025,0.5) -- (0.,0.) -- (0.,-1.) -- cycle;
						\draw[sty1] (0.,1.) -- (0.866025,0.5) -- (0.,0.) -- (-0.866025,0.5) -- cycle;
					\end{scope}
				\end{tikzpicture}
			\end{array}\\
			\hline
			w &-2 & -1 & 0 & 1 \\
		\end{array}$
		\caption{Local pictures for plane (3D) partitions}\label{loc_pic_plane}
	\end{center}
\end{figure}
\endgroup
We will call these configurations local pictures, and there are  {\bf 8} of them, shown in Fig.\ \ref{loc_pic_plane}.
We note that among the local pictures depicted in Fig.\ \ref{loc_pic_plane}, there are three pairs of plane partitions, (1,8), (2,7), and (3,6), that complement each other in the following sense.
The two plane partitions in a pair can combine into a $n\times n\times n$ cubic partition after one of them is reflected in all axes, namely, the hills of one complement the pits of the other.
(Using the same terminology as for ordinary partitions, we call two plane partitions in such a pair being transposed to each other: $\bpi\leftrightarrow\bpi^T$.)
In addition, the plane partitions (4) and (5) are self-transposed.
Therefore, if we further factor out the {\bf 8} plane partitions in Fig.\ \ref{loc_pic_plane} by the transposition symmetry, we finally arrive at {\bf 5} independent local pictures, coinciding with the classification given in \cite[Sec.~4.5]{Prochazka:2015deb}.

\section{\texorpdfstring{$\CN=(0,2)$}{N=(0,2)} 2d effective theory of D8-D0 branes} \label{s:gauge}
The effective theory of a D8-D0 brane system wrapping CY${}_4$ is given by an $\CN=(0,2)$ 2d effective theory \cite{Closset:2017yte}.
In this Appendix we summarize the basic features of this $\CN=(0,2)$ 2d theory, following \cite{Gadde:2013lxa}.

There are three types of multiplets:
\begin{itemize}
	\item[\textcolor{\myblue}{\textbullet}] Chiral multiplet:
	\begin{equation}
		\Phi=\phi+\sqrt{2}\theta^+\chi_+-\I\,\theta^+\bar\theta^+\,\p_+\phi\,.
	\end{equation}
	\item[\textcolor{\myblue}{\textbullet}] Fermi multiplet:
	\begin{equation}
		\Psi=\psi_--\sqrt{2}\theta^+G-\I\,\theta^+\bar\theta^+\,\p_+\psi_--\sqrt{2}\bar\theta^+E\,,
	\end{equation}
	where $E$ is a holomorphic function of the chiral fields $\phi$.
	\item[\textcolor{\myblue}{\textbullet}] Vector multiplet:
	\begin{equation}
		V=A_--2\I\,\theta^+\lambda_--2\I\,\bar\theta^+\bar\lambda_-+2\theta^+\bar\theta^+{\bf D}\,,
	\end{equation}
	where $A_\pm=\frac{1}{2}\left(A_0\pm A_1\right)$. 
\end{itemize}

The Lagrangian has five terms:
\begin{equation}
	\CL=\CL_{\rm gauge}+\CL_{\Phi}+\CL_{\Psi}+\CL_{\rm FI}+\CL_J\,,
\end{equation}
Here, the gauge Lagrangian is:
\begin{subequations}
	\begin{equation}
		\CL_{\rm gauge}=\frac{1}{8e^2}\int d^2\theta\; \bar\Lambda\Lambda=\frac{1}{e^2}\left(\frac{1}{2}F_{01}^2+\I\,\bar\lambda_-\p_+\lambda_-+\frac{1}{2}{\bf D}^2\right)\,;
	\end{equation}
	the chiral Lagrangian is:
	\begin{equation}
		\begin{split}
			\CL_{\Phi}&=-\frac{\I}{2}\int d^2\theta\;\bar\Phi_i\nabla_-\Phi^i=\\
			&=-\left|D_{\mu}\phi_i\right|^2+2\I\,\bar\chi_{+i}D_-\chi_+^i-\I q_i\, \bar\phi_i\lambda_-\chi_+^i+\I q_i\,\phi^i\bar\chi_{+i}\bar\lambda_-+q_i\,|\phi_i|^2{\bf D}\,;
		\end{split}
	\end{equation}
	the Fermi Lagrangian is:
	\begin{equation}
		\begin{split}
			\CL_{\Psi}&=-\frac{1}{2}\int d^2\theta\; \bar\Psi_a\Psi^a=\\
			&=2\I\,\bar\psi_{-a}D_+\psi_-^a+|G_a|^2-\left|E_a(\phi)\right|^2-\bar\psi_{-a}\frac{\p E^a}{\p \phi_i}\chi_{+i}-\bar\chi_{+i}\frac{\p \bar E_a}{\p\bar\phi_i}\psi_-^a\,;
		\end{split}
	\end{equation}
	and the Faye-Illiopolous term is:
	\begin{equation}
		\CL_{\rm FI}=\frac{t}{4}\int d\theta^+\;\Lambda\Big|_{\bar\theta^+=0}+{\rm c.c.}=-r{\bf D}+\frac{\theta}{2\pi}F_{01}\,.
	\end{equation}
	Finally, the $J$-term, which replaces the canonical F-term, is
	\begin{equation}
		\begin{split}
			&\CL_{J}=\int d\theta^+\;W_J(\Psi,\Phi)\Big|_{\bar\theta^+=0}+{\rm c.c.}=\int d\theta^+\;\Psi_aJ^a(\Phi)\Big|_{\bar\theta^+=0}+{\rm c.c.}\\
			=&G_aJ^a(\phi)+\frac{1}{2}\psi_{-a}\chi_{+i}\frac{\p J^a}{\p\phi^i}+{\rm c.c.}\,.
		\end{split}
	\end{equation}
\end{subequations}

The action is invariant (modulo boundary terms) under the following SUSY transformations:
\begin{equation}\renewcommand{\arraystretch}{1.5}
	\begin{array}{ll}
		\delta A_0=\dfrac{\I}{2}\bar\epsilon_-\lambda_-+\dfrac{\I}{2}\epsilon_-\bar\lambda_-\,,& \delta\phi^i=-\epsilon_-\chi_+^i\,,\\
		\delta A_1=-\dfrac{\I}{2}\bar\epsilon_-\lambda_--\dfrac{\I}{2}\epsilon_-\bar\lambda_-\,, & \delta \chi_+^i=2\I\,\bar\epsilon_-D_+\phi^i\,,\\
		\delta\lambda_-=\I\,\epsilon_-\left({\bf D}-\I F_{01}\right)\,, & \delta\psi_-^a=\epsilon_-G^a+\bar\epsilon_-E^a\,,\\
		\delta{\bf D}=-\bar\epsilon_-D_+\lambda_-+\epsilon_-D_+\bar\lambda_-\,, & \delta G^a=-2\I\,\bar\epsilon_-D_+\psi_-^a+\bar\epsilon_-\dfrac{\p E^a}{\p\phi^i}\chi_+^i\,.\\
	\end{array}
\end{equation}
The complete SUSY invariance imposes the following constraint on the $E$- and $J$-fields: 
\begin{equation}
	\sum\lm_aE_a(\Phi_i)J^a(\Phi_i)=0\,.
\end{equation}
And finally, there is an additional anomaly cancellation constraint:
\begin{equation}
	\Tr\;\gamma^3GG=\sum\lm_{i:\,{\rm chiral}}q_i^2-\sum\lm_{a:\,{\rm Fermi}}q_a^2=0\,.
\end{equation}

\bibliographystyle{utphys}
\bibliography{biblio}

\providecommand{\href}[2]{#2}\begingroup\raggedright\begin{thebibliography}{10}

\bibitem{Harvey:1995fq}
J.~A. Harvey and G.~W. Moore, ``{Algebras, BPS states, and strings},''
  \href{http://dx.doi.org/10.1016/0550-3213(95)00605-2}{{\em Nucl. Phys. B}
  {\bfseries 463} (1996) 315--368},
  \href{http://arxiv.org/abs/hep-th/9510182}{{\ttfamily arXiv:hep-th/9510182}}.

\bibitem{Harvey:1996gc}
J.~A. Harvey and G.~W. Moore, ``{On the algebras of BPS states},''
  \href{http://dx.doi.org/10.1007/s002200050461}{{\em Commun. Math. Phys.}
  {\bfseries 197} (1998) 489--519},
  \href{http://arxiv.org/abs/hep-th/9609017}{{\ttfamily arXiv:hep-th/9609017}}.

\bibitem{Kontsevich:2010px}
M.~Kontsevich and Y.~Soibelman, ``{Cohomological Hall algebra, exponential
  Hodge structures and motivic Donaldson-Thomas invariants},''
  \href{http://dx.doi.org/10.4310/CNTP.2011.v5.n2.a1}{{\em Commun. Num. Theor.
  Phys.} {\bfseries 5} (2011) 231--352},
  \href{http://arxiv.org/abs/1006.2706}{{\ttfamily arXiv:1006.2706 [math.AG]}}.

\bibitem{Ooguri:2009ijd}
H.~Ooguri and M.~Yamazaki, ``{Crystal Melting and Toric Calabi-Yau
  Manifolds},'' \href{http://dx.doi.org/10.1007/s00220-009-0836-y}{{\em Commun.
  Math. Phys.} {\bfseries 292} (2009) 179--199},
  \href{http://arxiv.org/abs/0811.2801}{{\ttfamily arXiv:0811.2801 [hep-th]}}.

\bibitem{Ooguri:2010yk}
H.~Ooguri, P.~Sulkowski, and M.~Yamazaki, ``{Wall Crossing As Seen By Matrix
  Models},'' \href{http://dx.doi.org/10.1007/s00220-011-1330-x}{{\em Commun.
  Math. Phys.} {\bfseries 307} (2011) 429--462},
  \href{http://arxiv.org/abs/1005.1293}{{\ttfamily arXiv:1005.1293 [hep-th]}}.

\bibitem{Li:2020rij}
W.~Li and M.~Yamazaki, ``{Quiver Yangian from Crystal Melting},''
  \href{http://dx.doi.org/10.1007/JHEP11(2020)035}{{\em JHEP} {\bfseries 11}
  (2020) 035}, \href{http://arxiv.org/abs/2003.08909}{{\ttfamily
  arXiv:2003.08909 [hep-th]}}.

\bibitem{Rapcak:2018nsl}
M.~Rapcak, Y.~Soibelman, Y.~Yang, and G.~Zhao, ``{Cohomological Hall algebras,
  vertex algebras and instantons},''
  \href{http://dx.doi.org/10.1007/s00220-019-03575-5}{{\em Commun. Math. Phys.}
  {\bfseries 376} no.~3, (2019) 1803--1873},
  \href{http://arxiv.org/abs/1810.10402}{{\ttfamily arXiv:1810.10402
  [math.QA]}}.

\bibitem{Rapcak:2021hdh}
M.~Rapcak, ``{Branes, Quivers and BPS Algebras},''
  \href{http://arxiv.org/abs/2112.13878}{{\ttfamily arXiv:2112.13878
  [hep-th]}}.

\bibitem{Nekrasov:2017cih}
N.~Nekrasov, ``{Magnificent four},''
  \href{http://dx.doi.org/10.4310/ATMP.2020.v24.n5.a4}{{\em Adv. Theor. Math.
  Phys.} {\bfseries 24} no.~5, (2020) 1171--1202},
  \href{http://arxiv.org/abs/1712.08128}{{\ttfamily arXiv:1712.08128
  [hep-th]}}.

\bibitem{Douglas:1995bn}
M.~R. Douglas, ``{Branes within branes},'' {\em NATO Sci. Ser. C} {\bfseries
  520} (1999) 267--275, \href{http://arxiv.org/abs/hep-th/9512077}{{\ttfamily
  arXiv:hep-th/9512077}}.

\bibitem{Douglas:1996sw}
M.~R. Douglas and G.~W. Moore, ``{D-branes, quivers, and ALE instantons},''
  \href{http://arxiv.org/abs/hep-th/9603167}{{\ttfamily arXiv:hep-th/9603167}}.

\bibitem{Awata:2005fa}
H.~Awata and H.~Kanno, ``{Instanton counting, Macdonald functions and the
  moduli space of D-branes},''
  \href{http://dx.doi.org/10.1088/1126-6708/2005/05/039}{{\em JHEP} {\bfseries
  05} (2005) 039}, \href{http://arxiv.org/abs/hep-th/0502061}{{\ttfamily
  arXiv:hep-th/0502061}}.

\bibitem{Kanno:2020ybd}
H.~Kanno, ``{Quiver matrix model of ADHM type and BPS state counting in diverse
  dimensions},'' \href{http://dx.doi.org/10.1093/ptep/ptaa079}{{\em PTEP}
  {\bfseries 2020} no.~11, (2020) 11B104},
  \href{http://arxiv.org/abs/2004.05760}{{\ttfamily arXiv:2004.05760
  [hep-th]}}.

\bibitem{Nekrasov:2018xsb}
N.~Nekrasov and N.~Piazzalunga, ``{Magnificent Four with Colors},''
  \href{http://dx.doi.org/10.1007/s00220-019-03426-3}{{\em Commun. Math. Phys.}
  {\bfseries 372} no.~2, (2019) 573--597},
  \href{http://arxiv.org/abs/1808.05206}{{\ttfamily arXiv:1808.05206
  [hep-th]}}.

\bibitem{Bonelli:2020gku}
G.~Bonelli, N.~Fasola, A.~Tanzini, and Y.~Zenkevich, ``{ADHM in 8d, coloured
  solid partitions and Donaldson-Thomas invariants on orbifolds},''
  \href{http://dx.doi.org/10.1016/j.geomphys.2023.104910}{{\em J. Geom. Phys.}
  {\bfseries 191} (2023) 104910},
  \href{http://arxiv.org/abs/2011.02366}{{\ttfamily arXiv:2011.02366
  [hep-th]}}.

\bibitem{Szabo:2023ixw}
R.~J. Szabo and M.~Tirelli, ``{Instanton Counting and Donaldson-Thomas Theory
  on Toric Calabi-Yau Four-Orbifolds},''
  \href{http://arxiv.org/abs/2301.13069}{{\ttfamily arXiv:2301.13069
  [hep-th]}}.

\bibitem{Kimura:2022zsm}
T.~Kimura, ``{Double Quiver Gauge Theory and BPS/CFT Correspondence},''
  \href{http://dx.doi.org/10.3842/SIGMA.2023.039}{{\em SIGMA} {\bfseries 19}
  (2023) 039}, \href{http://arxiv.org/abs/2212.03870}{{\ttfamily
  arXiv:2212.03870 [hep-th]}}.

\bibitem{Piazzalunga:2023qik}
N.~Piazzalunga, ``{The one-legged K-theoretic vertex of fourfolds from 3d gauge
  theory},'' \href{http://arxiv.org/abs/2306.12405}{{\ttfamily arXiv:2306.12405
  [hep-th]}}.

\bibitem{Nekrasov:2023nai}
N.~Nekrasov and N.~Piazzalung, ``{Global magni$4$icence, or: 4G Networks},''
  \href{http://arxiv.org/abs/2306.12995}{{\ttfamily arXiv:2306.12995
  [hep-th]}}.

\bibitem{Kimura:2023bxy}
T.~Kimura and G.~Noshita, ``{Gauge origami and quiver W-algebras},''
  \href{http://arxiv.org/abs/2310.08545}{{\ttfamily arXiv:2310.08545
  [hep-th]}}.

\bibitem{Cao:2017swr}
Y.~Cao and M.~Kool, ``{Zero-dimensional Donaldson\textendash{}Thomas invariants
  of Calabi\textendash{}Yau 4-folds},''
  \href{http://dx.doi.org/10.1016/j.aim.2018.09.011}{{\em Adv. Math.}
  {\bfseries 338} (2018) 601--648},
  \href{http://arxiv.org/abs/1712.07347}{{\ttfamily arXiv:1712.07347
  [math.AG]}}.

\bibitem{Cao:2018rbp}
Y.~Cao and M.~Kool, ``{Counting zero-dimensional subschemes in higher
  dimensions},'' \href{http://dx.doi.org/10.1016/j.geomphys.2018.11.004}{{\em
  J. Geom. Phys.} {\bfseries 136} (2019) 119--137},
  \href{http://arxiv.org/abs/1805.04746}{{\ttfamily arXiv:1805.04746
  [math.AG]}}.

\bibitem{Cao:2019tnw}
Y.~Cao and M.~Kool, ``{Curve counting and DT/PT correspondence for Calabi-Yau
  4-folds},'' \href{http://dx.doi.org/10.1016/j.aim.2020.107371}{{\em Adv.
  Math.} {\bfseries 375} (2020) 107371},
  \href{http://arxiv.org/abs/1903.12171}{{\ttfamily arXiv:1903.12171
  [math.AG]}}.

\bibitem{Cao:2023lon}
Y.~Cao and G.~Zhao, ``{Quasimaps to quivers with potentials},''
  \href{http://arxiv.org/abs/2306.01302}{{\ttfamily arXiv:2306.01302
  [math.AG]}}.

\bibitem{Franco:2015tna}
S.~Franco, D.~Ghim, S.~Lee, R.-K. Seong, and D.~Yokoyama, ``{2d (0,2) Quiver
  Gauge Theories and D-Branes},''
  \href{http://dx.doi.org/10.1007/JHEP09(2015)072}{{\em JHEP} {\bfseries 09}
  (2015) 072}, \href{http://arxiv.org/abs/1506.03818}{{\ttfamily
  arXiv:1506.03818 [hep-th]}}.

\bibitem{Franco:2015tya}
S.~Franco, S.~Lee, and R.-K. Seong, ``{Brane Brick Models, Toric Calabi-Yau
  4-Folds and 2d (0,2) Quivers},''
  \href{http://dx.doi.org/10.1007/JHEP02(2016)047}{{\em JHEP} {\bfseries 02}
  (2016) 047}, \href{http://arxiv.org/abs/1510.01744}{{\ttfamily
  arXiv:1510.01744 [hep-th]}}.

\bibitem{Franco:2019bmx}
S.~Franco and A.~Hasan, ``{Graded Quivers, Generalized Dimer Models and Toric
  Geometry},'' \href{http://dx.doi.org/10.1007/JHEP11(2019)104}{{\em JHEP}
  {\bfseries 11} (2019) 104}, \href{http://arxiv.org/abs/1904.07954}{{\ttfamily
  arXiv:1904.07954 [hep-th]}}.

\bibitem{Franco:2021elb}
S.~Franco and X.~Yu, ``{BFT$_{2}$: a general class of 2d$ \mathcal{N} $ = (0,
  2) theories, 3-manifolds and toric geometry},''
  \href{http://dx.doi.org/10.1007/JHEP08(2022)277}{{\em JHEP} {\bfseries 08}
  (2022) 277}, \href{http://arxiv.org/abs/2107.00667}{{\ttfamily
  arXiv:2107.00667 [hep-th]}}.

\bibitem{Szabo:2022zyn}
R.~J. Szabo and M.~Tirelli, ``{Noncommutative Instantons in Diverse
  Dimensions},'' \href{http://arxiv.org/abs/2207.12862}{{\ttfamily
  arXiv:2207.12862 [hep-th]}}.

\bibitem{Szendroi:2007nu}
B.~Szendroi, ``{Non-commutative Donaldson\textendash{}Thomas invariants and the
  conifold},'' \href{http://dx.doi.org/10.2140/gt.2008.12.1171}{{\em Geom.
  Topol.} {\bfseries 12} no.~2, (2008) 1171--1202},
  \href{http://arxiv.org/abs/0705.3419}{{\ttfamily arXiv:0705.3419 [math.AG]}}.

\bibitem{King}
A.~D. King, ``{Moduli of representations of finite dimensional algebras},''
  {\em The Quarterly Journal of Mathematics} {\bfseries 45} no.~4, (12, 1994)
  515--530.

\bibitem{Okounkov:2003sp}
A.~Okounkov, N.~Reshetikhin, and C.~Vafa, ``{Quantum Calabi-Yau and classical
  crystals},'' \href{http://dx.doi.org/10.1007/0-8176-4467-9_16}{{\em Prog.
  Math.} {\bfseries 244} (2006) 597},
  \href{http://arxiv.org/abs/hep-th/0309208}{{\ttfamily arXiv:hep-th/0309208}}.

\bibitem{Iqbal:2003ds}
A.~Iqbal, N.~Nekrasov, A.~Okounkov, and C.~Vafa, ``{Quantum foam and
  topological strings},''
  \href{http://dx.doi.org/10.1088/1126-6708/2008/04/011}{{\em JHEP} {\bfseries
  04} (2008) 011}, \href{http://arxiv.org/abs/hep-th/0312022}{{\ttfamily
  arXiv:hep-th/0312022}}.

\bibitem{MR2836398}
K.~Nagao and H.~Nakajima, ``Counting invariant of perverse coherent sheaves and
  its wall-crossing,'' \href{http://dx.doi.org/10.1093/imrn/rnq195}{{\em Int.
  Math. Res. Not. IMRN} {\bfseries 2011} no.~17, (2011) 3885--3938},
  \href{http://arxiv.org/abs/0809.2992}{{\ttfamily arXiv:0809.2992 [math.AG]}}.

\bibitem{MR2592501}
S.~Mozgovoy and M.~Reineke, ``On the noncommutative {D}onaldson-{T}homas
  invariants arising from brane tilings,''
  \href{http://dx.doi.org/10.1016/j.aim.2009.10.001}{{\em Adv. Math.}
  {\bfseries 223} no.~5, (2010) 1521--1544}.

\bibitem{Yamazaki:2010fz}
M.~Yamazaki, ``{Crystal Melting and Wall Crossing Phenomena},''
  \href{http://dx.doi.org/10.1142/S0217751X11051482}{{\em Int. J. Mod. Phys.}
  {\bfseries A26} (2011) 1097--1228},
\href{http://arxiv.org/abs/1002.1709}{{\ttfamily arXiv:1002.1709 [hep-th]}}.

\bibitem{OEIS}
``Sequence {A000293} at the {OEIS}.''
\newblock \url{https://oeis.org/A000293}.

\bibitem{Galakhov:2020vyb}
D.~Galakhov and M.~Yamazaki, ``{Quiver Yangian and Supersymmetric Quantum
  Mechanics},'' \href{http://dx.doi.org/10.1007/s00220-022-04490-y}{{\em
  Commun. Math. Phys.} {\bfseries 396} no.~2, (2022) 713--785},
  \href{http://arxiv.org/abs/2008.07006}{{\ttfamily arXiv:2008.07006
  [hep-th]}}.

\bibitem{Franco:2023tly}
S.~Franco, ``{4d Crystal Melting, Toric Calabi-Yau 4-Folds and Brane Brick
  Models},'' \href{http://arxiv.org/abs/2311.04404}{{\ttfamily arXiv:2311.04404
  [hep-th]}}.

\bibitem{Kenyon:2003uj}
R.~Kenyon, A.~Okounkov, and S.~Sheffield, ``{Dimers and amoebae},''
  \href{http://arxiv.org/abs/math-ph/0311005}{{\ttfamily
  arXiv:math-ph/0311005}}.

\bibitem{Dijkgraaf:2008ua}
R.~Dijkgraaf, D.~Orlando, and S.~Reffert, ``{Quantum Crystals and Spin
  Chains},'' \href{http://dx.doi.org/10.1016/j.nuclphysb.2008.11.027}{{\em
  Nucl. Phys. B} {\bfseries 811} (2009) 463--490},
  \href{http://arxiv.org/abs/0803.1927}{{\ttfamily arXiv:0803.1927
  [cond-mat.stat-mech]}}.

\bibitem{Prochazka:2015deb}
T.~Proch\'azka, ``{$ \mathcal{W} $ -symmetry, topological vertex and affine
  Yangian},'' \href{http://dx.doi.org/10.1007/JHEP10(2016)077}{{\em JHEP}
  {\bfseries 10} (2016) 077}, \href{http://arxiv.org/abs/1512.07178}{{\ttfamily
  arXiv:1512.07178 [hep-th]}}.

\bibitem{Closset:2017yte}
C.~Closset, J.~Guo, and E.~Sharpe, ``{B-branes and supersymmetric quivers in
  2d},'' \href{http://dx.doi.org/10.1007/JHEP02(2018)051}{{\em JHEP} {\bfseries
  02} (2018) 051}, \href{http://arxiv.org/abs/1711.10195}{{\ttfamily
  arXiv:1711.10195 [hep-th]}}.

\bibitem{Gadde:2013lxa}
A.~Gadde, S.~Gukov, and P.~Putrov, ``{(0, 2) trialities},''
  \href{http://dx.doi.org/10.1007/JHEP03(2014)076}{{\em JHEP} {\bfseries 03}
  (2014) 076}, \href{http://arxiv.org/abs/1310.0818}{{\ttfamily arXiv:1310.0818
  [hep-th]}}.

\end{thebibliography}\endgroup

\end{document}